\documentclass[11pt]{iopart}
\usepackage[dvips]{graphicx,psfrag}

\expandafter\let\csname equation*\endcsname\relax
\expandafter\let\csname endequation*\endcsname\relax
\usepackage{amssymb}
\usepackage{mathrsfs}
\usepackage{amsmath,amsfonts}
\usepackage{iopams}
\usepackage{color}
\input epsf

\definecolor{Blue}{rgb}{0.3,0.3,0.9}
\definecolor{Red}{rgb}{1,0,0}
\definecolor{Green}{rgb}{0,1,0}

\newcommand{\be}{\begin{equation}}
\newcommand{\ee}{\end{equation}}
\newcommand{\bea}{\begin{eqnarray}}
\newcommand{\eea}{\end{eqnarray}}
\newcommand{\lb}{\label}
\newcommand{\sign}{\mathrm{sign}}
\newcommand{\xmax}{x_{\mathrm{max}}}
\newcommand\mub{\bar{\mu}}
\newcommand\sQ{\sigma_{Q}^2}
\newcommand{\lpl}{\ell_{\mathrm{Pl}}}
\newcommand{\epsi}{\epsilon}
\newcommand{\Hc}{\mathscr{H}}

\def\be{\nopagebreak[3]\begin{equation}}
\def\ee{\end{equation}}
\def\ba{\nopagebreak[3]\begin{eqnarray}}
\def\ea{\end{eqnarray}}

\def\lp{\ell_{\rm Pl}}

\def\f{\frac}

\def\rcr{\rho_{\rm max}}

\def\dd{{\rm d}}

\def\e{\mathfrak{e}}

\def\a{\mathfrak{a}}

\def\K{\mathfrak{K}}

\def\S{\mathcal{S}}

\def\b{{\rm b}}
\def\B{{}_{\rm B}}
\def\L{{\cal L}}

\def\d{{\rm d}}

\def\e{\mathring{e}}

\def\ow{\mathring{\omega}}
\def\Tr{{\rm Tr\,}}

\def\l{\lambda}

\newcommand{\bark}{\bar{k}}
\newcommand{\barp}{\bar{p}}

\newcommand{\ff}[1]{(\ref{#1})}
\usepackage{ulem}

\begin{document}

\topical[Observational issues in LQC]{Observational issues in loop quantum cosmology}

\author{A. Barrau$^{1,2}$, T. Cailleteau$^3$, J. Grain$^{4,5}$ and J. Mielczarek$^{6,7}$}

\address{$^1$ Laboratoire de Physique Subatomique et de Cosmologie, UJF, CNRS/IN2P3, INPG, 53, av. des Martyrs, 38026 Grenoble cedex, France}
\address{$^2$ Institut des Hautes Etudes Scientifiques, Le Bois-Marie  35 route de Chartres, 91440 Bures-sur-Yvette, France}
\address{$^3$ Institute for Gravitation \& the Cosmos, Penn State, University Park, PA 16802, USA}
\address{$^4$ Institut d'Astrophysique Spatiale, UMR8617, CNRS, Orsay, France, F-91405}
\address{$^5$ Universit\'e Paris-Sud 11, Orsay, France, F-91405 }
\address{$^6$ Institute of Physics, Jagiellonian University, Reymonta 4, 30-059 Cracow, Poland}
\address{$^7$ Department of Fundamental Research, National Centre for
Nuclear Research,\\ Ho{\.z}a 69, 00-681 Warsaw, Poland}

\eads{\mailto{aurelien.barrau@cern.ch}, \mailto{thomas@gravity.psu.edu}, \mailto{julien.grain@ias.u-psud.fr} and \mailto{jakubm@fuw.edu.pl}}


\begin{abstract}
Quantum gravity is sometimes considered as a kind of metaphysical speculation. 
In this review, we show that, although still extremely difficult to
reach, observational signatures can in fact be expected. The early universe is an
invaluable laboratory to probe ``Planck scale physics". Focusing on Loop
Quantum Gravity as one of the best candidate for a non-perturbative and background-independant
quantization of gravity, we detail some expected features.\\

Invited topical review for {\it Classical and Quantum Gravity}.
\end{abstract}


\tableofcontents
\maketitle

\section{Introduction}

Building a quantum theory of gravity --that is a quantum model of spacetime itself-- is often
considered as the most outstanding problem of contemporary physics. It is usually thought this way
because  most scientist consider this is the unavoidable horizon for unification. This might not be that clear.
Unquestionably, unification has been an efficient guide for centuries: it worked with Kepler,
with Maxwell, with Glashow, Salam and Weinberg. Yet, is the World more ``unified" today than at the
end of the nineteenth century? Let's consider the macrocosm: stars, planets, comets, dust,
cosmic-rays, pulsars, quasars, white dwarfs, black holes, magnetohydrodynamics turbulence, galaxy
collisions... Is this a ``unified" firmament? Naturally, one should better consider the microcosm.
However, there are today about 120 degrees of freedom in the standard model (not even mentioning
possible supersymmetric ones), which is slightly {\it more} than the number of atoms in the old Mendeleev periodic
table. Of course, it might be more relevant to consider fundamental interactions instead of the 
matter content. But, once
again, grand unification is still missing and, to account for the accelerated expansion of the
universe, many cosmologists rely on a quintessence scenario. (This is obviously not the only
possibility for a true cosmological constant as advocated in \cite{carlo1}, is even possible although
the numerical value doesn't fit quantum mechanical expectations.) Quintessence means 
{\it quintus essentia}, that is ``fifth force". We will resist the temptation to elaborate here on the
string theory landscape \cite{land} which, interestingly, exhibits an 
unprecedented diversity within the realm of a tentative fully unified model. (For a more
philosophical investigation of diversity and unification in physics, one can refer to
\cite{aurel}.) The roads toward unification are probably much more intricate that usually thought: 
they might very well be organized as a kind of {\it rhizome}; refering here to the
philosophical concept of the ``image of thought", as suggested, within the so-called {\it French
Theory}, in \cite{deleuze}.

Does this mean that the idea of quantum gravity itself should be forgotten? After all, this has been
shown to be such an incredibly difficult theory to establish that the wise attitude could just 
be to withdraw from this apparently never-ending quest. Unfortunately, it is impossible to 
ignore the ``gravity-quantum" tension. The first reason is that the quantum world interacts with
the gravitational field. This, in itself, requires gravity to be understood in a quantum
paradigm, as can be demonstrated by appropriate thought experiments \cite{oriti}, though a possible way out is 
still conceivable in the framework of ``emergent" or
``entropic" gravity \cite{ver,jac}. The second reason is that, the other way round, gravity
requires quantum field theories to live in curved spaces. Just because of the equivalence
principle, it is easy to get convinced \cite{birell} that this cannot be
rigorously implemented without quantum general relativity. Basically speaking, the nonlinearity of gravity
frustrates all attempts to ignore quantum gravity: each time a strong gravitational field is involved the coupling
to gravitons should also be strong. The third reason is the existence of singularities: general
relativity predicts, by itself, its own breakdown (as, exhibited, {\it e.g.}, by the Penrose-Hawking theorems \cite{singul}). 
This is a truly remarkable feature. Although the first two reasons can, to some extent, be considered
as ``heuristic" motivations, the last one does imperatively requires a way out. Pure general relativity
ontologically fails. 

It is sometimes said either that we have too many candidate theories \cite{kiefer} for quantum gravity or that 
we do not have a single (convincing) one \cite{rovelli}. Although apparently contradictory, both statements
are in fact correct. This is a quite specific situation: many different approaches are investigated, all are promising, 
but none is fully consistent. At this stage, experiments are obviously missing to eradicate those theories
that are deeply on the wrong track and to improve those that might be correct. Unfortunately, 
quantum gravity is known to be out of reach of any possible experiment, recalling that the ratio of the Planck scale
to the LHC scale is roughly the same than the ratio of the human scale to the distance to the closest
star. We shall now underline that this might not be true. 

\section{The ideas of Loop Quantum Gravity}

In this section, we assume that the reader is unfamiliar with Loop Quantum Gravity (LQG). There are
several good introductions available (see {\it e.g.}, \cite{rovelli,lqg_review}). Here, we just try to
give a flavor -- heavily based on \cite{intro} -- of the basic ideas underlying LQG. We will go neither
into the details nor into modern spinfoam formulations. The results won't be demonstrated. We focus on
the concepts and main guidelines so that a cosmologist knowing nothing about LQG can understand
something about what the theory looks like. 

Basically, LQG is a tentative non-perturbative and background invariant quantization of General Relativity (GR). 
It does not require strong assumptions like supersymmetry or extra-dimensions. It is not a ``theory of everything", 
just a candidate theory of quantum geometry addressing, in particular, the issue of Planck-scale physics in the 
gravitational sector.

\subsection{Reminder on Hamiltonian mechanics}

The Hamiltonian formulation treats space and time in a different way. This is exactly what one does
not want when dealing with an explicitly covariant theory like GR. And this is precisely why we do not
learn GR in its Hamiltonian form at school. However, the Hamiltonian formalism is known for being the
royal road toward quantization. This is therefore probably the best framework to investigate quantum
gravity. Although the Hamiltonian formulation seems not to treat space and time on a equal footing, one
should of course not worry too much: the correct writing of GR in this formalism automatically
"compensates" for this and leads to results insensitive to the actual choice of a timelike direction.\\

\begin{figure}
\begin{center}
\includegraphics[width=110mm]{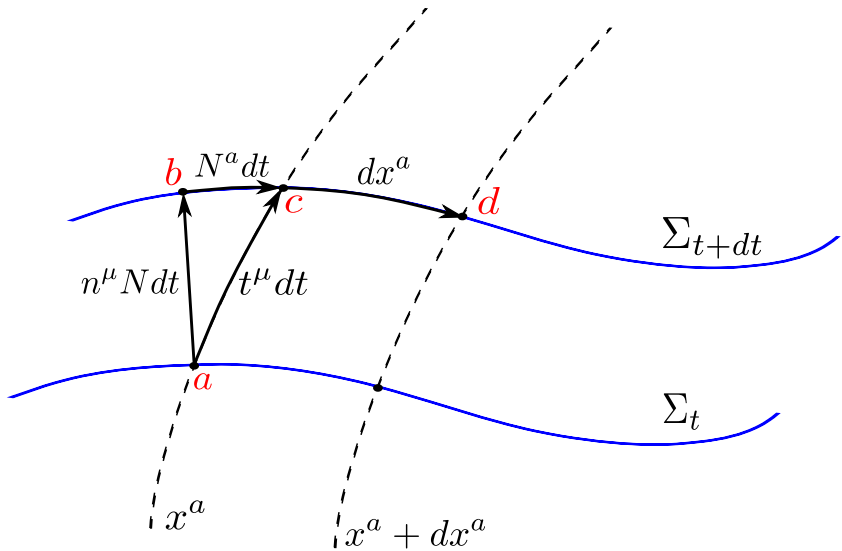}
\caption{Graphical interpretation of the lapse function $N$ and shift vector $N^a$.}
\label{fig1}
\end{center}
\end{figure}

The so-called ADM (Arnovitt-Deser-Misner) decomposition of GR splits the hyperbolic manifold in time and 
space parts and can be understood as a foliation of spacetime in space slices. The line elements reads as
\begin{equation}
ds^2=-(Ndt)^2+q_{ab}(N^adt+dx^a)(N^bdt+dx^b).
\end{equation}
The quantities $N$ and $N^a$ are known as the laps function and the shift vector. Although the detailed counting 
for a constrained theory turns out to be more subtle than this, we obviously have here as many degrees of 
freedom as in the 10 functions of the symmetric space-time metric $g_{\mu\nu}$: 1 quantity N,
3 quantities $N^a$ and 6 quantities for the symmetric space metric $q_{ab}$. The intuitive meaning of the lapse and 
shift is shown on Fig. \ref{fig1}. Let $\Sigma_t$ and $\Sigma_{t+dt}$ be a space hypersurface taken at times $t$ and 
$t+dt$. Let $n^{\mu}$ be a unit vector normal to $\Sigma_t$. Basically, if we start at a point $a=(t,x^a)$ on 
$\Sigma_t$, moving by $n^{\mu}Ndt$ leads to a point $b$ on $\Sigma_{t+dt}$ which, when shifted by $N^adt$ leads to $c$, 
that is the "time evolution" of $a$. We will see later that the lapse and shift turn out to be arbitrary functions, 
reflecting some fundamental freedom we have in the theory.

Although not very often met in GR outside of this context, one can define the extrinsic curvature of the surface. 
Intuitively, this is the curvature that it would have as seen from a higher dimensional euclidean manifold. 
For example, a cylinder has a vanishing intrinsic curvature but has obviously a nonzero extrinsic curvature. 
The other way round, a map of the Earth as drawn on a sheet of paper has a vanishing extrinsic curvature but has 
an intrinsic curvature. The extrinsic curvature is the Lie derivative of the space metric which is, 
itself, closely related to the time derivative of the space metric. In the Hamiltonian formulation of GR, 
the configuration variable will be the space metric  $q_{ab}$  and its conjugate momentum will be 
(closely related to) the extrinsic curvature $K_{ab}$.\\

In classical mechanics, one has a Lagrangian $L(q_i(t),\dot{q}_i(t))$ from which it is possible to define canonical 
momenta:
\begin{equation}
p_i(t)\doteq\frac{\partial L}{\partial \dot{q}_i}.
\end{equation}
The hamiltonian is defined by
\begin{equation}
H(q_i,p_i)=\sum_i p_i\dot{q}_i-L.
\end{equation}
The phase space is the space of pairs $(p_i,\dot{q}_i)$ while the space of $\dot{q}_i$ is the configuration space. 
The dynamics of the system is given  by:
\begin{equation}
\dot{q}_i=\frac{\partial H}{\partial p_i},~
\dot{p}_i=-\frac{\partial H}{\partial q_i},
\end{equation}
which can also be written as
\begin{equation}
\dot{q}_i=\left\{q_i,H\right\},~
\dot{p}_i=\left\{q_i,H\right\},
\end{equation}
with
\begin{equation}
\left\{f,g\right\}=\sum_i\frac{\partial f}{\partial q_i}\frac{\partial g}{\partial p_i}-
\frac{\partial f}{\partial p_i}\frac{\partial g}{\partial q_i}.
\end{equation}
This latter quantity is called the Poisson bracket. The important points are,
first, that
\begin{equation}
\left\{q_i,p_i\right\}=\delta_{ij},
\end{equation}
and, second, that the equations of evolution can be rewritten as
\begin{equation}
\dot{q}_i=\left\{q_i,H\right\},~
\dot{p}_i=\left\{p_i,H\right\}.
\end{equation}

In real life, many systems are constrained. It means that we use more variables 
than really needed. The best known example is the electromagnetic field which has 
only two degrees of freedom in vacuum whereas it is usually described either with 
a 4-components vector potential of a 6-components rank 2 symmetric tensor $F^{\mu\nu}$. In GR 
also, hidden constraints would decrease the 10 metric components to 2 independent
degrees of freedom in the vacuum. The constraints are fundamental relations 
between the variables: $f(p_i,q_i)=0$. This also means that $f$ is conserved (necessarily
due to an underlying symmetry). Constraints are handled by Lagrange multipliers. If
a system is described by $N$ canonical variables and $M$ constraints, one can add
to the Hamiltonian the constraints multiplied by Lagrange multipliers, leading to
the so-called "total Hamiltonian":
\begin{equation}
H_{tot}=H_{init}+\sum_{i=1}^M\lambda_if_i.
\end{equation}
The $N$ equations of motion and $M$ constraints define the dynamics of the system.
The
solutions found for given initial conditions $(q_i(0),p_i(0))$ will differ
depending on the arbitrary values of the Lagrange multipliers. But those solutions
will be gauge equivalent, related by a symmetry of the system. This framework can
be straightforwardly extended to field theories by considering  functional
derivatives and Poisson brackets between conjugate variables now typically written 
as
\begin{equation}
\left\{A_i(x),E^j(y)\right\}=\delta^j_i\delta(x-y).
\end{equation}

In gravity, the situation is a bit tricky because the Hamiltonian vanishes. This is
a totally constrained system. In such a system, there is no, strictly speaking,
time evolution. The dynamical variables are instead subject to a constraint and the
canonical transformations generated by this constraint are interpreted as gauge.
Points in the phase space related by canonical transformations generated by the
constraint are therefore physically equivalent. The curve joining those points is called a
gauge orbit and represents a true position in phase space. Physical observables are
such that they are constant along gauge orbits ({\it i.e.} have a vanishing Poisson
bracket with the constraint). Understanding how such a frozen dynamics can lead to
evolution is not easy. The main idea is to use a variable as an effective time with
respect to which gauge orbits are described in a relational way. 

\subsection{GR in Ashtekar variables}

The action of GR reads as
\begin{equation}
S=\frac{1}{2\kappa}\int \sqrt{-g} R d^4x.
\end{equation}
Going through the calculation shows that $g^{00}$ and $g^{0i}$ appear in the
expression of the Ricci scalar without time derivatives. This means that they are
in fact Lagrange multipliers related to the lapse and shift by $g=g^{00}=1/N^2$ and
$g^{0i}=N^i/N^2$. As in electrodynamics, the Lagrange multipliers are associated
with constraints. The constraints associated with the shift form a vector whose
flow is related to spatial diffeomorphisms. The constraint associated with
the lapse (called Hamiltonian constraint) reflects the possible deformations of the
spatial surface. There are six configuration degrees of freedom (the components of the
symmetric space metric) and four constraints: the theory has two degrees of
freedom. In LQG, one uses in fact different variables. At this stage they  just
allow a mathematically equivalent rewriting of the theory. The first set of such
variables are the densitized triads $E_i^a$. They are local vector fields
related to the metric by $\det(q)q^{ab}=E_i^aE_j^b\delta^{ij}$.
Intuitively, one can see them as the "square root" of the space metric. The second
set are $SU(2)$ connections $A_a^i$ related to the spin connections $\Gamma_a^i$ by
$A_a^i=\Gamma_a^i+\gamma K_a^i$, $\gamma$ being a free parameter called the
Barbero-Immirzi parameter. Its (non-vanishing) value does not affect the classical theory. 
The spin connections themselves are just the
objects used to define consistently the covariant derivative for objets having an
internal index, as require by the concept of triads (the $i$ and $j$ indices are
internal indices). Technically, if one defines the spin connection by
\begin{equation}
D_a V^i=\partial_q V^i + \Gamma_{aj}^iV^j,
\end{equation}
then $\Gamma_a^i=\Gamma_{ajk}\epsilon^{jki}$, and $K^i_a=\frac{K_{ab}E^{bi}}{\sqrt{\det(q)}}.$
Those two sets of variables are conjugate:
\begin{equation}
\left\{A_i^a(x),E^b_j(y)\right\}=8\pi G \gamma \delta^a_b
\delta^i_j\delta(x-y).
\end{equation}
The Lagrangian of GR then reads as
\begin{equation}
L=\int_{\Sigma} d^3x\left( \frac{1}{8\pi G \gamma}E^a_i\dot{A}^i_a-NC-N^aC_a-N^iC_i \right),
\label{lag}
\end{equation}
where 
\begin{eqnarray}
C &=& \frac{1}{16\pi G} \frac{E^a_iE^b_j}{\sqrt{|\det E|}} \left[ {\varepsilon^{ij}}_k F_{ab}^k -
2(1+\gamma^2)K^i_{[a} K^j_{b]} \right], \\
C_a &=& \frac{1}{8\pi G \gamma}(E^b_i F^i_{ab} - A^i_a C_i) \label{wiezDiff1}, \\
C_i &=&\frac{1}{8\pi G \gamma}\mathcal{D}_a E^a_i 
= \frac{1}{8\pi G \gamma} \left(\partial_aE^a_i+\epsilon_{ijk} A^j_a E^a_k \right),\\
\end{eqnarray}
and $F^i_{ab} =\partial_a A^i_b -\partial_b A^i_a+{\epsilon^i}_{jk} A^j_a A^k_b$ is the
curvature of the Ashtekar connection.

The important point here is that it is now clear that $E^a_i$ and $A^i_a$
are canonically conjugate and that the lapse $N$, the shift $N^a$ and the gauge
parameters $N^i$ are Lagrange multipliers. The terms multiplying the Lagrange
multipliers are called respectively the Hamiltonian constraint ($C$), the diffeomorphism
constraint ($C_a$) and the Gauss constraint ($C_i$). One can instead of the 
diffeomorphism constraint consider the momentum constraint which does not include 
contribution from the Gauss constraint. 

The total Hamiltonian of GR, $H_{G}[N^i,N^a,N]$ is therefore a combination of those constraints times Lagrange
multipliers without any other term, as can immediately be seen by performing a Legendre 
transformation of Eq. (\ref{lag}). So, as announced before, the Hamiltonian vanishes and 
gravity is a totally constrained system. Written with the Ashtekar variables, GR is a specific 
kind of $SU(2)$ gauge theory. This makes possible the use of quite a lot of the techniques 
developed to deal with Yang-Mills theories in standard quantum field theory.

In what follows it will be useful to work also with smeared constraints: 
\begin{eqnarray}
\mathcal{C}_1 &=& G[N^i] = \int_{\Sigma}d^3x\ N^i C_i, \\
\mathcal{C}_2 &=& D[N^a] =  \int_{\Sigma}d^3x\ N^a C_a, \\
\mathcal{C}_3 &=& S[N] =  \int_{\Sigma}d^3x\ N C,
\end{eqnarray}
that are such that  $H_{G}[N^i,N^a,N] = G[N^i] +D[N^a]+S[N]$. 
Because the Hamiltonian is weakly vanishing, $H_{G}[N^i,N^a,N] \approx 0$ at all times, 
the time derivative of the Hamiltonian constraint is also weakly vanishing, $\dot{H}_{G}[N^i,N^a,N] \approx 0$. 
The Hamilton equation $\dot{f}=\{f,H_{G }[M^i,M^a,M]\}$ therefore leads to
\begin{equation}
\left\{H_{G }[N^i,N^a,N], H_{G }[M^i,M^a,M]\right\} \approx 0, \label{HH}
\end{equation}
which, when explicitly written, means:
\begin{equation}
\left\{ G[N^i] +D[N^a]+S[N], G[M^i] +D[M^a]+S[M] \right\} \approx 0. \nonumber
\end{equation}
Due to the linearity of the Poisson bracket, one can straightforwardly find that the condition 
(\ref{HH}) is fulfilled if the smeared constraints belong to a first class algebra
\begin{equation}
\{ \mathcal{C}_I, \mathcal{C}_J \} = {f^K}_{IJ}(A^j_b,E^a_i) \mathcal{C}_K. \label{algebra}
\end{equation}
Because structure functions ${f^K}_{IJ}( A^j_b,E^a_i)$ dependent on the phase space 
(Ashtekar) variables  $(A^j_b, E^a_i)$, the algebra (\ref{algebra}) is therefore not a Lie 
algebra but rather a Lie algebroid. At the surface of the Gauss constraint $G[N^i] \approx 0$, 
the algebra of constraints (\ref{algebra}) reduces to 
\begin{eqnarray}
\left\{D[M^a],D [N^a]\right\} &=& D[M^b\partial_b N^a-N^b\partial_b M^a],  \\
\left\{D[M^a],S[N]\right\} &=& S[M^a\partial_a N-N\partial_a M^a],  \\
\left\{S[M],S[N]\right\} &=& s D\left[ q^{ab}(M\partial_bN-N\partial_bM)\right], 
\end{eqnarray} 
where $s$ is the spacetime metric signature ($s=1$ for Lorentzian signature and $s=-1$ for 
Euclidean signature). The requirement of closure of the algebra of constraints will be an important
guiding principle while taking into account quantum effects. 

\subsection{The ideas of the quantization}

The historic way to quantize GR is to consider the metric -- as the configuration variable --  and its
conjugate momentum, which is closely linked with the extrinsic curvature. Then, one constructs a wave functional
of the metric $\Psi(q_{ab})$ and hope to be able to define in a consistent way the probability of observing a given geometry.
With the Ashtekar variables, an equivalent procedure would be to consider a wave functional of the connection
$\Psi(A^i_a)$ and then to promote $A^i_a$ to be a multiplicative operator acting on $\Psi$. In this case,
$E^a_i$ would naturally be a derivative
operator. Although a reverse situation (as $A^i_a$ is associated with the extrinsic curvature and not with the metric)
this is very reminiscent to the usual Wheeler-deWitt approach.

As expected, the known problems of Wheeler-deWitt quantization of GR are still present at this stage. Without
going into the details, let us mention the inability to have a good control over the space of solutions
of the Gauss and diffeomorphism (quantum) constraints, the intricate geometrical interpretation of the 
Hamiltonian constraint, and, more importantly, the difficulty to define an inner product. The inner product 
is the key ingredient of the Hilbert space and is mandatory to compute transition probabilities.\\

Loops enter the game at this stage. To overcome those difficulties, it is convenient to go to the loop 
representation. Loops are constructed with holonomies. Holonomies are just parallel propagators along 
closed curves. And a parallel propagator is simply a notion which allows one to move ``as parallel as 
possible'' so as to generalize the concept of the circulation of a vector. One can show that it is a matrix 
which can be written as a "path ordered exponential", that is, if one considers $A_a$ over $\gamma(s)$: 
\begin{equation}
P\left[e^{\int_\gamma \dot{\gamma}^a(s)A_a(s)}\right],
\end{equation}
where the $P$ term means that factors are taken with larger values of $t$ appearing first.

An important theorem states that if one determines the trace of the holonomy 
of a connection for all loops on a manifold, then one determines all the gauge 
invariant information of the connection. It means that traces of the holonomies 
can be used as a basis on which a state $\Psi\left[A\right]$ can be expanded. 
Working with the coefficients of this development is known as working in the loop representation. \\

Loops are not pulled out of a hat. They are rooted in the deep groundings of Maxwell theory of electromagnetism. 
Of course the road is long from Faraday loops to Wilson loops but the main idea remains the same. In addition, 
they are particularly natural when dealing with gravity. The very notion of curvature implies the circulation over 
a closed loop so as to be measured. Remember the way the Riemann tensor is defined in GR. In addition, for 
technical reasons that will not be detailed here, the main difficulties associated with the old quantization scheme 
are indeed solved in loop quantum gravity.\\

The Ashtekar connection is $SU(2)$ valued. Of course, there are representations of $SU(2)$ in all dimensions. 
One can choose any representation to construct a connection and parallel transport it along a curve. The result 
is a matrix. If two curves carrying different representations intersect at one point, the indices can be tied up thanks 
to an object called ``intertwiner''. The resulting structure is a spin network, that is a graph with colored lines (the 
color corresponds to the dimensionality of the representation and therefore of the matrix used to parallel transport) 
and intersections.\\

\begin{figure}
\begin{center}
\includegraphics[width=90mm]{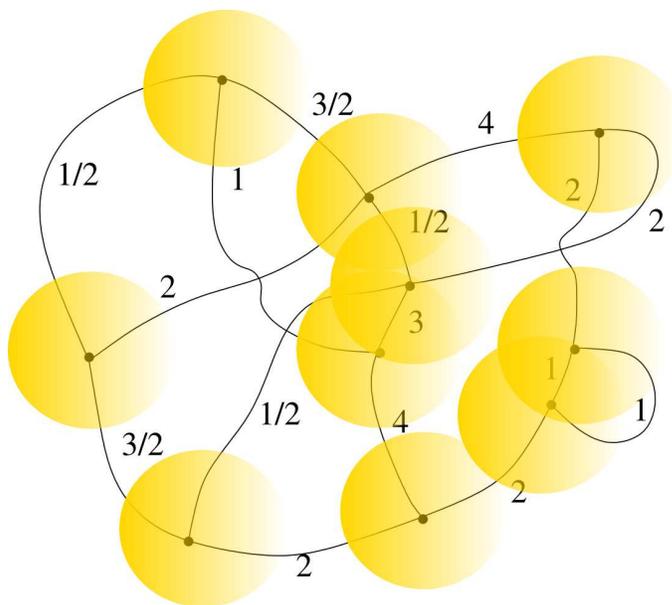}
\caption{Example of a spin network. The (yellow) balls represent quanta (grains) of space.} 
\end{center}
\end{figure}

This historical presentation is probably not the best possible one as covariant approaches are now better developed 
and more precisely defined. The key point is that they basically lead to the very same LQG theory. 
The spin-network Hilbert space is the same than the usual Hilbert space of a $SU(2)$ lattice Yang-Mills theory. 
The kinematics of GR can be  cast into the same form as the kinematics of $SU(2)$ Yang-Mills theory. The other way 
round, the Hilbert space of a SU(2) Yang-Mills lattice theory admits an interpretation as a description of quantized
geometries, formed by quanta of space. This interpretation is the fundamental content of the Penrose ``spin-geometry''
theorem (which generalizes an earlier theorem by Minkowski). This grounds the kinematics of LQG.
The most remarkable consequence is the discreteness of the geometry at the Planck scale, which appears here as a rather
conventional quantization effect: In GR, the gravitational field determines lengths, areas and volumes. Since the
gravitational field is a quantum operator, these quantities are given by quantum operators. Planck scale discreteness 
follows from the spectral analysis of those operators. The transition amplitudes are clearly defined and this makes
the theory expressed in a fully consistent quantum way (see Rovelli's Zakopane lectures in \cite{lqg_review}).

\section{Loop quantum cosmology: the general framework}
\label{sec:lqcgen}

\subsection{Basic ideas}

There are many good reviews available also on loop quantum cosmology (see, {\it e.g.}, \cite{lqc_review} and 
\cite{Calcagni:2012vw}). We, once again, just give here a flavor of the main con taint of the theory. If assumed, 
as supported by many observations, to be homogeneous and isotropic, the Universe is one of the easier system 
to describe. At large scales, it is extremely symmetric and is therefore one of the simplest possible objects. Why 
should we apply quantum gravity to the Universe whereas general relative describes it so well ? After all, it is 
highly probable that quantum gravity cannot help much solving the two main problems of cosmology: dark matter 
and dark energy. This is correct but this point of view forgets about another big -if not bigger- problem: the Big 
Bang itself. The Big Bang is a singularity. This is probably not a pathology of space-time itself but a pathology 
of the theory which describes it. General relativity tells us the history of the Universe, beginning by an event 
which is inconsistent. The story is neither complete nor reliable. This is precisely where quantum gravity is 
expected to play a role.\\

In principle, it would be necessary to find quantum states allowed by LQG and, then, to search for those that 
are (nearly) homogenous and isotropic. This would be the rigorous procedure but this is extremely difficult. 
Another way to proceed, much easier, is to first reduce the classical theory to homogeneity and isotropy and 
then quantize. In such a case, known as the mini-superspace approximation, the number of degrees of freedom 
is finite. Nothing ensures from the beginning that the results derived in this scheme will be correct. This is 
however a promising and usual approximation. In this framework, the full method of LQG does not apply directly. 
The idea is instead to introduce techniques that mimic some some important behaviors of LQG. As those 
techniques do violate some of the hypotheses of the Stone-Von Neumann uniqueness theorem, results different 
from the well-known conclusions of the Wheeler-de Witt quantum cosmology can be expected. This is fortunate 
as the  Wheeler-de Witt equation does not solve generically the Big Bang singularity.\\

Due to the symmetries, the Ashtekar variables can be chosen diagonal and written as
\begin{equation}
A^i_a=c\delta^i_a,~
E^a_i=p\delta^a_i.
\end{equation}
Here, $c$ and $p$ become the relevant conjugate variables. The first one is related with the scale factor 
($a^2=p$) and the second one is related with the Hubble parameter ($c$ is proportional to $\dot{a}$). 
The variables $c$ and $p$ are related by the symplectic structure of the theory (with the normalization 
chosen to agree with most LQC articles):
\begin{equation}
\left\{c,p\right\}=8\pi G \gamma /3.
\end{equation}
One can then add matter and write down the full Hamiltonian constraint. Through the evolution equations 
$\dot{p}=\left\{p,H\right\}$, one can easily recover, at the classical level, the Friedmann equation
\begin{equation}
H^2=\frac{8\pi G}{3}\rho.
\end{equation}
This is, as expected, in exact agreement with standard GR as, at this stage, only a different writing of GR 
has been used.\\

A first possible quantization is the Wheeler-De Witt one. The variables of the system, spanning the phase 
space, are $(c,p,\varphi, p_\varphi)$, where $\varphi$ is the matter field assumed to fill the Universe and 
$p_\varphi$ is its conjugate momentum. To quantize, one promotes $c$ and $p$ to be operators. The operator 
$\hat{c}$ acts as multiplication and the operator $\hat{p}$ acts as derivation. The Wheeler-De Witt equation 
is nothing else than the quantum version of the Hamiltonian constraint. This is a consistent solution but the 
uncertainty remains small all the way to the Big Bang and quantum effects do not generically remove the 
singularity. This theory remains, in general, singular.\\

Another possibility is precisely the LQC approach. Here, the operators are given by the holonomy of the 
connection (there is no operator associated with the connection) and the flux of the triads. As space is 
homogeneous, one can consider a fiducial cell that can be infinitely repeated. One can then construct the 
holonomy of $c$ along a loop around the cell. It is basically given by $h=\exp(i\lambda c)$. A kinematical 
Hilbert space in introduced so that $\hat{p}$ acts as a multiplicative self-adjoint operator. A quantum gravitational 
state can be expanded on the $|p_i\rangle$ basis which is orthonormal. A key point is that this space is 
not endowed with the usual inner product on square integrable function (one deals here with a Kronecker 
delta in the product between states, not with the Dirac delta). This looks like what happens on the usual 
loop states of LQG. The operators act as:
\begin{equation}
\hat{p}|p\rangle=p|p\rangle,~ \hat{h}|p\rangle=|p+\lambda\rangle.
\end{equation}
As $|p+\lambda\rangle$ is orthogonal to $|p\rangle$ for all non-vanishing $\lambda$, 
it is not possible to take the derivative of $\hat{h}$ with respect to $\lambda$, 
it is not even continuous. This is the hypothesis of the Stone-Von Neumann theorem 
which is violated and this is why some highly non-trivial features arise in LQC. \\

Although requiring some algebra, it is not difficult to write down the classical Hamiltonian constraint in the 
cosmological framework. Unfortunately, this expression cannot be turned into a quantum operator straightforwardly 
as there is precisely no operator associated with $c$. This is why another expression is usually used where 
the substitution $c \rightarrow \sin(\mu c)/\mu$ is performed. This coincide with the usual expression when 
$\mu \rightarrow 0$. This expression is meaningful as it can be representer as an operator: the sine function 
can be expanded in terms of holonomies. This substitution is called the holonomy correction and is one of the 
most important effective LQC correction. Another one, that will be introduced later is called the inverse-volume 
correction. Of course, it is not possible to take the full limit $\mu \rightarrow 0$ in the quantum theory as this 
would correspond to shrink the holonomy to a point. If LQC is to really account for LQG one can expect that 
the smallest possible value corresponds to the one of the first eigenvalue of the area operator of LQC.\\

If one makes this substitution ($c \rightarrow \sin(\mu c)/\mu$) in the classical Hamiltonian, one is left with an 
effective holonomy corrected LQC Hamiltonian. By writing $\dot{p}=\left\{p_i,H_{\text{eff}}\right\}$, it is not long to obtain:
\begin{equation}
H^2=\frac{8\pi G}{3}\rho\left(1-\frac{\rho}{\rho_{\text{c}}}\right),
\end{equation}
with where $\rho_{\text{c}}$ is if of order of the Planck density. This equation clearly shows that the Big Bang 
singularity is solved and replaced by a Big Bounce: when $\rho=\rho_{\text{c}}$, the Hubble parameter vanishes 
and changes sign. Bounces also appear in different scenarios, see  \cite{bounces} for examples.

\subsection{With slightly more details}

Without going into the actual calculations we give here a few more details, following \cite{agullo}. 

The classical theory is described by four variables. In the gravitational sector they are $c$ and $p$, so 
that $c=\gamma \dot{a}/N$, and $|p|=a^2$. In the matter sector, we still assume that the content is 
dominated by a scalar field described by its value and its momentum. As explained previously, when 
quantizing, the kinematical Hilbert space of LQC is inequivalent to the Wheeler-deWitt one. The quantization 
scheme, as close as possible to the one developed in LQG, is called polymeric. Importantly, this choice 
was shown to be unique when diffeomorphism invariance is taken seriously \cite{campi}. This whole 
approach could seem strange when compared with ordinary quantum mechanics but it is deeply grounded 
in the underlying symmetries. \\

the procedure is as follow. On starts with the full expression of the Hamiltonian 
constraint in terms of Ashtekar variables and uses the FLRW symmetries to reduce those 
variables to basically $c$ and $p$. In the polymeric Hilbert space, there is no $\hat{c}$ 
operator. Only exponential functions of the gravitational connection -- that is 
holonomies -- become well defined. These functions generate an algebra of almost periodic 
functions and the resulting kinematical Hilbert space is the space of square integrable 
functions on the compactification of the real line. In this space, the eigenstates of 
$\hat{p}$, labelled by $\mu$, satisfy $\langle\mu_1|\mu_2\rangle=\delta_{\mu_1,\mu_2}$ with a 
Kronecker delta instead of the usual Dirac distribution. Those states are normalized and 
constitute a basis for the kinematical Hilbert space.

%
%
%

To obtain the quantum constraint, the classical gravitational one has to be rewritten with the field strength $F_{ab}^i$:
\begin{equation} \label{eq:cgrav}
C_{\mathrm{grav}} = - \gamma^{-2} \int_{\cal
V} \d^3 x\,  \epsilon_{ijk} \,\f{E^{ai}E^{bj}}{\sqrt{|\det E|}}\,
F_{ab}^k.
\end{equation} 
Then, $F_{ab}^i$ is expressed in terms of holonomies and triads and quantized, whereas
the matter sector is quantized in the
standard way. By choosing correctly the lapse function, the gravitational constraint can be written as,
\be
\label{eq:cgrav2}
{\cal C}_{\mathrm{grav}} = - \gamma^{-2}   {\epsilon^{ij}}_{k} \,\e^{a}_{i}\e^{b}_{j}\,
F_{ab}^k,
\ee 
where $\e^{a}_{i}$ is the (un-densitized) triad defined on a fiducial cell.
As suggested previously, classically, the field strength can be written in terms of a trace of
holonomies over a square loop $\Box_{ij}$, considered over a face
of the cell, with its area shrinking to zero: 
\be
\label{F} F_{ab}^k\, = \, -2\,\lim_{Ar_\Box
 \rightarrow 0} \,\, \Tr\,
\left(\f{h^{(\lambda_c)}_{\Box_{ij}}-1 }{\lambda_c^2} \right)
\,\, \tau^k\, \ow^i_a\,\, \ow^j_b\, = \,\lim_{\lambda_c
 \rightarrow 0}   {\epsilon^k}_{ij}\, \ow^i_a\,\, \ow^j_b\,\left(
 \f{\sin^2{\lambda_c c}}{\lambda_c^2}\right),
\ee
with
\be
h^{(\lambda_c)}_{\Box_{ij}}=h_i^{(\lambda_c)} h_j^{(\lambda_c)}
(h_i^{(\lambda_c)})^{-1} (h_j^{(\lambda_c)})^{-1}\, , \ee

and $\ow^i_a$ is the co-triad compatible with the metric $q_{ab}$.
%
%
As explained previously, if LQC is to account for LQG, the loop $\Box_{ij}$ can be
shrunk at most to the area which is given by the minimum eigenvalue of the area operator: 
$\Delta = \tilde\kappa\, \lp^2$, with $\tilde\kappa$ of order one. Note that it has been 
standard in the LQC literature to choose $\tilde\kappa= 2 \sqrt{3} \pi \gamma$  \cite{abl}, 
but
it can also be taken as a parameter to be determined \cite{acs}.
The area of the loop with respect to the physical metric is $\lambda_c^2 |p|$. Requiring
the classical area of the loop $\Box_{ij}$ to have the quantum area gap
as given by LQG, we are led to set
$\lambda_c = \sqrt{\Delta/|p|}$. The action of $\exp(i \lambda_c(p) c)$ in volume ($\nu$) 
basis is  simple: it drags the state by a unit affine parameter.

One can then introduce the variable
$\b := \f{c}{|p|^{1/2}}$,
such that $\lambda_c c = \lambda_\b \b$ where $\lambda_\b := \sqrt{\Delta}$ is the
 affine parameter ($\b$ is  conjugate variable to $\nu$).
The volume operator acts as
\be
\hat V \, |\nu\rangle = 2 \pi \lp^2 \gamma |\nu| \, |\nu\rangle ~,
\ee
and the action of the exponential operator is:
\be
\widehat{\exp(i \lambda_c c/2)} \, |\nu\rangle ~ = ~\widehat{\exp(i \lambda_\b \b/2)}
\, |\nu\rangle ~ = ~ |\nu +  \lambda_\b\rangle ~.
\ee
The 
quantum constraint operator on wave functions $\tilde{\Psi}(\nu,\varphi)$ of $\nu$ and 
$\varphi$ is
\be \label{hc4}
\partial_\varphi^2\, \tilde{\Psi}(\nu,\varphi) = 3\pi G\, |\nu|\,
\f{\sin\lambda_\b\b}{\lambda_\b}\, |\nu|\, \f{\sin\lambda_\b\b}{\lambda_\b}\,
\tilde{\Psi}(\nu,\varphi).\, 
\ee

This simplifies to
\ba \label{hc5} \partial_\varphi^2 \,\tilde{\Psi} (\nu, \varphi) &=& 3\pi
G\, \nu\, \f{\sin\lambda_\b\b}{\lambda_\b}\, \nu\,
\f{\sin\lambda_\b\b}{\lambda_\b}\, \tilde{\Psi}(\nu,\varphi) \nonumber\\
&=&\f{3\pi G}{4\lambda_\b^2}\, \nu \left[\, (\nu+2\lambda_\b)
\tilde\Psi(\nu+4\lambda_\b) - 2\nu \tilde\Psi(\nu) + (\nu -2\lambda_\b)
\tilde\Psi(\nu-4\lambda_\b)\, \right]\nonumber\\
&=:& \Theta_{(\nu)}\, \tilde\Psi(\nu,\varphi)\, . \label{Quant-Const}
\ea
The geometrical part, $\Theta_{(\nu)}$, of the constraint is a
difference operator in steps of $4\lambda_\b$. 
\be C^+(\nu) \Psi(\nu + 4 \lambda_\b) +
C^0 \Psi(\nu) + C^- \Psi(\nu - 4 \lambda_\b) = \hat C_{\mathrm{matt}}
\, \Psi(\nu). \ee
The equivalent of the Wheeler-De Witt equation is in LQC a {\it difference} equation in
the geometrical variable, instead of a differential equation. This is the key theoretical 
feature.\\

\section{Main features of the background evolution in LQC}
\label{sec:bckg}

\subsection{A bouncing universe}
We focus here, still following \cite{agullo}, on the $k=0$ case. Simulations \cite{vt,aps2,apsv} have shown that effective equations
approximate very well the full quantum dynamics (in the sense of expectation values of
Dirac observables).  
The effective Hamiltonian constraint is, for $N$=1 corresponding to cosmic time $t$,
\be \label{effham}
{\cal C}_{\rm eff}=\f{3}{8 \pi G\gamma^2} \, \f{\sin2(\lambda_b\, \b)}{\lambda_b^2} V - 
{\cal C}_{\mathrm{matt}} \, ,
\ee
which leads to modified Friedmann and Raychaudhuri equations. Using this equation, one can show that the
energy density $\rho = H_{\mathrm{matt}}/V$ equals
$3 \sin2(\lambda_b\, \b)/(8 \pi G \gamma2 \lambda_b^2)$ which reaches its
maximum when $\sin2(\lambda_b\, \b)=1$. The density has a maximum given by
\be
\rho_{\rm c}= \f{3}{8 \pi G \gamma^2 \lambda_b^2}\, ,
\ee
which is
identical to the supremum $\rho_{\mathrm{sup}}$ for the density operator in LQC. The difference is that in the effective dynamics every trajectory undergoes a bounce and reaches the maximum
possible density, while in the quantum theory this is not true for every state. 

The dynamics defined by the effective Hamiltonian can be solved and the equations of motion
found by $\dot{O} = \{O,{\cal C}\}$ with $\dot{f}$ meaning derivative of $f$ with respect to cosmic time $t$.
The equation of motion for the volume is
\be
\dot V=\frac{3}{\gamma\lambda_b}V\sin{(\lambda_b\b)}\cos{(\lambda_b\b)}\, ,
\ee
leading to the modified Friedmann equation
\be
H^2:=\left(\frac{\dot{a}}{a}\right)^2=\left(\frac{\dot{V}}{3V}\right)^2=\frac{8\pi G}{3}\,\rho\, \left(1-\frac{\rho}{\rho_{\rm c}}\right)\, ,
\label{eff-fried}
\ee
where $\rho_{\rm c}=\frac{9}{2\alpha^2}\frac{1}{\lambda_b^2}$ is the critical energy density, which cannot be exceeded, 
set by the minimal area. This is precisely the energy density reached by the scalar field at the bounce.
For every trajectory there are quantum turning points at $\b =\pm\frac{\pi}{2\l}$, where $\dot{V}=0$,
corresponding to a bounce. It should be pointed out that the Hubble parameter is absolutely
bounded, indicating that the congruence of cosmological observers can never
have caustics, independently of the matter content. This maximum density was shown to be
explicitly given by 
\begin{equation}
\rho_{\rm c} = \frac{\sqrt{3}}{32\pi^2\gamma^3}m^4_{\rm Pl} \simeq 0.41 m^4_{\rm Pl},  \label{rhoc}
\end{equation}  
with $m_{\rm Pl}=1/\sqrt{G}$ the Planck mass. Similarly, the modified Raychaudhuri and Klein-Gordon equations can be written as:
\be
\f{\ddot{a}}{a}=-\f{4\pi G}{3}\,\rho\left(1-4\, \f{\rho}{\rho_{\rm c}}\right) -4\pi G\,P\left(
1-2\, \f{\rho}{\rho_{\rm c}}\right)\,
\ee
and
\be
\ddot{\varphi}+3H\dot{\varphi}+m^2\varphi=0.
\ee
For the scalar field, we have here assumed the simplest potential $V(\varphi)=\frac{1}{2}m^2\varphi^2$. 
More generally, we will base our reasonings on potential wells.


This effective dynamics encoding the first order quantum corrections captures the main features of the full 
quantum dynamics of the states sharply peaked around classical trajectories. In the large ``volume'' limit, 
$\rho$ is much smaller than $\rho_{\rm c}$ and one recovers the standard cosmological equations of motion, {\it i.e.}
\begin{equation}
	H^2=\frac{8\pi G}{3}\rho~~~\mathrm{and}~~~\frac{\ddot{a}}{a}=-\frac{4\pi G}{3}\left(\rho+3P\right).
\end{equation}
Nevertheless, in the small ``volume" limit, the energy density of the content of the Universe grows close to $\rho_c$ 
and the dynamics is drastically modified. The Big Bang singularity is now replaced by a regular bounce. Though the 
fundamental reason is rooted in the quantum dynamics of the sharply peaked state (preventing the transition to the 
singular state of null volume), the occurrence of the bounce can be understood and retrieved from the effective dynamics 
of the quantum universe. The Klein-Gordon equation can be recasted in the more usual form of energy 
conservation (we remind that for a scalar field $\rho=\frac{1}{2}(\dot\varphi)^2+V(\varphi)$ and $P=\frac{1}{2}(\dot\varphi)^2-V(\varphi)$):
\begin{equation}
	\dot{\rho}+3H(\rho+P)=0.
\end{equation}
From the above equation and the Friedmann equation, one can easily deduce the rate of variation of the Hubble 
parameter which takes the following form in the effective LQC framework:
\begin{equation}
	\dot{H}=-4\pi G (\rho+P)\left(1-2\frac{\rho}{\rho_c}\right).
\end{equation}
To understand better how the bounce is occurring, let us consider the simple case of a universe filled with 
dust-like matter, {\it i.e.} $P=0$. In the general relativistic framework, one obtains the following set of equations 
for the scale factor and the energy density:
\begin{eqnarray}
	\left(\frac{\dot{a}}{a}\right)^2&=&\frac{8\pi G}{3}\rho, \\
	\frac{\ddot{a}}{a}&=&-\frac{4\pi G}{3} \rho, \\
	\dot{\rho}&=&-3H\rho.
\end{eqnarray}
Considering a contracting universe, $H<0$ (that is looking at our expanding Universe {\it backward} in time), 
one easily understand that reaching $a=0$ is unavoidable as the total evolution is monotonous. Indeed, while 
the universe is contracting, $H$ is negative valued, $a$ is decreasing and $\rho$ is increasing. Starting from 
that and using the above dynamical equations, one figures out that the sign of $\dot{\rho}$, $\dot{a}$ and 
$\ddot{a}$ cannot change. In other words, $\left|H\right|$ keeps increasing and the contraction is never slowed
down nor reversed. Inversely, in the effective LQC framework, the above dynamical equations are modified as 
follows (still working with a dust-like content for simplicity):
\begin{eqnarray}
	\left(\frac{\dot{a}}{a}\right)^2&=&\frac{8\pi G}{3}\rho\left(1-\frac{\rho}{\rho_c}\right), \\
	\frac{\ddot{a}}{a}&=&-\frac{4\pi G}{3} \rho\left(1-4\frac{\rho}{\rho_c}\right), \\
	\dot{\rho}&=&-3H\rho.
\end{eqnarray}
Clearly, the situation drastically differs. Once $\rho=\rho_c/4$, the sign of $\ddot{a}$ changes. Similarly, 
the sign of $\dot{H}$ changes when $\rho=\rho_c/2$. (We stress that for a scalar field $-\rho<P<\rho$ and 
the condition for $\dot{H}$ to change its sign remains $\rho>\rho_c/2$.) After $\rho=\rho_c/2$, $\dot{H}$ 
becomes positive valued: $\left|H\right|$ is now decreasing and the contraction is slowed down. Finally, at 
$\rho=\rho_c$, the Hubble parameter vanishes while $a$ is {\it non vanishing}. As $H$ changes its sign 
at that moment, the Universe regularly transits from a contracting phase to an expanding phase, which 
corresponds to the Big Bounce.

Obviously the replacement of the Big Bang by a Big Bounce is the more important result of LQC.
On the one hand, this is a kind of minimum requirement if the theory is to solve the
singularity issue. On the other hand, it was not that obvious that the effective theory will
indeed exhibit this behavior. This is the basis of possible phenomenological
investigations. Huge repulsive effects from quantum geometry allow escaping the Big Bang
although classical gravity is extraordinary attractive in this regime.

\subsection{A bouncing and inflationary universe}

How does the scalar field evolve in a LQC bouncing Universe? The key question behind is the possible release of 
inflation shortly after the bounce. Scalar fields as matter content for the universe have been initially introduced to 
trigger a phase of accelerated expansion in the very early stages of the cosmic history. This phase of primordial 
accelerated expansion, dubbed inflation, is a crucial ingredient of the standard cosmological paradigm as it solves, 
among others, the horizon and flatness problems. Moreover,  quantum fluctuations of the vacuum are amplified during 
inflation and thanks to the dramatic increase of the scale factor, those fluctuations are stretched from microphysical 
sizes to astronomical sizes. Those amplified quantum fluctuations finally act as classical cosmological perturbations 
when the Universe exits inflation, and constitute the primordial seeds for galaxies and large-scale structures  latter on 
in the cosmic history. 

Though inflation solves many problems in standard cosmology, this mechanism brings with itself its own issues. 
As already underlined, standard matter or radiation cannot initiate a phase of inflation and one has to invoke a 
scalar field as the dominating content of the very early Universe. Except the Higgs field (that cannot play the role 
of the inflaton in simple models), no fundamental scalar field is known so far. In addition, the 'initial conditions' for inflation 
to start and be sufficiently long to solve for the horizon and flatness problems (that is an inflation with at least 60 
e-folds; the number of e-folds being the logarithm of the ratio of the scale factor at the end and beginning of 
inflation) could appear as rather fine-tuned. Roughly speaking, for inflation to start and the universe to stay on such 
a trajectory, the scalar field should be initially in a state such that its energy density is dominated by its potential energy and such that the field stays in this peculiar setting. This can be met by requiring two conditions to be fulfilled: 
\begin{equation}
\epsilon:=\frac{(\dot\varphi)^2}{(\dot{\varphi})^2+2V(\varphi)}\ll1~~~\mathrm{and}~~~\eta:=\frac{-\ddot{\varphi}}{3H\dot{\varphi}}\ll1.
\end{equation}
The first condition allows the universe to enter a phase of accelerated expansion as one easily sees that
$
	{\ddot{a}}/{a}\simeq({8\pi G}/{3}) V(\varphi)\left(1-2\epsilon\right)
$
is positive-valued for a positive-valued potential. Nevertheless, this condition is not sufficient for inflation 
to last long enough for the number of e-folds to be greater than 60. Considering the analogy with a massive 
pendulum, this first condition simply means that the pendulum is far from its equilibrium. If one lets the 
pendulum move starting from this far-from equilibrium position, it will certainly {\it rapidly} rolls down towards 
its equilibrium regime where the potential energy domination is not (constantly) met anymore. The second 
condition is precisely here to prevent such a fast rolling. The idea is in fact very simple: the Hubble drag 
just prevents the field to rapidly roll down along its potential. The key question is therefore to know if those 
conditions are easily met in the primordial universe. Assuming they are met at some point, a straightforward 
calculation, under this so-called slow-roll approximation scheme, shows that the number of e-folds during inflation is given by
\begin{equation}
N_\mathrm{inf}:=\displaystyle\int_{t_i}^{t_f}H(t)dt\simeq-8\pi G\displaystyle\int_{\varphi_i}^{\varphi_f}\left(\frac{V}{V_{,\varphi}}\right)d\varphi.
\end{equation}
Considering for simplicity a monomial potential (usually called a 'scaling solution'), $V(\varphi)=\lambda\varphi^\alpha$, 
the number of e-folds is therefore $N_\mathrm{inf}=4\pi G(\varphi^2_i-\varphi^2_f)/\alpha$. To reach the critical number 
of 60 e-folds, one easily figures out that initially the value of the scalar field should be bounded from below by:
\begin{equation}
\varphi_i\geq2.1\sqrt{\alpha}~m_\mathrm{Pl}.
\end{equation}
For concave potentials, $\alpha>1$ and initially the field should admit trans-Planckian values. Clearly, meeting those conditions in the standard 
paradigm is far from being obvious. (It is worth 
noticing that this does note mean that the energy density, roughly given by $V(\varphi_i)$ at the onset of inflation, 
will be trans-Planckian as the coupling constant will affect the value of $\rho_i$. As an example, the initial energy 
density for a massive scalar field is bounded by $4.5\times m^2\times m^2_\mathrm{Pl}$ which is smaller than the 
Planck energy density as long as $m<m_\mathrm{Pl}/\sqrt{4.5}$.)

\begin{figure}
\begin{center}
\includegraphics[scale=0.55]{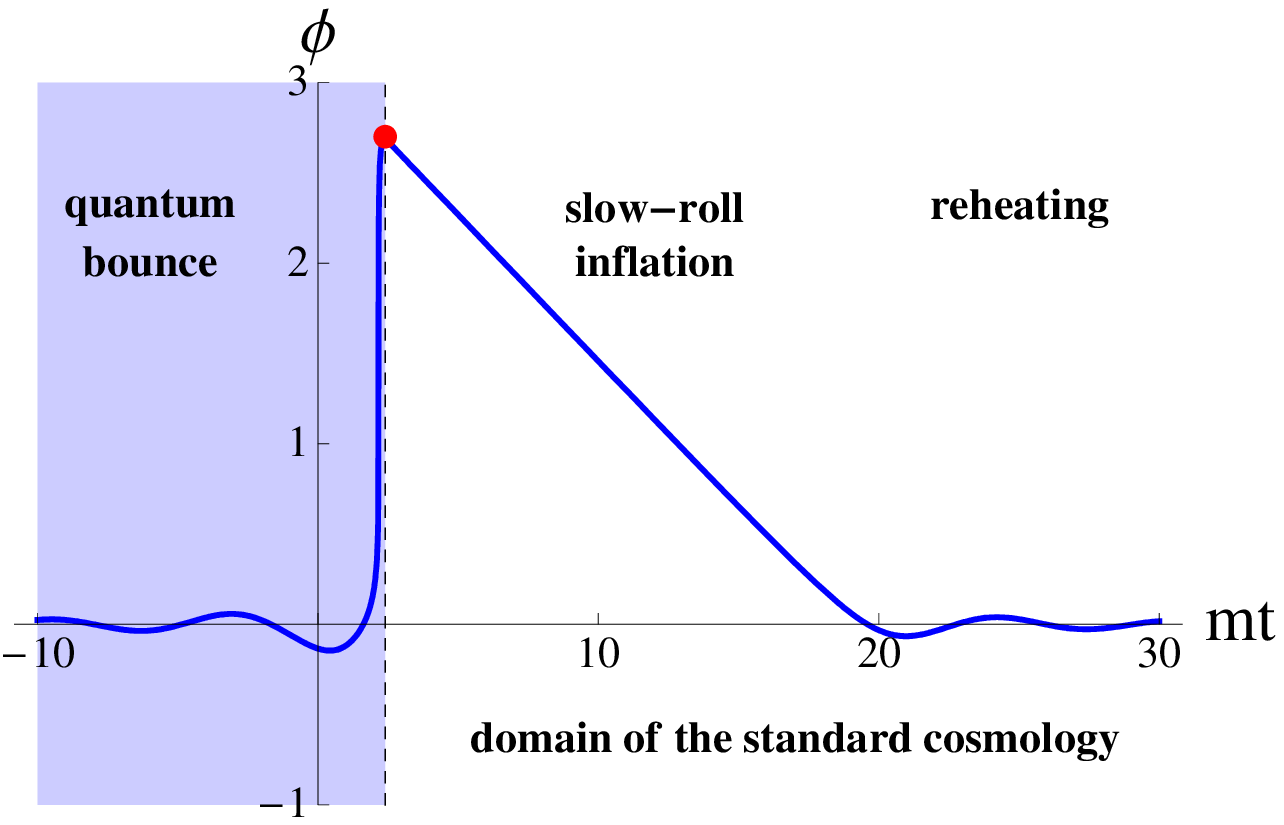} \includegraphics[scale=0.55]{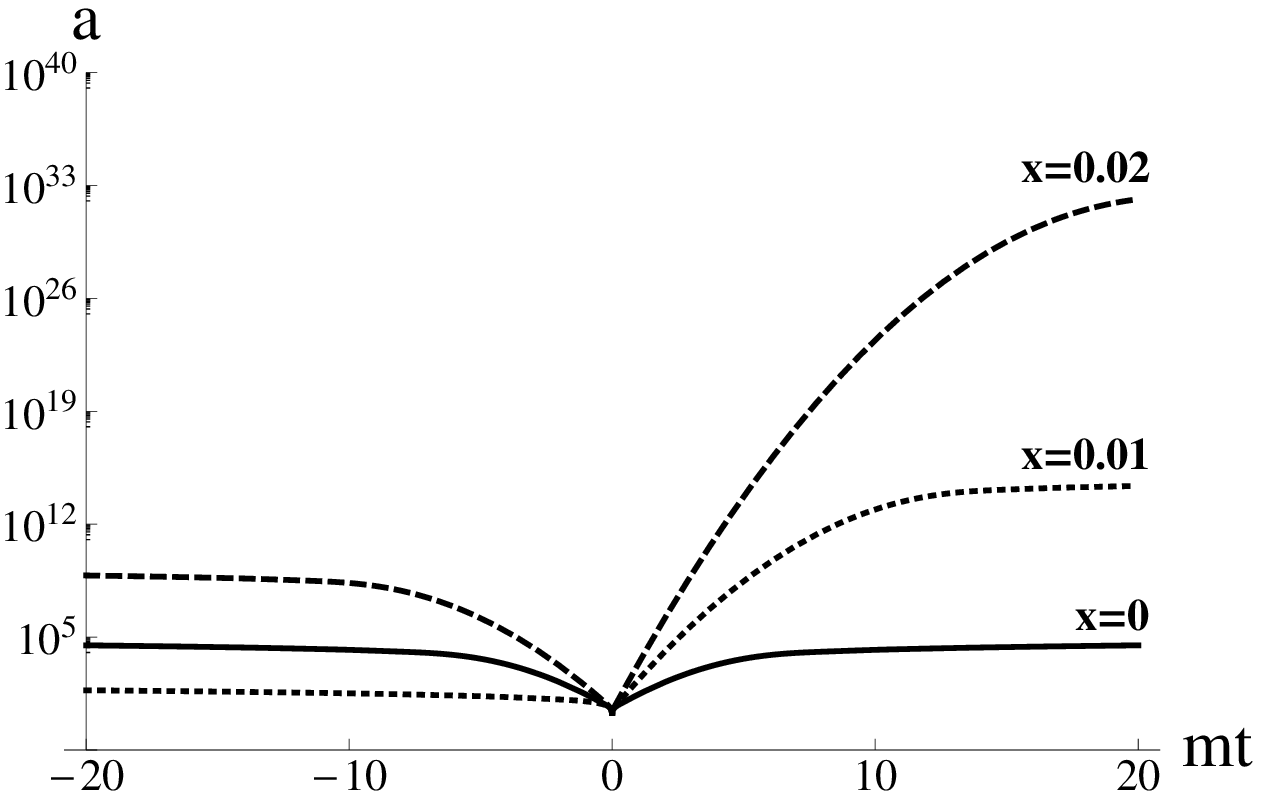}
\caption{{\it Left panel :} Evolution of the scalar field as a function of cosmic time in a bouncing universe. 
The field first experiences oscillations during contraction, then climbs up its potential to finally enters in a 
phase of linear variation with time, corresponding to slow-roll inflation. {\it Right panel :} Evolution of the 
scale factor as a function of cosmic time. The time $t=0$ corresponds to the bounce. (From Ref. \cite{jakub})}
\label{bckg1}
\end{center}
\end{figure}

What can LQC say about inflation? The first point to notice is that the contracting phase and the 
bounce can set the scalar field in the appropriate conditions for a phase of accelerated expansion 
to start after the bounce \cite{jakub}. This simply comes from the peculiar dynamics of the bouncing 
universe. Let us recall that in the Klein-Gordon equation, the Hubble parameter acts as a friction term 
during expansion. However, as $H$ is negative-valued during contraction, it acts on the scalar field 
as an {\it anti-friction} term. Suppose that in the remote past, deep in the contracting phase, the scalar 
field is only slightly away from its equilibrium position in the potential well. During contraction, the field 
will oscillates but because of anti-friction the amplitude of its oscillations will be amplified. As a consequence, 
the scalar field will climb up its potential well to finally reach the appropriate value for inflation to start 
after the bounce. Such a scenario is depicted on Fig. \ref{bckg1} for the case of a massive scalar field. 
The left panel shows the evolution of the scalar field which first experienced some oscillations during the
contraction, then climbed up its potential and finally evolved linearly with time as expected in slow-roll 
inflation, $\varphi\simeq\varphi_i-m(t-t_i)/\sqrt{12\pi G}$ with $\varphi_i$ the value of the field at the onset of 
inflation $t_i$ (denoted by the red dot on the graph). The right panel shows the evolution of the scale 
factor and clearly exhibits the bounce at $t=0$ (although not obvious from this graph, the behavior 
remains differentiable at the bounce). This is depicted for three values of the amount of potential energy 
at the bounce, encoded in the parameter $x_B=(V(\varphi_B)/\rho_c)^{1/2}=m\varphi_B/\sqrt{2\rho_c}$ with 
$\varphi_B$ the value of the field at the time of the bounce. This shows that a non-zero amount of potential 
energy is needed at the bounce for a sufficiently long phase of inflation to occur. Nevertheless, this 
amount only needs to be tiny (a mandatory condition for the above mentioned effective equations 
to be valid). It is interesting to notice that as the energy density is bounded from above, the total 
number of e-folds during inflation is also bounded from above: $N\leq(4\pi \rho_c)/(m\times m_\mathrm{Pl})^2$.

The fact that the bounce can set the scalar field in the suitable dynamical conditions for inflation 
to start still does not imply that inflation occurs {\it naturally} in LQC, for this peculiar behavior could 
be associated only with a tiny fraction of all the possible trajectories. The underlying question is to 
know wether or not the volume of trajectories with inflation having at least 60 e-folds is, according 
to a relevant measure, dominant. This question has been addressed in \cite{abhay} and more 
recently in Ref. \cite{linse} and the answer is in the affirmative. The strategy consists in exploring 
the volume of initial conditions and inspect which of those trajectories does effectively pass through 
a phase of inflation with at least 60 e-folds. Roughly speaking, the ratio of the volume of those
trajectories passing through inflation divided by the total volume of trajectories gives an hint on the 
probability for inflation to occur in LQC. This obviously raises the question of when to set the initial 
conditions and what kind of probability distribution function should be chosen to explore the volume
of initial conditions. In \cite{abhay}, the volume of initial conditions is explored at the bounce while 
in Ref. \cite{linse}, it is explored far in the contracting phase of the universe. Fortunately, as will be 
detailed in a dedicated section of this article, both approaches lead to similar conclusions: the 
cumulative probability to have a phase of inflation with at least 60 e-folds, $P(N_\mathrm{inf}>60)$, 
is very close to unity. This is in that precise sense that inflation is said to be naturally triggered in LQC.
It is fair to underline that this is basically a generic GR feature: inflation is an attractor. The specific
feature of LQC is to allow clear prediction, {\it e.g.} on the duration of inflation.\\

It should be noticed that the reason why effective questions of LQC are so accurate was recently explained in \cite{edcarlo}.

\section{Perturbations with holonomy corrections}
\label{sec:holo}

Extraction of the cosmological sector from  LQG still remains one the main open 
problems within the theory. While some progresses have been made in this direction in 
recent years, there is still no single cosmological prediction obtained from  full 
LQG. This deficiency has been filled mainly by the analysis of simpler situations where 
loop quantization is performed  after symmetry reduction. 
A nice feature of LQC is that its main features, as a bounce, can be 
captured within an effective framework. In this approach, the classical dynamics is deformed 
so that dynamical trajectories follow mean values in the given quantum state $\Psi$.
For example, the mean value of volume $\langle \hat{v} \rangle_{\Psi} \approx v_{\text{eff}}$.
The term $v_{\text{eff}}$ is obtained by solving effective equations of motion  based  on a
Hamiltonian being modified by quantum corrections. It must be stressed, 
however, that such an analysis is reliable only for states $\Psi$ being sufficiently sharply 
peaked on the effective trajectories.       
  
Thus section is mainly focused on the so-called holonomy corrections.  These 
corrections come from the fact that loop quantization is based on holonomies, i.e. 
exponentials of the connection, rather than direct connection components. For the flat 
FLRW model, the correct effective dynamics is obtained by the following replacement:
\begin{equation}
\bar{k} \rightarrow \mathbb{K}[n] := \frac{\sin (n \gamma \bar{\mu} \bar{k})}{n \gamma \bar{\mu}},
\label{generalcor}  
\end{equation}  
with $n=1$, at the level of the classical Hamiltonian. 

Intuitively, the motivation for considering such modifications comes from the
expression of the Ashtekar connection expressed in 
terms of the holonomy around a single loop.   
While classically $F^k_{ab} \sim \bar{k}^2$, the holonomy
regularization leads to   $F^k_{ab} \sim   \left( \frac{\sin(\gamma \bar{\mu}  \bar{k})}{\gamma \bar{\mu}}\right)^2$.
The classical expression is recovered for $\bar{\mu} \rightarrow 0$. 
However, not all $\bar{k}$ terms in classical expressions come 
from the curvature of the Ashtekar connection. For such terms, one 
can postulate that the holonomy correction has the
form given in Eq. (\ref{generalcor}),  where $n$ is some unknown integer. It should be an integer 
because, when quantizing the theory, the $e^{i\gamma \bar{k}}$ factor 
is promoted to be a shift operator acting on the lattice states.  
If $n$ was not an integer, the action of the  operator corresponding  
to $e^{i n\gamma \bar{k}}$ would be defined in a different basis. 
Another issue is related with the choice of  $\bar{\mu}$, which 
corresponds to the so-called lattice refinement. Models 
with a power-law parametrization $\bar{\mu} \propto \bar{p}^{\beta}$ 
were discussed in details in the literature. While, in general, 
$ \beta \in [-1/2,0]$, it was pointed out that the choice $\beta = -1/2$, 
called the $\bar{\mu}-$scheme (new quantization scheme), is the 
favored one \cite{Nelson:2007um}.  As we will discuss later in this section, the study of
scalar perturbations with holonomy corrections make such 
choice mandatory. 

By introducing the holonomy corrections, 
a new translational symmetry appears due to sine type dependence of the connection variable $\bar{k}$.  
Because of that, introduction of the holonomy correction can be interpreted as periodification 
of the connection part of the phase space. Namely, the initial phase space of the FLRW model 
$\Gamma_{\text{FRW}} = \mathbb{R} \times  \mathbb{R}$ becomes $\Gamma^Q_{\text{FRW}} 
= U(1) \times  \mathbb{R}$. The $U(1)$ symmetry is a reminiscence of the original $SU(2)$ 
connection. However, here, it is reduced due to symmetries of homogeneity and isotropy.   
  
For  homogeneous and isotropic models the holonomy corrections lead to the modified 
Friedmann equation (\ref{eff-fried}). This equation, due to its simplicity, played an important role in 
studying phenomenological aspects of loop quantum cosmology. However, predictions 
which can be derived based on that equation are limited to the singularity resolution issue, 
the presence of inflation and its duration, etc. In order to build a firm contact with 
the CMB physics, a theory of cosmological perturbations with the effects of holonomies included
is needed. It is therefore tempting to extrapolate the effective dynamics with holonomy 
corrections to the inhomogeneous cosmological sector, in particular to the perturbative 
inhomogeneities. 

Here, for pedagogical purpose, we will present a detailed analysis of the tensor perturbations 
(gravitational waves) with holonomy corrections. The full derivation for scalar perturbations 
is much more laborious, therefore we will restrict ourself to present final results only. The 
case of tensor perturbations is much simpler because 
the shift vector is vanishing, $N^a = 0$. This implies vanishing of the spatial 
diffeomorphism constraints (vector constraint).  Because of that the algebra of constraints 
reduces significantly and the issue of anomaly freedom is not relevant. This is in contrast 
with the case of scalar perturbations where the requirement of anomaly freedom introduces 
strong restrictions on the form of  quantum corrections. 

For the tensor modes the lapse function $N$ is not a subject of perturbations therefore 
$N = \bar{N}$. In this section, we choose $\bar{N} = \sqrt{\bar{p}} = a$.  Such a choice 
corresponds to the conformal time $\eta = \int \frac{dt}{a(t)}$. Therefore, in what follows 
time differentiations are done with respect to confromal time $\eta$ and will be denoted with a prime, {\it i.e.} $f'\equiv \partial_\eta f$. 

The spatial metric $q_{ab}=a^2(\delta_{ab}+h_{ab})$, where $|h_{ab}| \ll 1$ and 
is transverse $\partial^ah_{ab}=0$ and traceless $\delta^{ab}h_{ab}=0$.
Based on this,  perturbations of the densitized triad $E^a_i$ reads 
\begin{equation}
\delta E^a_i = - \frac{1}{2} \bar{p} h^a_i.  
\label{dEtensor}
\end{equation} 

While $E^a_i$ forms a conjugated pair with $A^i_a$, when studying cosmological 
perturbations it is convenient to use the extrinsic curvature $K^i_a = \frac{1}{\gamma}(A^i_a-\Gamma^i_a) $ 
instead of $A^i_a$.  Therefore, a perturbative expansion of the Hamiltonian in $\delta E^a_i $ 
and $\delta K^i_a$ variables is  performed. For the linear perturbations theory, the 
development is performed up to the quadratic terms. For the tensor perturbations the resulting
gravitational part of the Hamiltonian reads \cite{bojo1}:
\begin{eqnarray}
H_G &=& \frac{1}{2\kappa} \int_{\Sigma} d^3x \bar{N}\left[
-6\sqrt{\bar{p}}  \bar{k}^2-\frac{1}{2\bar{p}^{3/2}}\bar{k}^2
(\delta E^c_j \delta E^d_k \delta^k_c \delta^j_d)  \right. \nonumber \\
&+&\sqrt{\bar{p}}(\delta K^j_c \delta K^k_d \delta^c_k \delta^d_j) 
-\frac{2}{\sqrt{\bar{p}}}\bar{k}(\delta E^c_j \delta K^j_c ) \nonumber \\
&+&\left. \frac{1}{\bar{p}^{3/2}} (\delta_{cd} \delta^{jk} \delta^{ef} \partial_{e}
\delta E^c_j\partial_f\delta E^d_k) \right], 
\end{eqnarray}
where the first term in the square bracket is the FLRW background term. Let us now introduce 
holonomy corrections to this Hamiltonian.  Originally, it was firstly done in  \cite{bojo1} 
by replacing $\bar{k} \rightarrow \mathbb{K}[n]$, and fixing the values of $n\in \mathbb{N}$.
In such a situation, only the homogeneous part of the phase space is deformed.  The perturbation
part $\delta K^j_c$ of the connection is not subject to holonomy corrections at the lowest order. 
The obtained holonomy corrected Hamiltonian is \cite{bojo1}:
\begin{eqnarray}
H^{\text{old}}_G &=& \frac{1}{2\kappa} \int_{\Sigma} d^3x \bar{N}\left[
-6\sqrt{\bar{p}}  \mathbb{K}[1]^2-\frac{1}{2\bar{p}^{3/2}}\mathbb{K}[1]^2
(\delta E^c_j \delta E^d_k \delta^k_c \delta^j_d)  \right. \nonumber \\
&+&\sqrt{\bar{p}}(\delta K^j_c \delta K^k_d \delta^c_k \delta^d_j) 
-\frac{2}{\sqrt{\bar{p}}}\mathbb{K}[2](\delta E^c_j \delta K^j_c ) \nonumber \\
&+&\left. \frac{1}{\bar{p}^{3/2}} (\delta_{cd} \delta^{jk} \delta^{ef} \partial_{e}
\delta E^c_j\partial_f\delta E^d_k) \right] . 
\end{eqnarray}
Based on this Hamiltonian it was shown \cite{bojo1} that the propagation of gravitons is given 
by the following equation of motion:
\begin{equation}
\frac{1}{2}\left[{h^i_a}''+\left(\frac{\sin{(2\gamma\bar\mu\bar{k})}}{\gamma\bar\mu}\right){h^i_a}'
-\nabla^2h^i_a-2\gamma^2{\bar\mu}^2\left(\frac{\bar{p}}{\bar\mu}\frac{\partial\bar\mu}{\partial\bar{p}}\right)
\left(\frac{\sin{(\gamma\bar\mu\bar{k})}}{\gamma\bar\mu}\right)^4h^i_a\right]=\kappa \Pi^i_{Qa}, 
\label{equbojo}
\end{equation}
where the corrected source term is
\begin{equation}
\Pi^i_{Qa} = \left[  \frac{1}{3V_0}  \frac{\partial \bar{H}_{\text{m}} }{\partial \bar{p}} 
\left( \frac{\delta E^c_j \delta^j_a\delta^i_c}{\bar{p}} \right)\cos (2 \gamma \bar{\mu} \bar{k})
+\frac{\delta H_{\text{m}}}{\delta \delta E^a_i}\right].
\end{equation}
Here $H_{\text{m}}$ is the matter Hamiltonian and $\bar{H}_{\text{m}}$ is its homogeneous part. The factor $V_0$ is an 
IR regulator introduced by restricting the spatial integration of the flat universe to the finite volume   
\begin{equation}
V_0 = \int_{\Sigma} d^3x.   
\end{equation}
The value of $V_0$ has no physical meaning and should not appear in the final expressions.  Here, it cancels
out with the $V_{0}$ coming from the homogeneous part of the matter Hamiltonian $\bar{H}_{\text{m}}$. 

The equation (\ref{equbojo}) was a starting point for numerous phenomenological considerations 
within loop quantum cosmology. However, recent analysis of the anomaly freedom for the scalar 
perturbations \cite{Cailleteau:2011kr} indicate that the Hamiltonian $H^{\text{old}}_G$ should be 
equipped with some additional terms \cite{Cailleteau:2012fy}. These, so-called counterterms guarantee the
anomaly freedom of the algebra of constraints for  scalar perturbations. The presence of such terms 
cannot be inferred from considerations of the tensor modes only. This is because the requirement of 
 anomaly freedom in identically fulfilled here due to the vanishing of the vector constraint.  The new 
version of the holonomy corrected Hamiltonian unavoidably leads to  new equations of motion, 
significantly different from (\ref{equbojo}). Furthermore, the requirement of anomaly freedom for the 
scalar perturbations implies also that the quantization scheme is the such that  \cite{Cailleteau:2011kr}  
\begin{equation}
\bar{\mu} = \sqrt{\frac{\Delta}{\bar{p}}},
\end{equation}
where $\Delta$ is a constant of dimension of a length squared.  The new Hamiltonian is obtained from 
$H^{\text{old}}_G$ by adding the counterterms part $H_C$:
\begin{equation}
H^{\text{new}}_G =  H^{\text{old}}_G + H_C,
\end{equation}
where 
\begin{eqnarray}
H_C &=& \frac{1}{2\kappa} \int_{\Sigma} d^3x \bar{N}\left[
-\frac{1}{\bar{p}^{3/2}}
\left( 3\bar{k} \mathbb{K}[2]-2\mathbb{K}[1]^2-\bar{k}^2\Omega  \right)
(\delta E^c_j \delta E^d_k \delta^k_c \delta^j_d) \right. \nonumber \\
&+&\left.  (\Omega-1) \sqrt{\bar{p}}(\delta K^j_c \delta K^k_d \delta^c_k \delta^d_j) 
-\frac{2}{\sqrt{\bar{p}}}\left(\mathbb{K}[2]-\bar{k}\Omega \right)(\delta E^c_j \delta K^j_c ) 
 \right]. 
\end{eqnarray}
For  later convenience we  have introduced here $\Omega := \cos (2 \gamma \bar{\mu} \bar{k})$.
The counterterm does not contribute in the classical limit, namely $\lim_{\bar{\mu}\rightarrow 0} H_C =0$.
The ``new'' gravitational Hamiltonian is
\begin{eqnarray}
H^{\text{new}}_G &=& \frac{1}{2\kappa} \int_{\Sigma} d^3x \bar{N}\left[
-6\sqrt{\bar{p}}  \mathbb{K}[1]^2   -\frac{1}{2\bar{p}^{3/2}}
\left( 6\bar{k} \mathbb{K}[2]-3\mathbb{K}[1]^2-2\bar{k}^2\Omega  \right)
(\delta E^c_j \delta E^d_k \delta^k_c \delta^j_d) \right. \nonumber \\
&+&\Omega \sqrt{\bar{p}}(\delta K^j_c \delta K^k_d \delta^c_k \delta^d_j) 
-\frac{2}{\sqrt{\bar{p}}}\left(2\mathbb{K}[2]-\bar{k}\Omega \right)(\delta E^c_j \delta K^j_c ) \nonumber \\
&+&\left.  \frac{1}{\bar{p}^{3/2}} (\delta_{cd} \delta^{jk} \delta^{ef} \partial_{e}\delta E^c_j\partial_f \delta E^d_k) \right]. 
\end{eqnarray}
Based on this, the total Hamiltonian, which is the sum of the gravity and of the matter parts can be constructed 
\begin{equation}
H_{\text{tot}}= H^{\text{new}}_G+H_{\text{m}}.
\end{equation} 
In the following, we consider a self-interacting scalar field matter which is relevant for the 
purpose of generating  the inflationary phase. As the gravity sector,  the matter part 
is also subject to a perturbative treatment. Namely, both the scalar field $\varphi$ and its conjugated 
momentum $p_\varphi$ are decomposed for the homogeneous and perturbation part as follows: 
\begin{equation}
\varphi = \bar{\varphi}+\delta \varphi \ \ \text{and} \ \  p_\varphi = \bar{p}_\varphi+\delta{p}_\varphi. 
\end{equation}
Decomposition of the phase space for the homogeneous and perturbation part is 
reflected by the following splitting of the Poisson bracket for the model under 
consideration:
\begin{equation}
\{ \cdot , \cdot \} =  \{ \cdot, \cdot \}_{\bar{k},\bar{p}} + \{\cdot, \cdot \}_{\delta K, \delta E} 
+ \{\cdot,\cdot \}_{\bar{\varphi},\bar{p}_\varphi} + \{\cdot,\cdot \}_{\delta \varphi, \delta{p}_\varphi}.
\label{Poisson}
\end{equation}
The constituent brackets are  
\begin{eqnarray}
\{ \cdot, \cdot\}_{\bar{k},\bar{p}} &:=&\frac{\kappa}{3 V_0} \left[ \frac{\partial  \cdot}{\partial \bar{k}} 
\frac{\partial  \cdot}{\partial \bar{p}} 
-\frac{\partial  \cdot}{\partial \bar{p}} \frac{\partial  \cdot}{\partial \bar{k}}\right], \\
\{\cdot, \cdot \}_{\delta K, \delta E} &:=&\kappa \int_{\Sigma} d^3x \left[ \frac{\delta  \cdot}{\delta \delta K^i_a} 
\frac{\delta  \cdot}{\delta \delta E^a_i} 
-\frac{\delta  \cdot}{\delta \delta E^a_i} \frac{\delta  \cdot}{\delta \delta K^i_a}\right], \\
\{\cdot, \cdot \}_{\bar{\varphi},\bar{p}_\varphi} &:=&\frac{1}{V_0} \left[ \frac{\partial  \cdot}{\partial \bar{\varphi}} 
\frac{\partial  \cdot}{\partial\bar{p}_\varphi} 
-\frac{\partial  \cdot}{\partial \bar{p}_\varphi} \frac{\partial  \cdot}{\partial \bar{\varphi}}\right], \\
\{\cdot, \cdot \}_{\delta \varphi, \delta{p}_\varphi} &:=&\int_{\Sigma} d^3x  \left[ \frac{\delta  \cdot}{\delta \delta \varphi} 
\frac{\delta  \cdot}{\delta \delta{p}_\varphi} 
-\frac{\delta  \cdot}{\delta \delta{p}_\varphi} \frac{\delta  \cdot}{\delta \delta \varphi}\right]. 
\end{eqnarray} 
Having defined the Poisson bracket, the Hamilton equation for any phase space function $f$
takes the form
\begin{equation}
{f'}= \{f, H_{\text{tot}} \}.
\end{equation}
In particular, equations of motion for the homogeneous part of the gravitational sector are  
\begin{eqnarray}
{\bar{p}}' &=& \left\{ \bar{p} , H_{\text{tot}} \right\} = \bar{N}2\sqrt{\bar{p}} (\mathbb{K}[2]),   
\label{dotpHameq} \\
{\bar{k}}'  &=& \left\{ \bar{p} , H_{\text{tot}} \right\}   =- \frac{\bar{N}}{\sqrt{\bar{p}}}\left[ 
\frac{1}{2} (\mathbb{K}[1])^2+\bar{p}\frac{\partial}{\partial \bar{p}}  (\mathbb{K}[1])^2 \right] 
+ \frac{\kappa}{3V_0} \left(\frac{\partial \bar{H}_{\text{m}}}{\partial \bar{p}} \right).  
\label{dotkHameq}
\end{eqnarray}
Based on Eq. (\ref{dotpHameq}), the conformal Hubble factor can be expressed as 
\begin{equation}
\mathcal{H} := \frac{{\bar{p}}'}{2\bar{p}} = \mathbb{K}[2],
\label{ConfHubb}
\end{equation}
where we used $\bar{N}=\sqrt{\bar{p}}$. This equality will allow us to write the final equations,
dependent on $\mathbb{K}[2]$, in a more convenient form.  

Let us now derive the equation of motion for the tensor modes $h^a_i$. The first step will be to differentiate 
expression (\ref{dEtensor}) with respect to conformal time
\begin{equation}
\left(\delta {E^a_i}\right)'  = -\frac{1}{2}\left( {\bar{p}}' h^a_i+\bar{p} {h^a_i}'  \right).
\label{ddE}
\end{equation}
On the other hand, the time derivative $(\delta {E}^a_i)'$ can be derived from the equations of motion 
\begin{eqnarray}
(\delta {E}^a_i)'  = \left\{ \delta E^a_i, H_{\text{tot}} \right\} = 
- \frac{\bar{N}}{2} \left[ 2\sqrt{\bar{p}}\Omega  \delta K^j_c \delta^a_j \delta^c_i
-\frac{2}{\sqrt{\bar{p}}}\left(2\mathbb{K}[2]-\bar{k}\Omega \right) \delta E^a_i \right], 
\label{dEEOM}
\end{eqnarray}
where we used the fact that $\frac{\delta H_{\text{m}}}{\delta \delta K^j_c}=0$. Using Eq. (\ref{ddE}),
with Eq. (\ref{dEEOM}) and  Eq. (\ref{dotpHameq}), one obtains the expression for the perturbation of the 
extrinsic curvature
\begin{equation}
\delta K^i_a = \frac{1}{2} \left[ \frac{1}{\Omega} {h^i_a}' +\bar{k} h^i_a \right].
\label{dK}
\end{equation}
By differentiating $\delta K^i_a$, given by  Eq. (\ref{dK}), with respect to conformal time, we obtain     
\begin{equation}
(\delta{K}^i_a)' = \frac{1}{2} \left[ -\frac{{\Omega'}}{\Omega^2} {h^i_a}'+\frac{1}{\Omega} {h^i_a}'' 
+{\bar{k}}' h^i_a+\bar{k} {h^i_a}' \right]. 
\label{ddK}
\end{equation}   
As in the case of ${E^a_i}'$, one can now write the equation of motion for $\delta K^i_a$:
\begin{eqnarray}
(\delta {K}^i_a)' &=& \left\{ \delta K^i_a, H_{\text{tot}} \right\} = \frac{\sqrt{\bar{p}}}{2} \left[
 -\frac{1}{\bar{p}^{3/2}}
\left( 6\bar{k} \mathbb{K}[2]-3\mathbb{K}[1]^2-2\bar{k}^2\Omega  \right)
(\delta E^c_j \delta^i_c \delta^j_a)  \right.  \nonumber \\ 
&-&\left. \frac{2}{\sqrt{\bar{p}}}\left(2\mathbb{K}[2]-\bar{k}\Omega \right)\delta K^i_a 
- \frac{2}{\bar{p}^{3/2}}  \delta^i_c \delta^j_a \nabla^2 \delta E^c_j \right] 
+ \kappa \frac{\delta H_{\text{m}}}{\delta \delta E^a_i}.
\label{dKEOM}
\end{eqnarray}
Using Eq. (\ref{ddK}) with (\ref{dKEOM}) and Eqs. (\ref{ddK}), (\ref{dotkHameq}) and (\ref{dEtensor}) 
one finds the following equation governing the dynamics of  tensor modes:
\begin{equation}
\frac{1}{2} \left[  {h^i_a}''+\left( 2 \mathbb{K}[2]- \frac{{\Omega'}}{\Omega} \right) {h^i_a}'
-\Omega \nabla^2 h^i_a \right] = \kappa \Pi^i_a,  
\label{TensorEOM}
\end{equation}
where 
\begin{equation}
\Pi^i_a = \Omega \left[  \frac{1}{3V_0}  \frac{\partial \bar{H}_{\text{m}} }{\partial \bar{p}} 
\left( \frac{\delta E^c_j \delta^j_a\delta^i_c}{\bar{p}} \right)+\frac{\delta H_{\text{m}}}{\delta \delta E^a_i}\right].
\end{equation}
In case of  scalar matter, the source term is vanishing $\Pi^i_a=0$ and the equation (\ref{TensorEOM})
simplifies to \cite{Cailleteau:2012fy}
\begin{equation}
{h^i_a}''+2\left(\mathcal{H}-\varepsilon \right) {h^i_a}'-c_s^2\nabla^2 h^i_a = 0,  
\end{equation}
where we also have used Eq. (\ref{ConfHubb}).  We have also defined the quantum correction 
to the friction term 
\begin{equation}
\varepsilon :=  \frac{1}{2} \frac{{\Omega'}}{\Omega} =  3 \mathcal{H} \left(\frac{\rho+P}{\rho_c-2\rho} \right),
\end{equation}
and the squared velocity
\begin{equation}
c_s^2 := \Omega = 1 - 2\frac{\rho}{\rho_{\text{c}}} \label{cs2},
\end{equation}
where $\rho$ and $P$ are the energy density and the pressure of the scalar field respectively.  
In the derivation of second equality in (\ref{cs2}), the homogeneous part of the Hamiltonian 
constraint has been used. In the classical limit, when $\rho_{\text{c}} \rightarrow \infty$ the 
correction term $\varepsilon \rightarrow 0$ and $c_s^2 \rightarrow 1$ recovering the 
classical equation of motion for the tensor modes:
\begin{equation}
{h^i_a}''+2\mathcal{H} {h^i_a}'-\nabla^2 h^i_a = 0.  
\end{equation}

However, when $\rho \rightarrow \frac{\rho_{\text{c}}}{2}$, quantum effects become dominant. 
In particular, since $c_s^2=0$ at $\rho = \frac{\rho_{\text{c}}}{2}$, the spatial derivatives are suppressed. 
Furthermore, the friction term diverges because $\epsilon \rightarrow \infty$ for $\rho \rightarrow \frac{\rho_{\text{c}}}{2}$.
At  energy densities $\rho \in (\rho_{\text{c}}/2, \rho_{\text{c}}]$, $c_s^2$ becomes negative and 
changes the type of the equation to the elliptic form. This behavior can be interpreted in terms of a 
metric signature change \cite{Bojowald:2011aa, Mielczarek:2012pf}. 

\subsection{Summary of results}

The theory of cosmological perturbations with holonomy corrections can be considered as fully derived. 
However, implications in the various cosmological scenarios, {\it e.g.} for the different scalar field potentials,
are awaiting to be explored. Here, to make our results more transparent, we collect derived equations 
which can be directly used for different cosmological applications.  The set of equations is relevant to study the
generation of  inflationary perturbations with  quantum holonomy effects. 

The cosmological scenario with holonomies can be split in background and perturbations parts. The background
dynamics is governed by the modified Friedmann equation 
\begin{equation}
\mathcal{H}^2 = a^2 \frac{\kappa}{3} \rho \left(1-\frac{\rho}{\rho_{\text{c}}}\right),   
\end{equation}
which is obtained by employing the homogeneous part of the scalar constraint $\bar{H}^{\text{new}}_G
+\bar{H}_{\text{m}}  \approx 0$ and Eq. (\ref{dotkHameq}). The energy density of the scalar matter field is
\begin{equation}
\rho = \frac{1}{2a^2} \left(\bar{\varphi}'\right)^2+V(\bar{\varphi}),
\end{equation}
and the Klein-Gordon equation in the FLRW background reads as
\begin{equation}
{\bar{\varphi}}''+2\mathcal{H} {\bar{\varphi}}'+ a^2 \frac{dV(\bar{\varphi})}{d\bar{\varphi}}=0.
\end{equation}
This equation is not affected directly by  holonomy corrections. 

Let us now consider perturbations. As we have already showed in the previous 
subsection the equation of motion for the tensor modes with holonomy correction is \cite{Cailleteau:2012fy}
\begin{equation}
{h}''_{ab}+2\left(\mathcal{H}-\frac{1}{2}\frac{{\Omega'}}{\Omega} \right) {h}'_{ab}-\Omega \nabla^2 h_{ab}= 0.  
\end{equation}
The vector modes, due to the scalar nature of the matter content, do not contribute and are 
identically equal to zero \cite{Bojowald:2007hv, Mielczarek:2011ph}.

The most interesting perturbations from the observational point of view are perhaps the scalar ones. For those
perturbations, the anomaly free formulation was found in Ref. \cite{Cailleteau:2011kr}. It was demonstrated that due to 
the holonomy corrections not only the scalar constraint is modified but also the expression of the gauge invariant
variables. Namely it was found that the gauge-invariant variables (Bardeen potentials) are:
 \begin{eqnarray}
\Phi &=& \phi +\frac{1}{\Omega} ({B'}-{E''}) +\left( \frac{\mathcal{H}}{\Omega}-\frac{{\Omega'}}{\Omega}\right)  (B-E'),  \\
\Psi &=& \psi - \frac{\mathcal{H}}{\Omega} (B-E'), \\
\delta \varphi^{GI} &=& \delta \varphi + \frac{\bar{\varphi}'}{\Omega} (B-E').
\end{eqnarray}
The $E,B,\phi$ and $\psi$ are the classical scalar perturbation functions. With use of the above gauge-invariant variables,
the  quantum corrected Mukhanov-Sasaki variable can be defined as
\begin{equation}
v_{\rm{S}} = a(\eta) \left( \delta \varphi^{GI} + \frac{\bar{\varphi}'}{\mathcal{H}} \Psi \right).
\label{MukhClass}
\end{equation}
The equation of motion for this variable is given by \cite{Cailleteau:2011kr}
\begin{equation}
{v}''_{\rm{S}} -\Omega \nabla^2 v_{\rm{S}}  - \frac{z''}{z} v_{\rm{S}}  = 0,
\end{equation}
where 
\begin{equation}
z = a(\eta) \frac{{\varphi}'}{\mathcal{H}}.
\end{equation}
                         
In the case of the  longitudinal gauge the equation of motion for $\phi$ takes the following form                          
\begin{eqnarray}
{\phi}''+2\left[\mathcal{H}- \left(\frac{\bar{\varphi}''}{\bar{\varphi}'}
+\varepsilon \right)\right]{\phi'} +2\left[ \mathcal{H}'-\mathcal{H} 
\left( \frac{\bar{\varphi}''}{\bar{\varphi}'}+\varepsilon\right) \right]\phi-c_s^2\nabla^2\phi=0,  
\label{finaleq}
\end{eqnarray}
where $\epsilon$ and $c_s^2$ are expressed in the same way as in the case of  tensor modes. 
The form of Eq. (\ref{finaleq}) agrees with the one found in Ref. \cite{WilsonEwing:2011es} 
using an alternative approach of introducing holonomy corrections, based on lattice states. 
In that approach, space is discretized by a cubic lattice, where each cube 
is described by the FLRW metric. Because of that, standard LQC methods 
can be applied to each cell whereas inhomogeneities are described in terms 
of interactions between neighboring cells.

\subsection{Holonomy corrected algebra of constraints}

The requirement of anomaly freedom of the algebra of constraints was a crucial 
ingredient of the construction of the effective theory of cosmological perturbations 
with holonomy corrections. This requirement ensures that  dynamical trajectories 
remain on the surface of constraints during the whole evolution, making the theory 
defined consistently. The closure of the algebra implies also that the algebra is of 
the first class, for which the constraints are becoming generators of gauge transformations. 
Furthermore, this requirement was proven to be very convenient, because it allowed 
one to fix all initial ambiguities related with the introduction of quantum corrections. 
As we have found, the requirement of anomaly freedom requires also that  the obtained 
algebra of effective constraints is deformed with respect to the classical one:
\begin{eqnarray}
\left\{D[M^a],D [N^a]\right\} &=& D[M^b\partial_b N^a-N^b\partial_b M^a],  \\
\left\{D[M^a],S^Q[N]\right\} &=& S^Q[M^a\partial_a N-N\partial_a M^a],  \\
\left\{S^Q[M],S^Q[N]\right\} &=& D\left[\Omega q^{ab}(M\partial_bN-N\partial_bM)\right]. 
\end{eqnarray} 
The term $\Omega= \cos (2 \gamma \bar{\mu} \bar{k}) =1-2\frac{\rho}{\rho_{\text{c}}}$ is 
a deformation function equal to one for the classical spacetime with Lorentzian signature.  

Classically, the algebra of constraints is encoding general covariance, a cornerstone of 
general relativity. Therefore, it is a very fundamental object. As it was shown in Ref. 
\cite{Hojman:1976vp},  Einstein equations can be regained from the algebra of constraints 
by finding its representation. This was proven only for the classical algebra where $\Omega=\pm1$.
Here, $``+''$ corresponds to the Lorentzian metric signature and $``-''$ to Euclidean one. 
So, only in this two cases does the standard metric description of spacetime geometry become 
relevant. In the case where $\Omega\neq \pm1$, a new description of the space-time 
geometry is required. 

An amazing consequence of the deformation of the algebra of constraints by the
holonomy  corrections is that the correction function $\Omega$ interpolates between
the two classical geometries, Lorentzian and Euclidean ones. Namely, in the low energy 
density limit  $\frac{\rho}{\rho_{\text{c}}} \rightarrow 0$, the $\Omega \rightarrow 1$
and the classical Lorentizan spacetime is recovered correctly. However, while the
energy density approaches its maximal value  $ \rho \rightarrow \rho_{\text{c}}$, then 
$\Omega \rightarrow -1$, describing a classical Euclidean spacetime. The effect of 
the holonomy corrections can  therefore be seen as a sort of dynamical Wick rotation.
However, it must be stressed that the for $-1 < \Omega < 1$, no classical space-time 
picture can be used in general.  There is however an exception for $\Omega=0$, corresponding to $\rho 
= \rho_{\text{c}}/2$. In this case the algebra of constraints reduces to the ultralocal 
form and becomes a Lie algebra. In this stage, the spatial derivatives are 
suppressed, leading to a state of asymptotic silence, where no information can 
propagate between spatial points \cite{Mielczarek:2012tn}. This state corresponds to 
the Carollian $c\rightarrow0$ limit studied at the level of the Poincar\'e algebra. 

\subsection{Holonomy corrections to the inflationary power spectra}

As an application of the equations of motion we will present holonomy corrections to the 
inflationary scalar and tensor power spectra \cite{Mielczarek:2012pf, Mielczarek2013}. We 
will focus on the slow-roll inflationary model driven by a single scalar field $\bar{\varphi}$ with potential 
$V(\bar{\varphi})$ occurring for $\rho \ll \rho_{\text{c}}/2$. The slow-roll parameters with the holonomy corrections are: 
\begin{eqnarray}
\epsilon :=  \frac{m_{Pl}^2}{16 \pi} \left( \frac{V_{,\bar{\varphi}}}{V} \right)^2\frac{1}{(1-\delta_H)}
\ \ \textrm{and} \ \ \eta :=  \frac{m_{Pl}^2}{8 \pi} \left( \frac{V_{,\bar{\varphi}\bar{\varphi}}}{V} \right)  \frac{1}{(1-\delta_H)}, 
\nonumber
\end{eqnarray}
where $\delta_H :=V/\rho_c$.

The derivation of the scalar and tensor power spectra is based on the application of the standard techniques 
of  quantum field theory on curved spaces. In calculations we have neglected the  pre-inflationary phase 
and normalized modes so that they agree with the WKB solution  
\begin{equation}
v_{\mathrm{S(T)},k} = \frac{1}{\sqrt{2k\sqrt{\Omega}} }e^{-ik\int^{\tau}\sqrt{\Omega(\tau')}d\tau'},
\label{WKBHolo}
\end{equation}
at  short scales ($\sqrt{\Omega}k \gg \mathcal{H}$). The solution (\ref{WKBHolo}) is automatically
satisfying Wronskian condition. For $\Omega=1$, the solution (\ref{WKBHolo}) reproduces the standard Minkowski 
vacuum.  

The obtained spectra of scalar and tensor (gravitational waves)  perturbations are
\begin{equation}
\mathcal{P}_{\text{S}}(k) = A_{\text{S}}\left(\frac{k}{aH}\right)^{n_{\text{S}}-1} \ \   \textrm{and} \ \ 
\mathcal{P}_{\text{T}}(k) = A_{\text{T}} \left(\frac{k}{aH}\right)^{n_{\text{T}}}, \nonumber
\end{equation}
where amplitudes and spectral indices are given as follows: 
\begin{eqnarray}
A_{\text{S}} &=& \frac{1}{\pi \epsilon} \left(\frac{H}{m_{Pl}} \right)^2  \left(1+2\delta_H \right) 
\ \ \textrm{and} \ \ n_{\text{S}} =1+2\eta-6\epsilon + \mathcal{O}(\delta_H^2), \nonumber \\
A_{\text{T}}  &=& \frac{16}{\pi} \left(\frac{H}{m_{Pl}} \right)^2  \left(1+\delta_H\right) 
\ \ \textrm{and} \ \ n_{\text{T}} =-2\epsilon+\mathcal{O}(\delta_H^2). \nonumber
\end{eqnarray}
Furthermore, the consistency relation reads as
\begin{equation}
r := \frac{A_{\text{T}}}{A_{\text{S}}} \simeq16\epsilon  \left(1-\delta_H  \right). 
\end{equation}

In the limit $\rho_c \rightarrow \infty$, the classical power spectra for  slow-roll inflation 
are recovered. The corrections are introduced through the factors $\delta_H$, which are of the 
order of $10^{-12}$ for typical values of the parameters. The corrections due to holonomies
are, in this approach, smaller than those expected from the inverse volume corrections.
Furthermore, it is worth stressing that for the chosen normalization, the spectral indices are 
not corrected in the leading order.  This makes the observation of these effects even harder. 
However, in the simplified analysis performed here, the effects due to the violent quantum dynamics in vicinity 
of  the transition point $\rho=\rho_c/2$ were not taken into account. Those effect will unavoidably 
lead to deformations of the inflationary predictions (small values of $k$). Furthermore, other possible
UV normalizations can be chosen, leading to linear (in $\delta_H$) corrections of the spectral indices 
\cite{Mielczarek2013}.\\

It should be noticed that interesting results about the matter bounce scenario and the Ekpyrotic universe were also obtained within the LQC framework \cite{ewe}. 

\section{Perturbations with inverse volume corrections}
\label{sec:iv}
The other main correction which can be studied in the effective theory is called the inverse volume (or, more 
generally, inverse-triad) correction. The inverse volume corrections are due to terms in the Hamiltonian constraint 
which cannot be quantized directly but only after being re-expressed as a Poisson bracket. Inverse volume, as 
defined, is associated with inverse powers of the determinant of the densitized triad, as seen in Eq. \ff{eq:cgrav}. 
As the volume operator can have zero as an eigenvalue, its inverse is consequently not densely defined and one 
way to overcome this difficulty is to re-express it through the so-called \textit{Thiemann's trick} \cite{Thiemann:1996aw}, 
in terms of the connection (and then holonomies) and positive volume powers,
\begin{equation} \lb{Thiemannstrick}
\{A^k_c,V\} = \frac{E^a_i E^b_j}{\sqrt{det(q)}} \epsilon^{ijk} \epsilon_{abc},
\end{equation}
where $det(q)$ is the determinant of the spatial metric, related to the densitized triad by the relation $ det(q) 
\cdot q^{ab} = E^a_i E^b_j \delta^{ij}$.
After using quantization tools inspired by LQG, the spectrum of the operator $\widehat{\frac{E^a_i E^b_j}{\sqrt{det(q)}}}$ 
can be derived. However, in the effective approach, we use an analytic expression which is exported into classical 
equations, so as to get an idea of the main consequences of this operator. Functional forms of the inverse-volume 
correction, usually called $\alpha$ in the gravitational sector, and $\nu$ and $\sigma$ in the matter sector,
\begin{equation}
\frac{1}{\sqrt{E}^n} \rightarrow \alpha \times \frac{1}{\sqrt{E}^n} 
\end{equation}
are in principe computable from the operators. This has been done for exactly isotropic models \cite{Bojowald:2001vw} 
and for regular lattice states with inhomogeneities \cite{Bojowald:2006zi}, with explicit parametrizations of the quantization 
ambiguities affecting the values of the parameters \cite{Bojowald:2002ny} \cite{Bojowald:2004ax}.\\
For now on, most studies on  inverse volume corrections have been done considering the semiclassical limit where 
quantum corrections are small and are generically defined as 
\begin{equation} \lb{semiclassshape}
\alpha \sim  1 + \alpha_0 \delta_{inv} \hskip0.5truecm \nu \sim  1 + \nu_0 \delta_{inv},
\end{equation}
with 
\begin{equation}
\delta_{inv} \sim a^{-\sigma},
\end{equation}
where $a$ is as usual the scale factor. The constants $\alpha_0$, $\nu_0$ and $\sigma$ encode the 
specific features of the model which will have an impact on the cosmological observables. It is therefore 
relevant to explore their allowed values.  This range strongly depends on the physical interpretation of the
model. There are mainly two views on this issue, one is the \textit{minisuperspace parametrization} and 
the other is the \textit{lattice refinement parametrization}.
\begin{itemize}
\item  The minisuperspace parametrization assumes an homogeneous and isotropic universe, and the 
calculations are performed within a finite volume, or fiducial cell, $V_0$. In this parametrization, 
\begin{equation}
\delta_{inv} \sim \left( \frac{l_{\text{Pl}}}{V_0} \right)^{-\frac{\sigma}{3}} a^{-\sigma},
\end{equation}
and one get \cite{Calcagni:2008ig} the natural choice:
\be \lb{alpha0nu0}
\sigma = 6,\qquad \alpha_0=\frac{1}{24}\approx
0.04\,,\qquad \nu_0=\frac{5}{36}\approx 0.14\,, 
\ee 
where $\sigma = 6$ corresponds to the \textit{improved quantization scheme} \cite{Ashtekar:2006wn}, 
or $\bar{\mu}-scheme$.

Phenomenologically, at the current level of precision, the most significant parameter is $\sigma$, which is 
not that much affected by different choices of the minisuperspace scheme. A possibly puzzling aspect is however 
the dependence of $\delta_{inv}$ upon $V_0$,  which will affect the equations of motion. For this picture to 
make sense, the final results should not depend on $V_0$.  As explained in \cite{Bojowald:2010me}, due to 
this dependence on the volume, a fully consistent derivation of the inverse volume correction might still be 
missing. In particular, as discussed in \cite{Bojowald:2007ra} and \cite{Bojowald:2008ik}, while the improved 
scheme does take into account refinement for holonomies, this scheme is not effective for the inverse volume 
corrections. A perfectly well-defined Hamiltonian is still missing. This difficulty to represent inverse volume 
effects can be seen as a serious limitation of pure minisuperspace models. Nevertheless, as it will become 
clear in the following, this limit can be overcome when taking into account inhomogeneities.
 
\item The lattice refinement parametrization is associated with inhomogeneities. In loop quantum cosmology, 
the spacetime is discrete and the dynamics can be seen as involving a patchwork of $\mathcal{N}$ volumes of different 
dynamical sizes. Instead of considering the whole volume $V = a^3 V_0$ as the place where to perform calculations, 
one assumes only the "mean" volume of a discrete patch,
\begin{equation}
v = \frac{V}{\mathcal{N}}.
\end{equation}
By construction, $v$ is now independent of the size of the region, since both $V$ and $\mathcal{N}$ scale in 
the same way when the size of the region is changed. Therefore, one can consider now
\begin{equation}
\delta_{inv} \sim \left(\frac{l_{\text{Pl}}^3}{v} \right)^{\frac{m}{3}} = \left( l_{\text{Pl}}^3 \frac{\mathcal{N}}{V}\right)^{\frac{m}{3}},
\end{equation}
where $m>0$. The picture can be improved by considering the dynamical \textit{lattice-refinement} behavior 
\cite{Bojowald:2006qu} where $\mathcal{N}(t)$ is time-dependent and written as
\begin{equation}
\mathcal{N} = \mathcal{N}_0 a^{-6x},
\end{equation}
where $\mathcal{N}_0$ is some coordinate-dependent and $V_0$-dependent parameter. The power $x$ describes different 
qualitative behaviors of  lattice changes and might be tuned to allow the model to match phenomenology 
\cite{Bojowald:2011iq}. The exponent has to be negative for the number of vertices $\mathcal{N}$ to increase with 
the volume. Comparing with the minisuperspace parametrization, one sees that $\sigma = (2 x + 1)m $, and  
that $v \sim a^{3(1+2x)}$ so that to be consistent $-1/2 \leq x \leq 0$ and consequently $ \sigma \geq 0$. 
If $\sigma < O(1)$, it can also be shown that the background inflates \cite{Bojowald:2011iq}.\\
In this scheme, the regions of size $v$ are given by an underlying discrete state and therefore present quantum 
degrees of freedom which do not exist classically. Moreover, this discretization of the geometry leads naturally to 
the presence of inhomogeneities which are thus unavoidable. The volume $V$, which could initially be seen as 
homogeneous in a first approximation, is in fact a region of the universe coarse grained into smaller regions of volume 
$v$ where the time-dependent inhomogeneities are small. The only reference to the physical space considered is 
now done through $\mathcal{N}_0$, which is general, and not anymore through a time-dependent volume $V(a)$. 
As long as one treats almost scale-invariant linear perturbations, the lattice refinement scheme might therefore, in 
this specific sense, be better suited to study  cosmological perturbations.  
\end{itemize}  
The inverse volume corrections can consequently be made physically meaningful. However, without fully knowing their 
subtle relation with the 
details of the underlying full quantum theory, their behavior is more relevant than their magnitude. Moreover, it has 
recently been claimed in \cite{Bojowald:2011iq} that, from precise \textit{semiclassical} arguments in the lattice 
picture (for instance, due to isotropy, one can  no more consider $SU(2)$ valued variables but $U(1)$ 
instead),  the minisuperspace parametrization leads to some tensions in situations where the lattice refinement 
picture seems, from this specific point of view, consistent. In this former regime, the inverse volume corrections 
depend only on the triad variables, $\alpha(E^a_i)$, and not on the connection (this was regarded before as a technical 
assumption without physical motivations). The usual minisuperspace parametrization of FLRW LQC seems to be in 
tension with the anomaly cancellation in inhomogeneous LQC, as well as with the simplest power-law solution. 
The lattice refinement parametrization was designed to overcome some of those problems, leading to much larger 
quantum corrections. It might  however lead to intricate situations, in particular due to the possible superluminal 
propagations of the perturbations. 
It will be shown in the next section that this entire view has to be refined and the ``non-dependence'' 
on the connection in the lattice picture might not be true anymore in the minisuperspace parametrization. 
This opens interesting perspectives.\\

Nevertheless, contrary to the Wheeler-DeWitt model, one new feature resulting from the fact that quantum corrections 
enter directly in the expression of the constraints is that now the background equations of motion undergo also the effect 
of the corrections. For instance, considering a massive scalar field $\bar{\varphi}$ as the main content of the Universe, 
the Friedmann equation in this framework is now modified such that 
\begin{equation}
H^2 = \left( \frac{\dot{a}}{a} \right)^2 = \frac{\kappa}{3} \bar{\alpha} \left( \frac{\dot{\bar{\varphi}}^2}{2 \bar{\nu}} + V(\bar{\varphi})\right),
\end{equation}
with the Klein-Gordon equation
 \begin{equation}
\ddot{\bar{\varphi}}+3H \left(1-\frac{d \ln \bar{\nu}}{d \ln \barp} \right)\dot{\bar{\varphi}}+ \bar{\nu} \partial V_{\bar{\varphi}} =0, 
\end{equation} 
where $\bar{X}$ stands for the purely homogeneous part of the variable $X$.
One should notice here that neither quantum backreaction nor the holonomy corrections are  included at the 
present stage. The inverse-volume correction was studied at the semiclassical limit, where it is hard to make 
comparisons with holonomy corrections. Nevertheless, one will see in the next section that a \textbf{consistent model} 
can be constructed considering both corrections simultaneously.
\vskip0.2truecm
As said before for the case where  holonomy corrections were taken into account, when perturbations and quantum 
corrections are considered, the closure of the effective constraint algebra must be imposed for consistency. The cancellation 
of anomalies is obtained by introducing counterterms. 
 After some early works based on toy models where the constraint algebra was not  explicitly closed 
 \cite{Bojowald:2002nz,Tsujikawa:2003vr,Bojowald:2004xq,Hossain:2004wm,Calcagni:2006pr,cope,Shimano:2009tn}, 
 the full set of constraints with small inverse-volume corrections was derived for vector \cite{Bojowald:2007hv}, tensor 
 \cite{bojo1}, and scalar modes \cite{Bojowald:2008gz,Bojowald:2008jv}. The associated gravitational wave spectrum 
 was studied in \cite{Calcagni:2008ig,cope2,grain_iv_ds}, while the scalar spectrum and the full set of linear-order 
 cosmological observables were found in \cite{Bojowald:2010me}. Observational consequences and subsequent 
 constraints on the quantum corrections were finally studied in \cite{Calcagni:2012vw,Bojowald:2011iq,Bojowald:2011hd,Calcagni:2011xj}. 

The requirement of consistency in a given scheme uniquely relates the counterterms, all vanishing at the classical 
limit, to the primary correction functions, but also restricts the range of allowed values for $\alpha$ and $\nu$. For 
the inverse-volume correction, considering  scalar perturbations, the first attempt to get an anomaly-free algebra 
was derived in \cite{Bojowald:2008gz}. In fact, as noticed in \cite{Cailleteau:2012fy},  scalar perturbations are the 
most general case and allow one to derive the full  underlying set of counterterms, fulfilled by all  
perturbations. (It is however worth to point out that one counter-term, usually called $\alpha_9$ in 
\cite{Cailleteautoappear}, as to be different depending on the kind of perturbations. This counterterm is important 
since it defines the speed of propagation of the tensor perturbations.  Its expression, obtained for the scalar perturbations, 
can vary. In the case where only  tensor perturbations are considered, it is not possible -- due to the properties of 
the tensor perturbations -- to derive an expression for $\alpha_9$ : there are too many degrees of freedom and 
one is led to guess what the expression of $\alpha_9$ could be.) 
The scalar constraint taking into account the inverse volume correction is written as
\begin{equation}
\mathcal{C}[N] \sim \int d^3 x N [ \bar{\alpha} C_g + \bar{\nu} C_{\pi} + \bar{\sigma} C_{\nabla} + C_V ].
\end{equation}
In first attempts, the anomaly functions for both inverse-volume corrections and counterterms were linearized. 
For instance, it was assumed that
\begin{equation} \lb{linear}
(1+f)(1+h) = 1 + f + h + O(l_{\text{Pl}}^2), 
\end{equation}
and an extra consistency condition between 2 counterterms was obtained
\be
\lb{consistencyIV1}
2\frac{d f_3}{d \ln p}+3(f_3-f)=0. 
\ee
Eq. \ff{consistencyIV1} is important because it lead to relations between the 
parameters \cite{Bojowald:2010me}: 
\be\label{extra}
\alpha_0\left(\frac{\sigma}6-1\right)-\nu_0\left(\frac{\sigma}6+1\right)\left(\frac{\sigma}3-1\right)=0\,.
\ee
In this inverse-volume case, the algebra shape is now modified and given by:
\begin{equation}
\{ \mathcal{C}[N_1],\mathcal{C}[N_2] \} = D[\alpha^2 \left( N_1 \partial^a N_2 - N_2 \partial^a N_1 \right)].
\end{equation}
The deformation algebra differs from what was obtained in the holonomy case. As pointed out in 
\cite{Calcagni:2012vw}, the somewhat unexpected possibility that LQC quantum corrections be 
large even during inflation might reflect of the way these corrections enter the game: the structure 
of spacetime itself is deformed by quantum effects, via the effective constraints. The theory is 
diffeomorphism invariant with respect to the consistently associated transformations. Gauge 
transformations belonging to a deformed algebra no longer correspond to ordinary coordinate 
transformations on a manifold. Thus, to take the new gauge structure into account one might better 
rely only on gauge-invariant perturbations. This philosophy (first quantize the classical system, then 
cast it in gauge-invariant variables) is embodied in the Mukhanov-Sasaki  equations. One might wonder 
whether one would obtain the same results by fixing the gauge before quantizing. Gauge fixing 
and quantization do not commute in general because the latter deeply affects the very notion of 
gauge invariance. Whenever gauge-ready variables can be constructed after quantizing, the 
gauge-invariant approach might be preferred. Although it is still possible to find some counterexamples 
to this strategy. There are many debates around this issue. In the following we try to avoid ``too early'' 
gauge fixing as long as possible.\\

In the presence of small inverse-volume corrections, after anomaly cancellation, the system of perturbed 
equations for scalar and tensor modes (vector modes are damped during inflation) reduces to 2 equations 
\cite{Bojowald:2010me}. The first one is the modified Mukhanov-Sasaki equation for scalar perturbations,
\begin{equation} \lb{scalarIVboj}
v''_\mathrm{S} - \left(s^2 \Delta + \frac{z''}{z}\right) v_\mathrm{S} = 0, 
\end{equation}
where prime refers to the conformal time (in which case the Hubble parameter is written $\mathcal{H}$), and 
\begin{equation}
s^2 = \alpha^2 (1-f_3)= 1 + \chi \delta_{inv},
\end{equation}
with 
\begin{equation}
\chi = \frac{\sigma \bar{\nu}_0}{3} \cdot \left( \frac{\sigma}{6}+1\right) + \frac{\bar{\alpha}_0}{2} \cdot \left( 5-\frac{\sigma}{3} \right),
\end{equation}
and 
\begin{equation}
z \,\,\, \dot{=} \,\,\,a(\eta) \frac{\bar{\varphi}'}{\mathcal{H}} \left[1 + \left( \frac{\alpha_0}{2} - \nu_0\right) \delta_{inv} \right]. 
\end{equation}

On a general quasi-de Sitter background, the power spectrum derived from Eq. \ff{scalarIVboj} has been 
calculated in \cite{Bojowald:2010me}:
\be\label{scasp}
\phantom{\Biggl(}{\cal P}_{\rm S} = \frac{G}{\pi}\frac{\mathcal{H}^2}{a^2\epsilon}\left[1+\left( \nu_0\left(\frac{\sigma}{6}+1\right)+\frac{\sigma\alpha_0}{2\epsilon}-\frac{\chi}{\sigma+1} \right) \delta_{\rm Pl} \right],
\ee
where the spectral index is given by: 
\be
n_{\rm s}-1 = 2\eta-4\epsilon +\sigma \left(  \a_0-2\nu_0+\frac{\chi}{\sigma+1} \right)\delta_{\rm Pl}  \phantom{\Biggl(},
\ee
which corresponds to an almost scale-invariant power spectrum. \\ 
As recalled in \cite{Bojowald:2010me}, and generic in cosmology, the fact that  scalar perturbations reduce to just 
one degree of freedom $u$ obeying a closed equation is related to the conservation of the gauge-invariant comoving 
curvature perturbation $\mathcal{R} = \frac{v_\mathrm{S}}{z}$ on large scales. A failure of the algebra closure would presumably 
immediately spoil also this property.\\
Moreover, the equation of motion for the tensor perturbations $h(\eta)$ can be written as
\begin{equation} \lb{oldeqIV}
v''_\mathrm{T} + \left(\bar{\alpha}^2 \Delta- \frac{\tilde{a}''}{\tilde{a}}\right) v_\mathrm{T} = 0,
\end{equation}
with 
\begin{equation}
\tilde{a} = a \left( 1 - \frac{\alpha_0}{2} \delta_{inv} \right),
\end{equation}
and $v_\mathrm{T}(\eta) = \tilde{a}(\eta)  h(\eta)$.\\ 

\begin{figure}
\begin{center}
\includegraphics[scale=0.5]{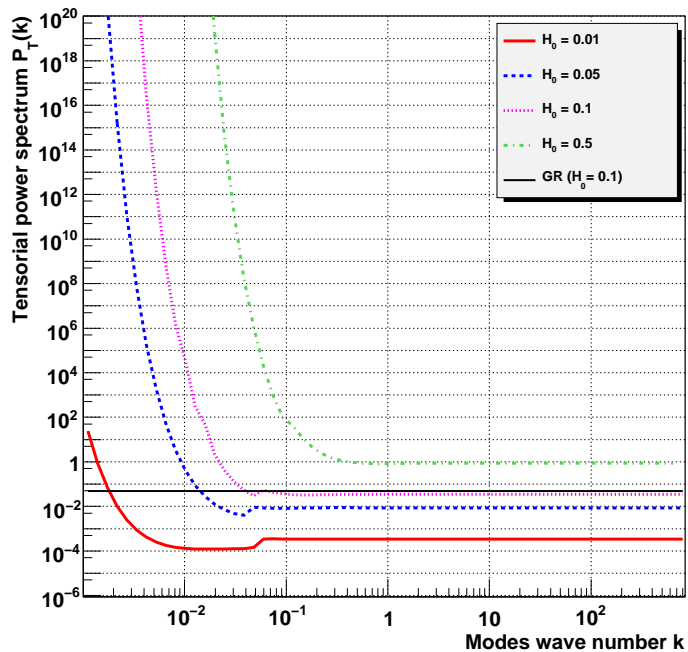}
\caption{Examples of tensor power spectrum modified by the inverse-volume correction without modifying 
the background \cite{grain_iv_ds}}
\label{fig:IVgrain}
\end{center}
\end{figure}

To have a first idea of the effect of inverse-volume corrections on the propagation of modes, a first attempt was 
\cite{grain_iv_ds} deliberately ignoring LQC corrections to the background. The power spectrum was shown to 
have the following IR limit: 

\begin{equation}
\mathcal{P}_{\mathrm{T}}^\mathrm{(IR)}=\left(\frac{\lpl}{\ell_0}\right)^2
\left(\frac{2^{\frac{3}{2}}\Gamma\left(3/2+\epsilon\right)}{\pi}\right)^2\left[2Z(1+\epsilon)\right]^{-\frac{3}{2}-\epsilon}
k^3\exp{\left(\frac{\pi\sqrt{2Z}(1+\epsilon)}{2k}\right)} \lb{scalpowgrain}.
\end{equation}
The full numerical computation is shown in Fig. \ref{fig:IVgrain}.

Nevertheless, an early work was also performed in \cite{Calcagni:2008ig} and both attempts were 
improved in \cite{Bojowald:2010me} where the tensor power spectrum is finally given by: 
\be
\phantom{\Biggl(}{\cal P}_{\rm T} = \frac{16G}{\pi}\frac{\mathcal{H}^2}{a^2} 
\left[1+\left( \frac{\sigma-1}{\sigma+1}\a_0\right) \delta_{\rm Pl} \right]\phantom{\Biggl(}.
\ee

Recently, a new approach \cite{Cailleteautoappear} to derive an anomaly-free algebra for the 
inverse-volume correction has been considered: this model is in fact based on the attempt to 
derive an anomaly-free algebra for both corrections in such a way that, contrary to previous 
works, the anomalies were kept non linear. Basically, Eq. \ff{linear} now becomes
\begin{equation}
(1+f)(1+h) = 1 + f + h + f h. 
\end{equation}
New expressions for the counterterms were obtained and a new algebra of deformation should 
therefore be considered. In this approach, the minisuperspace parametrization was considered 
without assuming a specific shape for the corrections, and the dependence of the corrections on the connection or/and the triad was let free. As a consequence, somehow surprisingly, it appeared that only the zeroth order in the corrections needs to be considered in the modification of the constraints.
Eq. \ff{consistencyIV1} is of course recovered, but it is no more a consistency condition: it rather 
allows one to understand relations between both counterterms. Not assuming the specific dependence of the inverse volume corrections on the connection or on the triad allows one to determine one counterterm which could not be derived otherwise and which will be of great importance when both corrections will be considered. 

\[\]
Therefore,  in the general case, without any assumptions, the structure function will be given by 
\begin{eqnarray}\lb{omegaIVTOT}
\Omega &=& \left(\frac{\partial^2}{\partial\bark^2}\bark^2\bar{\alpha}\right)\Sigma_{\barp}f_1[\barp],
\label{OmegaIV}
\end{eqnarray}
where $f_1[\barp]$ is a function determined by finding the expression for $ \alpha_3 = -1+\frac{f_1[\barp]}{\bar{\alpha}}$ (see \cite{Cailleteautoappear} for more details) and $\Sigma_{\barp}$ is the result of a differential equation on $\bar{\nu}$. \\
Different cases can be considered after some further assumptions. First, the simplest one in the case where $\bar{\alpha} = \bar{\alpha}(\barp)$ only. Then, 
\begin{equation}\lb{omegap}
\Omega= f_1[p] \times \sqrt{\gamma_0}.
\end{equation}
So far, nothing has been assumed on the shape of $f_1[p]$ except that it has to go to 1 at the classical limit. One could of course assume naturally that $ f_1[p] = \frac{1}{\sqrt{\gamma_0}} $ such that $\Omega=1$ and  $\alpha_3 = -1+\frac{1}{{\gamma_0}^{\frac{3}{2}}}$. \\
In the case where, for instance, $\bar{\alpha} = \bar{\alpha}\left[\frac{\bark}{\sqrt{\barp}}\right]$, the correction depending now also on the connection, an expression can be found for the correction if one assumes also that $f_1 = \Omega = 1$, such that 
\begin{equation} \lb{solIVkp}
\bar{\alpha} =1+ D_1 \frac{\sqrt{\barp}}{\bark} + D_2 \frac{\barp}{\bark^2},
\end{equation} 
where $D_i$ are constants. This diverges at the classical limit as long as $D_i \neq 0$ (is the opposite case, there are no inverse volume corrections anymore). Nevertheless, the interesting part comes from the matter sector, where an analytical expression for the correction can then be found, 
\begin{equation}\lb{solnu}
\nu_0=2\frac{\barp^3}{C_1^2}\left(\frac{C_1}{p^{3/2}}-\ln\left[1+\frac{C_1}{p^{3/2}}\right]\right),
\end{equation}
where $C_1$ is an unknown constant. Of course, this case is more a toy model than a real prediction. 

Finally, one could mix both case and consider the more general case where 
$\Omega=\Omega[\barp]$, whose solution for the correction is given by
\begin{equation}\label{solgamma}
\gamma_0=\frac{\Omega}{2\Sigma_{\barp}f_1[\barp]}+\frac{f_2[p]}{\bark}+\frac{f_3[p]}{\bark^2},
\end{equation}
for some unknown functions $f_i$ which leads to the good classical limit $\Omega=1$. In this case, nothing really probing can be said, more information on the unknown functions are required. Nevertheless, one can derived the modified Friedmann equation and discover surprisingly that a bounce is also possible (with just the inverse-volume correction), with a dynamical bounce time, depending on the value of the scale factor given by $\barp$ : 
\begin{equation}\lb{bounceIV}
H^2 = c_1[\barp] \left( \frac{\kappa}{3} \rho[\barp] - c_3[p] \right),
\end{equation}
where $c_i =g(f_i)$. A bounce is possible when $\frac{\kappa}{3} \rho[\barp] = c_3[p]$. \\
This new approach derived in  \cite{Cailleteautoappear} is therefore not yet fully conclusive but shows that, with more hints, the results can be made more consistent with the full theory where the correction should be also a function of the holonomy and therefore should depend also on the connection.  Moreover, it is possible to see that the case $\Omega =1$ is consistent which shows that having a perturbed and quantum-corrected constraint whose corrections depend possibly on both the connection and the flux of densitized triad does not necessarily lead to a deformed algebra as suggested in some previous work.  \\
Even at this stage, it is possible to derive the equations of motion for the different kind of perturbation, namely the tensor and scalar ones. For the tensor perturbations, the equation is easily derived 
\begin{eqnarray}
0 &=& h'' + h' \left( 2\Hc\left(1+\frac{2\barp}{f_1[\barp]}\frac{df_1[\barp]}{d\barp}\right) -\frac{{\Omega}'}{\Omega}\right) \nonumber \\
&&-\frac{\Omega}{2\Sigma_{\barp}}\left(1+\frac{2\barp}{f_1[\barp]}\frac{df_1[\barp]}{d\barp}\right)\nabla^2 h, \lb{tensorIValone}
\end{eqnarray}
which is not really the case for the scalar perturbations as, when using the generic approach first developed for general relativity in \cite{Langlois:1994ec} and then adapted to LQC in \cite{Cailleteau:2011mi}, some rather complicated and long functions of derivatives of the corrections appear. Therefore, no real expression can be given but as soon as one will be able to derive the final expressions for the corrections, using the references cited above, one will be able to obtain the correct equation of motion.
It is nevertheless possible to see quickly that the speed of propagation of both kind of perturbations will be different. This deserves further investigations.\\
Most of the relevant cosmological studies performed in \cite{Bojowald:2010me} are still to be performed 
in this framework. \\
One could wonder if instead of using the minisuperspace parametrization, the lattice refinement parametrization 
should not be preferred so as to obtain a meaningful interpretation of the inverse-volume correction. However, 
the expression of the quantum corrections expressed in terms of $\alpha$ has to fulfill an equation where no 
explicit parametrization was used. The approach therefore remains in fact valid and one can also rely 
on the model in the lattice picture.

\[\]

\section{Fully corrected effective-LQC algebra}
\label{sec:all}

\subsection{First attempt}

Although the curvature regimes for which inverse-volume and holonomy corrections play their dominant roles might 
not be exactly the same, there are obviously times in the early Universe when both are simultaneously relevant. 
It is therefore interesting to try to account for both simultaneously. A first attempt has been performed in this direction 
in \cite{grain_iv_holo_ds}. However, in this early study, the correct algebra was not yet available. In addition, the background 
was assumed to be the standard inflationary one. After quite a lot of algebra, it was shown that in the slow-roll approximation 
$a(\eta) = l_0 |\eta|^{-1-\epsilon}$, one is led to the effective Schr\"{o}dinger equation 
$\left[ \frac{d^2}{d \eta^2} + E_k(\eta) - V(\eta)   \right] \phi_k(\eta) = 0,$ with, at first order,
\begin{eqnarray}
E_k(\eta) &=& S^2 k^2 = \left[1+ 2\lambda_s \left( \frac{l_{\text{Pl}}}{l_0} \right)^s |\eta|^{s(1+\epsilon)} \right] k^2, \\
V(\eta) &=& \frac{2+3 \epsi}{\eta^2} +\frac{6}{\kappa} \frac{1}{\rho_c}  \frac{(1+4 \epsi)}{l_0^2} |\eta|^{-2 (1-\epsi)}  \nonumber \\
&+& \lambda_s \left(\frac{l_{\text{Pl}}}{l_0} \right)^s  \left[ - \frac{12}{\kappa} \frac{1}{\rho_c} \frac{(1+4 \epsi)}{l_0^2}
|\eta|^{s-2 + \epsi (s+2)} + s(1+2\epsi) |\eta|^{s(1+\epsi)-2}  \right. \nonumber \\
&-& \left. \frac{1}{2} s(s-1+\epsi(2s-1)) |\eta|^{s(1+\epsi)-2} \right],\nonumber \\
\end{eqnarray}
using the effective parametrization $S=1 + \lambda_s (q)^{-\frac{s}{2}}$,
$q=(a/l_{\text{Pl}})^2$ for the inverse-volume correction.\\

The analytical investigation shows that the UV limit is equivalent to the GR case whereas the IR limit is given by
\begin{equation}
\mathcal{P}_\mathrm{T}^\mathrm{(IR)}(k) = 16 \pi^3 \left(\frac{l_{\text{Pl}}}{l_0} \right)^2 (Z(1-4\omega))^{-\frac{3}{2}} k^3 e^{\pi \sqrt{\frac{Z}{8}}
\frac{(1-4\omega)}{k}},
\label{spectrumboth}
\end{equation}
leading to a similar power spectrum as the one depicted in Fig. \ref{fig:IVgrain}.
The holonomy and inverse-volume
corrections alone (still in the standard inflationary background assumed in this section) lead to very different 
spectra. The result shown here underlines that the power spectrum is increasing at the IR limit, in exact agreement 
with what was obtained with the inverse-volume correction alone. This proves that, under the standard inflationary 
background evolution hypothesis, the inverse-volume term strongly dominates over the holonomy one. This is 
to be contrasted with the background evolution in the very remote past where the holonomy term alone leads 
to the replacement of the singularity by a bounce.

\subsection{Full treatment}

These first attempts showed that both corrections act multiplicatively with respect to each other: this mainly 
remains true in the recent study deriving the exact algebra for the complete case \cite{Cailleteautoappear}. 

In this former approach, a model of anomaly-free algebra where both corrections are considered simultaneously 
was obtained building on the recent lessons learned when deriving the algebra for either holonomy or inverse-volume 
corrections, in particular when the anomalies are assumed to remain non-linear and when all counter terms are 
considered. Not so surprisingly, as for the holonomy correction, there is a quasi-unique way to obtain the expression 
of all the counterterms: the degeneracy observed for the inverse-volume only case is partly removed  due to the action of 
the holonomy corrections. A counterterms, in the gravitational sector, which was considered as vanishing in the case 
of the holonomy correction, but not in the case of the inverse-volume correction, has now to be taken into account.  
As this counterterm  plays an important  role in the propagation of  perturbations, the final results for the equations 
of motion may be substantially different with respect to previous studies. This might lead to specific observational 
signatures. Nevertheless, as stated previously for the case where only the inverse volume correction was considered, the lack of information on some unknown functions $f_i$ prevents us from finding the fully determined solution.

The main conclusion from the derivation of the full anomaly-free algebra is the same as mentioned previously: 
both corrections act mostly independently without deforming each other and the result is mostly similar to what 
would have been obtained if both 
corrections were taken separately. This is however not fully true: in order to remove all the anomalies, the 
inverse-volume correction has to depend on the connection and therefore on the holonomy. This has been claimed 
previously (based on the semiclassical approach were both correction were decoupled) but the derivation done in 
\cite{Cailleteautoappear} is intended at being more general. This connection-modification applies only to 
$\alpha$ and possibly to $\sigma$, but not to $\nu$ which depends only on the triads: as the matter sector does 
not depend on the connection $K$, only the gravitational one is affected, as shown previously. It was understood 
with the holonomy correction  that the matter counterterms should not depend on the connection (except for one 
specific counterterm). Consequently, 
only one non-vanishing counterterm was needed to define a consistent theory. If one considers the inverse-volume 
correction alone, the fact that this correction depends only on the triads doesn't prevent the matter counterterms to 
play a physical role and many of them are non-vanishing. However, when both corrections
are considered simultaneously, it can be seen that almost all  counterterms have to depend on the triads. It has to be 
underlined that, in principle, the holonomy correction is not expected to be modified by the inverse-volume correction. 
It is however not true the other way round, see Eq. \ff{Thiemannstrick}. One should therefore obtain the same global 
structure as the one found considering the holonomy correction, say typically the $\bar{\mu}$-scheme. 
It was shown 
\cite{Cailleteautoappear} that the connection-modification of the inverse-volume correction has  to fulfill some complicated 
and non-linear differential equations which will not be recalled here, except the more important ones.
 The counterterms were obtained following the same 
procedure as explained previously, fixing the ambiguities by taking the different required limits (classical, holonomy 
and inverse-volume).

\[\]

 All the counterterms have the correct ({\it i.e.} vanishing) classical limit, and as for the inverse volume correction, only the zeroth order of the correction is relevant. In the general approach, the structure function of the algebra is now given by:
\begin{eqnarray}
\Omega = \left(\frac{\partial^2}{\partial\bark^2}\gamma_0\ \mathbb{K}[1]^2\right)\Sigma_{\barp} f_1[\barp]\lb{OmegaIVHolo},
\end{eqnarray}
and as before, in the case where $\Omega= \Omega[\barp]$ (from which $\Omega = 1$ is an obvious subcase), an analytic expression for the correction can also be found:
\begin{equation}
\alpha =\frac{1}{ \mathbb{K}[1]^2}\left(\frac{\bark^2\Omega}{2\Sigma_{\barp}f_1[\barp]}+\bark f_2[\barp]+f_3[\barp]\right).
\end{equation}
It is interesting to notice that the expression of the correction has a really similar shape as the one given by Eq. (\ref{solgamma}) for the inverse-volume correction. Moreover, in order to understand the consequences of the correction on the related phenomenology, it is necessary to consider it with the constraint densities. An interesting consequences arises: in the gravitational constraint density, the term in the numerator is now given by $\K{1}^2$ which cancels exactly the same term in the correction, and only the terms in $\bark$ remains. In other words, after including this correction in the constraint, the final constraint will have exactly the same expression as the one derived previously for the inverse-volume correction only. Therefore, this case is not interesting, the holonomy correction being cancelled by the inverse-volume correction, even if the idea of the bounce remains. One can consequently assume that the interesting cases are obtained in a similar way as when the holonomy correction was considered, that is to say when the structure function $\Omega$  depends not only on $\barp$, but also on $\bark$. This question is left opened for further investigations.\\

As far as the background is concerned, the modified Klein-Gordon equation is the same 
as the one obtained for the inverse-volume only case. In cosmic time it reads as:
 \begin{equation}
\ddot{\bar{\varphi}}+3H \left(1-\frac{d \ln \bar{\nu}}{d \ln \barp} \right)\dot{\bar{\varphi}}+ \bar{\nu} \partial V_{\bar{\varphi}} =0,
\end{equation}
and the modified Friedmann equation is now given by 
\begin{equation}
H^2=\frac{\kappa}{3}\bar{\alpha} \rho\left(1-\frac{\rho}{\rho_c \bar{\alpha} }+\Gamma_{\bark[2]}+\frac{1}{4}\Gamma_{\bark[1]}^2\right),
\end{equation}
with 
\begin{eqnarray}
\Gamma_{\bark[1]} &:=& \frac{ \mathbb{K}[1]}{\gamma_0}\frac{\partial \gamma_0}{\partial \bark}, \\
\Gamma_{\bark[2]} &:=& \frac{ \mathbb{K}[2]}{\gamma_0}\frac{\partial \gamma_0}{\partial \bark}.
\end{eqnarray}

The equation of motion for the tensor perturbations would be the same as Eq. (\ref{tensorIValone}) but due to the lack of knowledge on some expressions, the one for the scalar perturbations is not fully determined.
When the connection correction will be completely determined, presumably from the full theory, it will be easy to 
derive the equation of motion for  scalar perturbations using the approach given in \cite{Cailleteau:2011mi}. 
It should be noticed that when both corrections are taken into account, the negative $\Omega$ phase, 
possibly associated with a change of signature or at least a change of regime, is not anymore unavoidable. It is not even clear that it remains possible.
The asymptotic silence phase described in the next sections of this article can however remain, 
depending on the expression of the (connection) correction that one should derive from the full theory.\\
The early attempt of \cite{grain_iv_holo_ds} has led to  Eq. \ff{spectrumboth} for the tensor power spectrum. 
Of course, as for the scalar perturbations, the background has also to be quantum-corrected as in \cite{Bojowald:2010me}, 
but this example shows that the infrared divergence due to the inverse volume corrections overcomes the 
contributions from the holonomies, and this effect may also be seen when deriving the power spectrum from the equations of motions. \\
The picture drawn here is far more consistent with the full theory than what was derived before. When one 
considers the ``quantized'' constraints, both corrections appear simultaneously and some effects cancel each 
other because of the subtle intertwining. When the corrections are taken separately, this ``counter action'' 
is not anymore present. 

It should be stressed that the work is in progress on this issue of the fully corrected algebra and results given
in this section should be considered as a first guess only.\\

Some concerns were expressed about the ``uniqueness" theorem of Hojman, Kuchar, and Teitelboim for the 
whole ``deformation algebra'' approach: if the phase space is the same as in GR, should not we end up with the 
very same algebra? In fact there is no inconsistency here as both the constraints and the algebra are deformed 
simultaneously. In addition, the phase space is not exactly the one of GR due to the way the truncation is performed. 
Although this is obviously not the only possible approach to investigate observational consequences of loop quantum 
gravity, the derivation of an anomaly-free deformed algebra and the subsequent calculation of power spectra is 
unquestionably a promising road.

\section{Dressed metric approach}
\label{sec:aan}

In a recent serie of papers, Refs. \cite{agullo1,agullo2,agullo3}, a different approach to derive the dynamics of 
cosmological perturbations propagating in a quantum background has been developped. The roadmap adopted there differs from the roadmap presented in the previous sections and the derived equations 
of motion for both scalar and tensor perturbations admit a different form than the previoulsy derived equations of motion.

In the previous sections, the derivation of the equations of motion roughly works as follows. First, the classical 
phase space of general relativity is reduced by  homogeneity and  isotropy symmetries. Loop quantization 
is then performed on such a symmetry reduced phase space $\Gamma_\mathrm{FLRW}$. For the case of the 
sharply peaked states, an effective dynamics for this symmetry reduced phase space is derived, leading to the 
above-defined modified Friedmann equation. The perturbations are then added on top of that effective description 
of the quantum universe in the following way. One starts from the classical Hamiltonian containing both the 
homogeneous and isotropic degrees of freedom ({\it i.e.} the background) and the first order inhomogeneous 
degrees of freedom ({\it i.e.} the cosmological perturbations). In this Hamiltonian, the homogeneous and isotropic 
degrees of freedom are then replaced by some effective functions of them accounting for either holonomy corrections 
or inverse-volume corrections, or both. This modification is performed by requiring the closure of the algebra. The dynamics 
of the perturbations are then given by the second order of this effective Hamiltonian restricted to the square of the 
first order perturbations ({\it i.e.} discarding second order perturbations). Finally, the inhomogeneous degrees of 
freedom are quantized using the techniques of quantum field theory in curved spaces, the background space being 
the effective background of LQC.

Another path is adopted in Refs. \cite{agullo1,agullo2,agullo3}. The starting point is not the reduced phase space 
of strictly homogeneous and isotropic background $\Gamma_\mathrm{FLRW}$ but the reduced phase space of 
the {\it perturbed} FLRW space, that is containing both the homogeneous and isotropic degrees of freedom and 
the inhomogeneous degrees of freedom at first order in perturbation, 
$\tilde\Gamma=\Gamma_\mathrm{FLRW}\times\Gamma_\mathrm{pert}$. 
The quantization is then directly performed starting from $\tilde\Gamma$. Starting from such a 'perturbed' phase 
space, the quantum states can be written as a tensor product 
$\Psi(\nu,v_\mathrm{S(T)},\varphi)=\Psi_\mathrm{FLRW}(\nu,\bar\varphi)\otimes\Psi_\mathrm{pert}(v_\mathrm{S},v_\mathrm{T},\bar\varphi)$ 
with $\nu$ the homogeneous and isotropic degrees of freedom and $v_\mathrm{S(T)}$ the degrees of freedom for 
perturbations. The quantization of the background, homogeneous and isotropic geometry is performed using  
loop quantization techniques, as explained in Sec. \ref{sec:lqcgen}. There exists many states being interpreted as 
the quantum geometry of the background but this is those coherent states sharply peaked around classical solutions 
that are of interest for cosmology. Those states are mainly considered in Refs. \cite{agullo1,agullo2,agullo3} 
though the developed framework can be applied to any background states. For such a quantum background geometry, 
it is possible to define a metric operator
\begin{equation}
\hat{g}_{\mu\nu}dx^\mu dx^\nu=\hat{H}^{-1}_\mathrm{FLRW}\ell^6\hat{a}^6(\bar\varphi)\hat{H}^{-1}_\mathrm{FLRW}d\bar\varphi^2-\hat{a}^2d\vec{x}\cdot d\vec{x},
\end{equation}
with $\hat{H}_\mathrm{FLRW}=\hbar\sqrt{\Theta_{(\nu)}}$ the Hamiltonian operator of the isotropic and homogeneous 
background, and $\ell^3$ the volume of the fiducial cell. (We recall that $\Theta_{(\nu)}$ is the difference operator 
introduced in Sec. \ref{sec:lqcgen}.) The quantum dynamics of the perturbations is then given by the second order part 
of the total Hamiltonian (still restricted to the square of the first order perturbations) raised to an operator. However, this 
operator acting on the perturbation part of the Hilbert space also depends on the time dependent scale factor of the 
{\it background}, meaning, as one can expect, that the perturbations are affected by the quantum background. To perform 
such a quantization and derive the quantum equation of motion for perturbations evolving in the quantum background, 
the authors of Refs. \cite{agullo1,agullo2,agullo3} make use of the techniques developed in \cite{ashkalew} allowing 
them to quantize a test scalar field evolving in a quantum geometry. This second part of the quantization roughly works 
as follows. First, one applies the full Hamiltonian operators (therefore operating both on the homogeneous and isotropic 
part of the quantum states and the perturbed part of the quantum states), {\it i.e.} 
$-i\hbar\partial_{\bar\varphi} \Psi(\nu,v_\mathrm{S(T)},\bar\varphi)=\left[\hat{H}_\mathrm{FLRW}+\hat{H}_\mathrm{pert}\right]\Psi(\nu,v_\mathrm{S(T)},\bar\varphi)$ 
and then switch to the interaction picture by performing the transformation
\begin{equation}
\Psi_\mathrm{int}=\exp\left[-\frac{i}{\hbar}\hat{H}_\mathrm{FLRW}({\bar\varphi}-{\bar\varphi}_B)\right]\left(\Psi_\mathrm{FLRW}(\nu,{\bar\varphi})\otimes\Psi_\mathrm{pert}(v_\mathrm{S(T)},{\bar\varphi})\right).
\end{equation}
Because the interaction picture makes use of a specific time variable, {\it i.e.} the relational time ${\bar\varphi}$, the operator 
$\hat{H}_\mathrm{pert}$ acting on the perturbed degrees of freedom has to be build using a specific choice for the 
lapse function to be in agreement with the choice of the relational time ${\bar\varphi}$. Second, the factor ordering of 
$\hat{H}_\mathrm{pert}$, also composed of the background metric operator, is chosen to be consistent with the factor 
ordering of the $\hat{g}_{\mu\nu}dx^\mu dx^\nu$ operator. The quantum dynamics in the interaction picture then reads 
(for tensor perturbations)
\begin{eqnarray}
\Psi_\mathrm{FLRW}\otimes i\hbar\partial_{\bar\varphi}\Psi_\mathrm{pert}= \nonumber \\
\frac{1}{2}\displaystyle\int\frac{d^3k}{(2\pi)^3}\left\{32\pi G \left[\hat{H}^{-1}_\mathrm{FLRW}\Psi_\mathrm{FLRW}(\nu,{\bar\varphi})\right]\otimes\left[\left|\hat\pi_{\mathrm{T},\vec{k}}\right|^2\Psi_\mathrm{pert}(v_\mathrm{S(T)},{\bar\varphi})\right]\right. \\
\left.+\frac{k^2}{32\pi G}\left[\hat{H}^{-1/2}_\mathrm{FLRW}\hat{a}^4({\bar\varphi})\hat{H}^{-1/2}_\mathrm{FLRW}\Psi_\mathrm{FLRW}(\nu,{\bar\varphi})\right]\otimes\left[\left|\hat{v}_{\mathrm{T},\vec{k}}\right|^2\Psi_\mathrm{pert}(v_\mathrm{S(T)},{\bar\varphi})\right]\right\}, \nonumber
\end{eqnarray}
with $\left(\hat{v}_{\mathrm{T},\vec{k}},\hat{\pi}_{\mathrm{T},\vec{k}}\right)$ the configuration and momentum 
operators for the perturbations degrees of freedom. Taking the scalar product of the above equation with 
$\Psi_\mathrm{FLRW}$ finally leads to the Schr\"odinger equation for the perturbation part of the wave function
\begin{eqnarray}
	i\hbar\partial_{\bar\varphi}\Psi_\mathrm{pert}&=&\frac{1}{2}\displaystyle\int \frac{d^3k}{(2\pi)^3}\left\{32\pi G\left<\hat{H}^{-1}_\mathrm{FLRW}\right>\left|\hat\pi_{\mathrm{T},\vec{k}}\right|^2\Psi_\mathrm{pert}\right. \\
	&&\left.+\frac{k^2}{32\pi G}\left<\hat{H}^{-1/2}_\mathrm{FLRW}\hat{a}^4({\bar\varphi})\hat{H}^{-1/2}_\mathrm{FLRW}\right>\left|\hat{v}_{\mathrm{T},\vec{k}}\right|^2\Psi_\mathrm{pert}\right\}. \nonumber
\end{eqnarray}
In the above, $\left<\cdot\right>$ means the quantum expectation value of background operators on the background state, {\it i.e.} $\left<\hat{H}^{-1}_\mathrm{FLRW}\right>:=\left<\Psi_\mathrm{FLRW}\right|\hat{H}^{-1}_\mathrm{FLRW}\left|\Psi_\mathrm{FLRW}\right>$.

Let us now consider the quantization of cosmological perturbations  evolving in a {\it classical} FLRW 
background. Using a Schr\"odinger approach to perform this quantization, the quantum dynamics of 
perturbations in a classical background reads
\begin{equation}
i\hbar\partial_{\bar\varphi}\Psi_\mathrm{pert}=\frac{1}{2}\displaystyle\int \frac{d^3k}{(2\pi)^3}\left\{32\pi G(\bar{p}_{\varphi})^{-1}\left|\hat\pi_{\mathrm{T},\vec{k}}\right|^2\Psi_\mathrm{pert}+\frac{k^2}{32\pi G}(\bar{p}_{\varphi})^{-1}{a}^4({\bar\varphi})\left|\hat{v}_{\mathrm{T},\vec{k}}\right|^2\Psi_\mathrm{pert}\right\},
\end{equation}
with $a$ the classical scale factor and $\bar{p}_{\varphi}$ the classical conjugate momentum of the scalar field. 
The above equation is obtained using a scalar field as the time variable, leading to the following 
expression of the classical metric
\begin{equation}
g_{\mu\nu}dx^\mu dx^\nu=(\bar{p}_{\varphi})^{-2}\ell^6 a^6({\bar\varphi}) d{\bar\varphi}^2-a^2({\bar\varphi}) d\vec{x}\cdot d\vec{x}.
\end{equation}
There is a formal analogy between the quantum dynamics of perturbations evolving on a classical 
background and the quantum dynamics of the perturbations evolving in a quantum background. As 
a consequence, the dynamics of perturbations in a quantum background can be formally described 
as the dynamics of perturbations in a classical background but using a {\it dressed} metric, {\it i.e.}
\begin{equation}
	i\hbar\partial_{\bar\varphi}\Psi_\mathrm{pert}=\frac{1}{2}\displaystyle\int \frac{d^3k}{(2\pi)^3}\left\{32\pi G(\tilde{p}_{\varphi})^{-1}\left|\hat\pi_{\mathrm{T},\vec{k}}\right|^2\Psi_\mathrm{pert}+\frac{k^2}{32\pi G}(\tilde{p}_{\varphi})^{-1}\tilde{a}^4({\bar\varphi})\left|\hat{v}_{\mathrm{T},\vec{k}}\right|^2\Psi_\mathrm{pert}\right\},
\end{equation}
using the identification
\begin{eqnarray}
	(\tilde{p}_{\varphi})^{-1}=\left<\hat{H}^{-1}_\mathrm{FLRW}\right> &~\mathrm{and}~&\tilde{a}^4=\frac{\left<\hat{H}^{-1/2}_\mathrm{FLRW}\hat{a}^4({\bar\varphi})\hat{H}^{-1/2}_\mathrm{FLRW}\right>}{\left<\hat{H}^{-1}_\mathrm{FLRW}\right>}.
\end{eqnarray}
This dressed metric $\tilde{g}_{\mu\nu}$ is {\it not} equal to the classical metric. More importantly, 
it is also {\it a priori} not equal to the metric traced by the peak of the sharply peaked background 
state. In other words, there is no reason for $\tilde{p}_{\bar\varphi}$ and $\tilde{a}$ to be solution of the 
modified, effective Friedmann equation given in Sec. \ref{sec:bckg}. (This is inevitable as the quantum 
operators do not commute.) Thanks to this identification, the final quantization of perturbations can 
be done using the standard techniques of quantum field theory in curved spaces but using the dressed 
metric instead of the classical one.

Such a framework presented for tensor modes has been widely developed in Ref. \cite{agullo2} for the 
case of a free scalar field and is extended to incorporate scalar perturbations. This has subsequently 
been amended to the case of a massive scalar field and finally applied to the case of a bouncing and 
inflationary LQC universe in order to compute the impact of LQC on the primordial power spectra. This 
has been mainly done in Refs. \cite{agullo1,agullo3}. Using the standard techniques of quantum field 
theory in curved spaces, the quantum states for perturbations are given by the knowledge of the mode 
functions solutions to the fields equations of motion. The main result is that those equations of motion for 
both scalar and tensor perturbations in a quantum background have the same form as in the classical 
theory of cosmological perturbations but replacing the classical metric by the dressed metric, {\it i.e.}
\begin{eqnarray}
&&Q''_k+2\left(\frac{\tilde{a}'}{\tilde{a}}\right)Q'_k+\left(k^2+\tilde{U}\right)Q_k=0, \\
&&h''_k+2\left(\frac{\tilde{a}'}{\tilde{a}}\right)h'_k+k^2h_k=0,
\end{eqnarray}
with $Q_k$ a gauge-invariant variable for scalar, related to the Mukhanov-Sasaki variables via 
$Q_k=a^{-1}\times v_{\mathrm{S},k}$; $\tilde{U}$ is a dressed potential-like term given by
\begin{equation}
\tilde{U}({\bar\varphi})=\frac{\left<\hat{H}^{-1/2}_\mathrm{FLRW}\hat{a}^2({\bar\varphi})\hat{U}({\bar\varphi})\hat{a}^2({\bar\varphi})
\hat{H}^{-1/2}_\mathrm{FLRW}\right>}{\left<\hat{H}^{-1/2}_\mathrm{FLRW}\hat{a}^4({\bar\varphi})\hat{H}^{-1/2}_\mathrm{FLRW}\right>},
\end{equation}
the quantum counterpart of
\begin{equation}
U({\bar\varphi})=a^2\left(fV({\bar\varphi})-2\sqrt{f}\partial_{\bar\varphi} V+\partial^2_{\bar\varphi} V\right),
\end{equation}
with $f=24\pi G (\dot{\bar\varphi}^2/\rho)$, the fraction of kinetic energy in the scalar field. The above expression 
for the potential-like term is valid for self-interacting scalar field (notice that $U=0$ for a free scalar field). 
The shape of the primordial power spectrum obtained in this approach will be described in the following 
section. We however already mention two important facts about the calculation, highlighting the 
differences with the approach presented in the preceding sections. First, as the background state is chosen 
to be a coherent state sharply peaked around classical trajectories, it appears that the dressed metric 
reduces to the metric traced by the peak of the state. As a consequence, it is sufficient to use the scale 
factor solution of the effective modified Friedmann equation, instead of using the full dressed metric, to 
propagate perturbations. In a certain sense, this gives the evolution of perturbations on the effective 
background. However, the resulting equations of motion do not coincide with the ones derived in Sec. 
\ref{sec:holo}. Second, the initial conditions for perturbations were chosen at the time of the bounce. 
However, the primordial power spectra computed using the anomaly-free approach developed in the 
previous sections of this paper are obtained by setting the initial conditions for perturbations far in the 
remote past of the contracting phase. For those two reasons, the primordial power spectra predicted 
using the dressed metric approach differs from the ones predicted using the anomaly-free approach.

\section{Primordial power spectra and CMB angular power spectra}

The main astronomical probe of the physics at play in the primordial Universe is the cosmic microwave background 
anisotropies. This statistically isotropic background radiation released at the time of recombination is described by a 
black body law at a temperature of $\sim$2.75 K, therefore peaking in the submilliter range. It has been observed since 
four decades and is now precisely mapped allowing for  sharp constraints on the cosmological parameters describing 
our Universe. The CMB exhibits tiny fluctuations, named anisotropies, which are originally sourced by the inhomogeneities 
of the universe produced during inflation and, in our case, during the pre-bounce contraction and the bounce. Those 
anisotropies come into three flavors: temperature anisotropies and two modes of polarized anisotropies, dubbed $E$ and 
$B$ modes. Their source being the quantum primordial perturbations of the Universe, the CMB anisotropies are fundamentally 
stochastic (more precisely, they are the unique realization of a stochastic process) and the physical informations they carry 
is contained in their statistical properties. A possible way to characterize the probability distribution function of those anisotropies 
is to compute the entire (and a priori infinite) set of  statistical moments. Fortunately, the stochastic process at their origin 
is the quantum fluctuations of the vacuum which is a gaussian process. The CMB anisotropies being fluctuations, they have 
a zero mean value and all the information is encoded in their two points correlation functions thanks to the Isserli theorem.
(This argument is valid at first order in perturbation theory and adding higher order will 
inevitably leads to departure from gaussianity. Nevertheless, those non-gaussianities will be small in amplitude as compared 
to the gaussian part of the statistics. The gaussian nature of the CMB anisotropies has been tested on the observed maps of 
the submilliter sky which shows that non-gaussianities are indeed tiny.) Those two points correlation functions are most easily 
treated in the space of multipoles (the harmonic space on the celestial sphere) making use of the angular power spectrum, 
$C^{XY}_\ell$ with $X,~Y=T,~E$ and $B$. The current strategy to relate LQC modeling of the primordial universe and astronomical 
data is grounded in those angular power spectra: on the one hand, they can be easily predicted from the knowledge of the 
primordial power spectrum at the end of inflation; on the other hand, they can be estimated from the reconstructed maps of the 
CMB anisotropies.

\subsection{From primordial power spectra to angular power spectra}

\paragraph{General discussion--} There are a priori six angular power spectra to be computed.
However, as long as parity invariant processes are considered as the only processes at play 
in the early universe, the $TB$ and $EB$ correlations are vanishing. The four remaining angular 
power spectra are related to the primordial power spectrum of scalar and tensor perturbations 
using the line-of-sight solution of the Boltzmann equation (see {\it e.g.} \cite{zalda_harari,zalda_seljak})
\begin{equation}
C^{XY}_\ell=\displaystyle\int^\infty_0 dk\int^{\eta_0}_{\eta_e}d\eta \left[\Delta^{X,S}_\ell(k,\eta)
\Delta^{Y,S}_\ell(k,\eta)\mathcal{P}_\mathrm{S}(k)+\Delta^{X,T}_\ell(k,\eta)\Delta^{Y,T}_\ell(k,\eta)
\mathcal{P}_\mathrm{T}(k)\right], \label{eq:los}
\end{equation}
with the time integration a priori performed from the end of inflation, $\eta_e$, up to now. In 
the above, $\mathcal{P}_\mathrm{S(T)}$ stands for the primordial power spectra for scalar 
(tensor) perturbations sourcing the CMB anisotropies. They simply corresponds to the Fourier 
representation of the 2-points correlation functions of the scalar and tensor perturbations 
computed at the end of inflation, {\it i.e.} 
\begin{equation}
	\left<v_{\mathrm{S(T)}}(\vec{x},\eta_e)v_{\mathrm{S(T)}}(\vec{x}+\vec{r},\eta_e)\right>=\displaystyle\int\mathcal{P}_{\mathrm{S(T)}}(k)\times\frac{\sin (kr)}{kr}\times d(\ln k),
\end{equation}
where statistical homogeneity is assumed. The knowledge of those primordial spectra is 
obtained by solving the equations of motion for the Mukhanov-Sasaki variables as derived 
in Secs. \ref{sec:holo}, \ref{sec:iv}, \ref{sec:all} and \ref{sec:aan}. These equations are 
integrated from the contracting phase to the end of inflation, or from the bounce to the end 
of inflation (depending at which time the initial conditions are set) and the primordial power 
spectra are defined as
\begin{eqnarray}
\mathcal{P}_\mathrm{S}(k)&=&\frac{k^3}{2\pi^2}\left|\frac{v_{\mathrm{S},k}(\eta_e)}{z(\eta_e)}\right|^2, \\
\mathcal{P}_\mathrm{T}(k)&=&\frac{32 G k^3}{\pi}\left|h_k(\eta_e)\right|^2.
\end{eqnarray}
The functions $\Delta^{X,S(T)}_\ell$ are the so-called transfer functions encoding  the radiative 
transfer integrated from the end of inflation up to now, and for the projections of three-dimensional 
inhomogeneities on the two-dimensional celestial sphere. For example, the function $\Delta_\ell^{E,S}$ 
encodes how scalar primordial perturbations contribute to the polarized anisotropies of $E$-type. 

The information about the history of the universe is encoded in both the transfer functions and 
the primordial power spectra, each of those two ingredients containing informations on two different 
parts of the cosmic history. Roughly speaking, all the information about the contraction, the bounce 
and inflation is gathered in the primordial power spectra (in the standard inflationary paradigm, the 
primordial power spectra contain the information on inflation only) while the transfer functions carry 
the information on the cosmic history after inflation. It is worth mentioning that the transfer functions 
depends on cosmological parameters relative to the background, that is cosmological parameters 
corresponding to the isotropic and homogeneous evolution of the Universe. Inversely, the primordial 
power spectra contain informations on inflation in a more indirect way: they depend on those 
cosmological parameters as functions of {\it the statistical properties of the perturbations} evolving in a 
specific inflationary background. In other words, any constraint set on {\it e.g.} inflation are obtained 
assuming a shape of $\mathcal{P}_\mathrm{S(T)}$. There are two necessary conditions allowing 
for the use of this approach in the LQC framework. First of all,the history of the LQC universe has to be 
identical to its standard history starting from the end of inflation. This ensures the transfer functions to 
capture the information about the post-inflationary cosmic history. This condition is fulfilled in LQC since 
the pre-inflationary dynamics of the LQC universe precisely set the universe in the appropriate conditions 
for inflation to start and the quantum corrections are fully negligible at the end of inflation. Second, the use of the above mentioned formalism is made possible if the 
content of the Universe {\it before} the end of inflation is completely dominated by the scalar field. 
This ensures that all the informations on the pre-inflationary era and inflation is solely contained in the 
primordial power spectra.

To understand how LQC could affect the angular power spectra of CMB anisotropies, let us rewrite 
the above equation as follows
\begin{eqnarray}
C^{XY}_\ell&=&\displaystyle\int^\infty_0  dk\int^{\eta_0}_{\eta_e}d\eta \left[\Delta^{X,S}_\ell(k,\eta)\Delta^{Y,S}_\ell(k,\eta)
F^\mathrm{LQC}_S(k)\mathcal{P}^\mathrm{STD}_\mathrm{S}(k)\right. \\
&+&\left.\Delta^{X,T}_\ell(k,\eta)\Delta^{Y,T}_\ell(k,\eta)F^\mathrm{LQC}_T(k)\mathcal{P}^\mathrm{STD}_\mathrm{T}(k)\right], \nonumber
\end{eqnarray}
with $\mathcal{P}^\mathrm{STD}_\mathrm{S(T)}$ the red-tilted primordial power spectra as predicted in 
the standard inflationnary scenario, and $F^\mathrm{LQC}_{S(T)}=
\mathcal{P}^\mathrm{LQC}_\mathrm{S(T)}/\mathcal{P}^\mathrm{STD}_\mathrm{S(T)}$ two functions encoding 
for the distortions of the primordial power spectra due to the peculiar evolution of the universe as modelled in LQC. 
As a heuristic approach to infer how the angular power spectra of the CMB anisotropies are affected by LQC, the functions $F^\mathrm{LQC}_{S(T)}$ can be factored out of the integral over wavenumbers. 
This rough approximation is heuristically relevant and is made possible as the previous derivations of the primordial 
power spectra in LQC show that in the UV limit the standard inflationary spectra are recovered, and that a characteristic 
length scale, dubbed $k_\star$, is usually introduced. In other word, the LQC primordial power spectra can be modelled as
\begin{equation}
\mathcal{P}^\mathrm{LQC}_\mathrm{S(T)}=F^\mathrm{LQC}_{S(T)}(k/k_\star)\times \mathcal{P}^\mathrm{STD}_\mathrm{S(T)}(k),
\end{equation}
with $F^\mathrm{LQC}_{S(T)}(k/k_\star)\simeq1$ for $k\gg k_\star$. The red-tilted inflationary power spectra are 
given by the standard power-law parametrized by an amplitude and a spectral index
\begin{eqnarray}
\mathcal{P}^\mathrm{STD}_\mathrm{S}(k)&=&A_s\left(\frac{k}{k_0}\right)^{n_S-1}, \\
\mathcal{P}^\mathrm{STD}_\mathrm{T}(k)&=&\left({T}/{S}\right)\times A_s\left(\frac{k}{k_0}\right)^{n_T},
\end{eqnarray}
with $A_S$ the amplitude of the scalar mode at $k=k_0$, $T/S=A_T/A_S$ the tensor-to-scalar 
ratio and, $n_S$ and $n_T$ the so-called tilt of the scalar 
and tensor modes respectively. (The tensor-to-scalar ratio, defined in terms of {\it primordial power spectra}, is 
mostly denoted $r$ in the literature. However, we denote it $T/S$ to avoid any confusion with 
the parameter $R$ introduced later on.) The wavenumber $k_0$ is the pivot scale, usually set equal to 
$0.002$~Mpc$^{-1}$. This is a reference scale defined to normalize the inflationary primordial 
power spectra and there is a priori no reason for this chosen scale to be equal to the characteristic 
scale imposed by the peculiar LQC dynamics of the Universe. The characteristic {\it length} scale 
$k_\star$ translates into a characteristic {\it angular} scale on the CMB anisotropies, roughly given 
by $\ell_\star=k_\star/k_H$ with $k_H\sim2.3\times10^{-4}$~Mpc$^{-1}$ the Hubble wavenumber 
today, since the transfer functions are roughly peaking at $\ell\sim k/k_H$. A very rough approximation 
therefore consists in supposing that the transfer functions are indeed sharply peaked around 
$\ell\sim k/k_H$ allowing for the $F^\mathrm{LQC}_{S(T)}$ to be factored out of the $k$ integral 
replacing $k/k_\star$ by $\ell/\ell_\star$. The line-of-sight integral is therefore performed using the 
standard inflationary-predicted primordial power spectra as initial conditions, leading to
\begin{equation}
	C^{XY}_\ell\sim F^\mathrm{LQC}\left(\ell/\ell_\star\right)\times\tilde{C}^{XY}_\ell,
\end{equation}
with $\tilde{C}^{XY}_\ell$ the angular power power as obtained in the standard inflationary paradigm. 
In the above, we have assumed that scalar perturbations and tensor perturbations are distorted by 
the same $F^\mathrm{LQC}$ function to simplify the final approximation. 

This approximation could serve as a heuristic guide to infer the distortion of the CMB angular power 
spectra due to LQC, as compared to the standard prediction of general relativity. Let us suppose, as 
an example that for $k>k_\star$, the primordial power spectra in LQC are roughly equal to the 
primordial power spectra as obtained in general relativity. A suppression (boost) of the primordial 
power spectra for $k<k_\star$, as compared to the general relativistic prediction, then translates into 
a suppression (boost) of the CMB angular power spectra for $\ell<\ell_\star$. The possible ``observation'' 
of such a suppression (boost) in the $C_\ell$'s is conditioned by the value of $k_\star$. We remind that 
$k_H$ corresponds to the largest length scale observable today. If $k_\star\ll k_H$, the characteristic 
scale set by LQC is still super-horizon today and cannot be observed today. From the prediction side, 
the key question to know if LQC distortion could be observed or not (that is to be able to distinguish 
LQC from the standard general relativistic, inflationary paradigm) would therefore be to know if $k_\star$ 
is greater or smaller than $k_H$. The value of $k_\star$ will basically depends on two types of parameters. 
First of all, it depends on the underlying LQC physics setting this characteristic length scale. Second, 
this typical length scale is obviously set in the cosmic history {\it before} inflation, simply because once 
inflation starts, the Universe is already in its classical regime and LQC corrections have no influence anymore. 
This means that this characteristic scale is subsequently stretched by inflation and all stages of cosmic expansion 
up to now. Let us take an example to explain this point. We assume that this typical scale is set at the 
time of the bounce, fixing the value of $k_\star(t_B)$. Because of the cosmic expansion from the bounce 
untill today, this scale is observable in the CMB angular power spectra if
$
k_\star(t_B)\times\exp\left(-N(t_B,t_0)\right)>k_H,
$
with $N(t_B,t_0)$ the total number of e-folds from the bounce to today. From the end of inflation 
up to today, the number of e-folds roughly varies from $\sim50$ to $\sim65$, depending on the 
value of the reheating temperature assumed to vary from $10^{10}$~GeV to $10^{16}$~GeV 
and, from the bounce to the onset of inflation, this number roughly varies from 0.6 for $m=1$ to 
7 for $m=10^{-8}$. The above inequality is therefore given by $k_\star(t_B)\times\exp\left(-N_\mathrm{inf}-\tilde{N}\right)>k_H$,
with $N_\mathrm{inf}$ the number of e-folds during inflation and $\tilde{N}$ varying from $\sim$50 
to $\sim$70. (In principle, 
the number of e-folds from the bounce to the onset of inflation and the number of e-folds during inflation 
are solely determined by the mass of the inflaton field and its relative amount of potential energy at 
the time of the bounce. Those two numbers are therefore not independent. We however choose to 
let them independent from a phenomenological perspective.) In this specific example, the scale is in the observable range if  
$N_\mathrm{inf}<\ln(k_\star(t_B)/k_H))-\tilde{N}$. If this typical scale is set at the time of the bounce, 
the associated wavenumber $k_\star(t_B)$ will be of the order of the inverse of the Planck length. 
This finally set an upper bound on the value of the number of e-folds during inflation for such a scale 
to be observed of the order of $N_\mathrm{inf}<90$ for $\tilde{N}=50$ and $N_\mathrm{inf}<70$ for 
$\tilde{N}=70$. If inflation lasts long enough for the number of e-folds to exceed $\sim90$, the 
characteristic scale of LQC is so stretched that the part of the primordial power spectra accessible 
thanks to the observation of CMB anisotropies is the one with a shape similar to the standard prediction of inflation. 
In such cases, distinguishing between the standard inflationary paradigm and the LQC paradigm is 
impossible via CMB anisotropies.

\begin{figure}
\begin{center}
\includegraphics[scale=0.5]{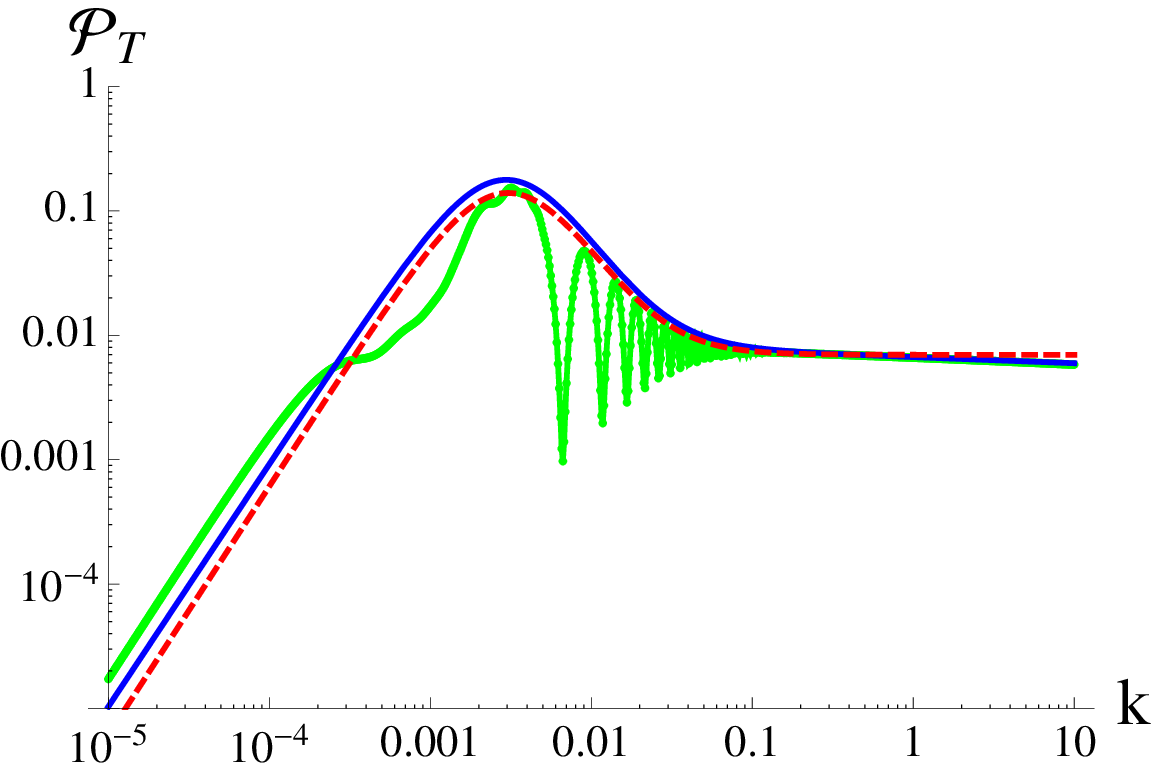}\includegraphics[scale=0.5]{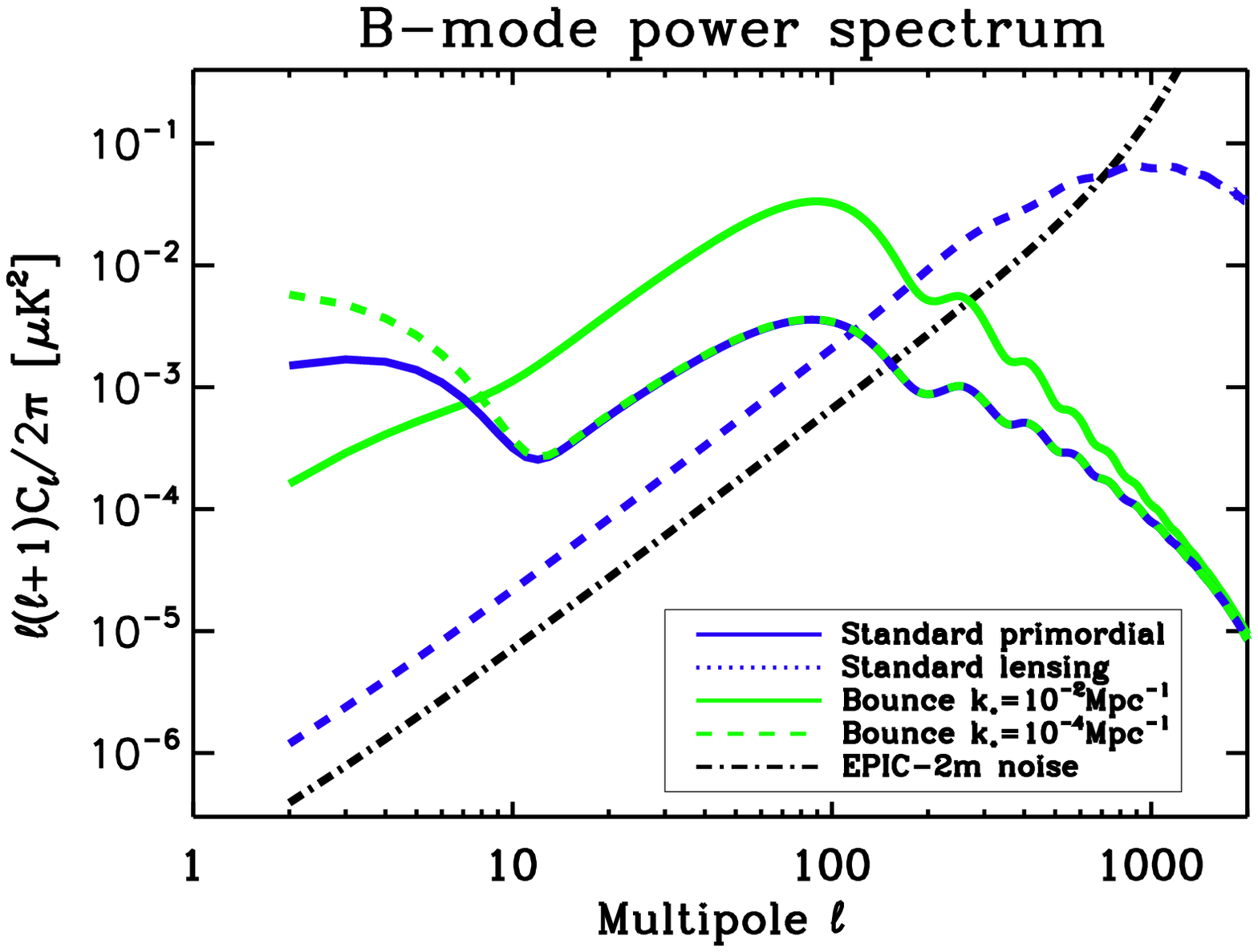}
\caption{{\it Left panel:} Exact primordial power spectrum of tensor modes (green curve) and its phenomenological 
parametrization (blue curve). This last parametrization is obtained by multiplying the standard inflationary power-law 
with a function (the dashed-red curve) accounting for LQC corrections (taken from \cite{jakub}). {\it Right panel:} 
Angular power spectrum of $B$-modes of the CMB polarized anisotropies as predicted in LQC for $R=10$ and 
$k_\star=10^{-2}$~Mpc$^{-1}$ (solid-green) and $k_\star=10^{-4}$~Mpc$^{-1}$ (dashed-green). The solid-blue 
curve shows the primary component of CMB $B$-modes as obtained in the standard inflationary paradigm for 
$n_T=-0.012$ and $r=0.05$, and the dashed-blue curve stands for the lensing-induced $B$-modes. The 
dashed-dotted-black curve shows a  typical noise level of a possible future satellite mission dedicating to the measurement
CMB polarized anisotropies (taken from \cite{grain_cmb}).}
\label{fig:cl}
\end{center}
\end{figure}

\paragraph{A specific example--} This is obviously a rough estimate which does  take into 
account neither the detailed shape of the predicted primordial power spectrum nor the precise 
dynamics of the bouncing universe. Moreover, the previous discussion assumes that for $k>k_\star$, 
the LQC-predicted primordial power spectra converge toward the standard prediction of inflation. 
This simplification has however been shown to be heuristically relevant in \cite{grain_cmb} where 
the $\Omega$-correction was not introduced yet. A quantitative estimate of the impact of LQC on
the CMB anisotropies requires to solve exactly the line-of-sight integral, Eq. (\ref{eq:los}), plugging 
in the primordial power spectra as derived in the LQC framework and making use of so-called 
Boltzmann codes as {\it e.g.} {\sc camb} \cite{camb} or {\sc class} \cite{class}. As an example, 
we will consider the phenomenological description proposed in \cite{jakub} for the specific case of 
tensor perturbations
\begin{equation}
\mathcal{P}^\mathrm{LQC}_\mathrm{T}(k)=
\frac{\mathcal{P}^{\mathrm{STD}}_\mathrm{T}(k) }{1+(k_\star/k)^2} \left[1+\frac{4R-2}{1+(k/k_\star)^2} \right].
\label{eq:specpheno}
\end{equation}
As compared to the red-tilted power law predicted in standard inflation, this primordial power spectrum 
shows three different regimes
\begin{itemize}
\item for $k\ll k_\star$, it is suppressed;
\item for $k\sim k_\star$ it shows a bump with a height defined by 
$R=\mathcal{P}^\mathrm{LQC}_\mathrm{T}(k_\star)/\mathcal{P}^{\mathrm{STD}}_\mathrm{T}(k_\star)$, 
$R$ being greater than unity;
\item for $k\gg k_\star$ it converges towards the standard inflationary spectrum.
\end{itemize}
This phenomenological description captures the main features of the exact, but numerically computed, 
primordial power spectrum of tensor modes. The agreement is depicted on the left panel of Fig. \ref{fig:cl} and 
it has been shown in \cite{grain_cmb} that the additional complexity (such as oscillations) in the exact 
$\mathcal{P}_\mathrm{T}$ has a negligible impact on the CMB angular power spectrum. In addition to 
the tensor-to-scalar ratio $T/S$ and the tensor tilt $n_T$, this parametrization introduces two additional 
cosmological parameters probing LQC distortions: the height of the bump $R$ and its 'position' $k_\star$. 

The above phenomenological description has been implemented in Boltzmann codes in \cite{grain_cmb,ma_zhao} 
to evaluate the impact of LQC on CMB anisotropies. The contribution of tensor perturbations to the CMB angular 
power spectra of type $TT,~EE$ and $TE$ is subdominant with respect to the contribution of scalar perturbations. 
However, only tensor mode can generate $B$-type polarization. We will therefore consider the case of the $BB$ 
angular power spectrum. (More precisely, only tensor perturbations can generate {\it primary} anisotropies of $B$-mode. 
In addition to that primary source, the lensing of CMB anisotropies due to large scale structures will 'distort' the primary 
anisotropies and transfer $E$-modes into $B$-modes. However, the primary and secondary anisotropies of $B$-modes 
peak at two different angular scales, primary $B$-modes being dominant at the largest scales, up to $\ell\sim100$, 
and secondary, lensing-induced, $B$-modes being dominant at the smallest angular scales, $\ell>100$.) Those 
$B$-modes angular power spectra are displayed in Fig. \ref{fig:cl}. The solid-blue curve shows the primary component 
of CMB $B$-modes as obtained in the standard inflationary paradigm for $n_T=-0.012$ and $r=0.05$ (the other 
cosmological parameters are set equal to the WMAP 7-yrs best fit \cite{komatsu}), and the dashed-blue curve stands 
for the lensing-induced $B$-modes. The green curves correspond to primary $B$-modes as obtained in the LQC 
framework for $R=10$ and $k_\star=10^{-2}$~Mpc$^{-1}$ (solid-green) and $k_\star=10^{-4}$~Mpc$^{-1}$ (dashed-green). 
The dashed-dotted-black curve shows a  typical noise level of a possible future satellite mission dedicating to 
CMB polarized anisotropies measurements (taken from \cite{bock}). This quantitative results confirm the heuristic 
inferences previously proposed: for $k_\star=10^{-2}$~Mpc$^{-1}$ marking the position of the bump in 
$\mathcal{P}^\mathrm{LQC}_\mathrm{T}$ corresponds a bump in the $C^B_\ell$'s roughly located at 
$\ell_\star\sim k_\star/k_H\sim100$; for $\ell<\ell_\star$ roughly corresponding to $k<k_\star$, the angular power spectrum 
is suppressed; and for $\ell>\ell_\star$, roughly corresponding to $k>k_\star$, the $C^B_\ell$'s tends toward the 
standard prediction of inflation. The situation is slightly different for $k_\star=10^{-4}$~Mpc$^{-1}$. In such a case, 
$k_\star$ is of the same order than $k_H$ and $\ell_\star$ is $\sim1$. The first few values of the multipoles $\ell$ (up 
to $\ell\sim10$) correspond to length scales such as $k$ is slightly greater than $k_\star$. As a consequence, one 
only ``sees'' the tail of the bump around $k_\star$ popping up at the largest angular scales. This explain why the 
dashed-green curve shows a bump for $\ell<10$. For higher values of the multipole, this corresponds to length scales 
such as $k\gg k_\star$ and the angular power spectrum predicted in LQC does not show any departure from the standard 
inflationary prediction.

\paragraph{Expected features in some selected cases--} The above results should not be considered as the generic 
impact of LQC on the CMB angular power spectrum of $B$-modes as it assumes a peculiar shape of the primordial 
power spectrum sourcing those anisotropies. In the above peculiar example, the primordial power spectra were obtained 
assuming holonomy corrections only and without considering the $\Omega$-correction from the closure of the algebra. 
Let us briefly review what could be the specific features of LQC in the CMB angular power spectra as expected from 
different LQC frameworks. We will focus here on tensor perturbations as a paradigmatic case.

\begin{figure}
\begin{center}
\includegraphics{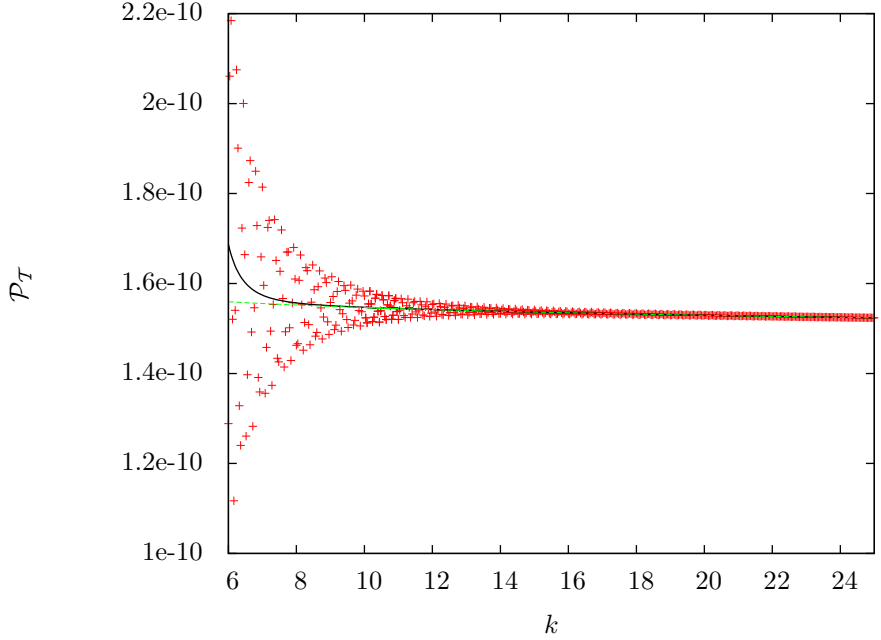}
\caption{Primordial power spectrum for tensor modes in the approach of cosmological perturbations in LQC 
developed in Refs. \cite{agullo1,agullo2,agullo3} (from \cite{agullo3}).}
\label{fig:pkagullo}
\end{center}
\end{figure}

We first consider the case of holonomy corrections only. In such a case, two approaches have been developed 
to derive the equation of motion of cosmological perturbations propagating in a quantum-corrected background. 
The first approach under scrutiny has been developed in Refs. \cite{agullo1,agullo2,agullo3} and summarized in Sec. \ref{sec:aan}. In this approach, 
the gravitational and scalar field phase space is extended to include the inhomogeneous degrees of freedom in a 
perturbative setting. The loop quantization is subsequently performed on that extended phase space. The basic 
result is that the equations of motion for gauge invariant perturbations admit the same form as in general relativity 
replacing the background quantities by effective, quantum-dressed background quantities. For the tensor mode, the 
equation of motion for the Mukhanov-Sasaki variable reads
\begin{equation}
	\frac{d^2v_{\mathrm{T},k}}{d\eta^2}+\left(k^2-\frac{\tilde{a}''}{\tilde{a}}\right)v_{\mathrm{T},k}=0,
\end{equation}
with $\tilde{a}$ a quantum-dressed scale factor accounting for both the quantum mean and the quantum fluctuations 
of the scale factor as an expectation value computed on some quantum states of the background part of the wave 
functional (see {\it e.g.} Eq. (2.13) of Ref. \cite{agullo3}). Following \cite{agullo3}, this effective dressed scale factor 
is estimated over the sharply peaked state of the background and can therefore be approximated by the scale factor 
traced by the peak of the wave function, that is the scale factor as determined by using the effective, modified Friedmann 
equation. Starting from that equation, the primordial power spectrum is computed choosing a 4-th order WKB vacuum 
{\it at the time of the bounce} as initial conditions (see \cite{birell} for a definition of WKB vacua). The resulting power 
spectrum for the tensor mode is depicted in Fig. \ref{fig:pkagullo}, taken from \cite{agullo3}. It shows a departure from the 
standard inflationary prediction for the largest scales where the primordial power spectrum is boosted. The typical length 
scale set by LQC in this framework is universal and given by $\sqrt{8\pi G \rho_c}\sim3.2$ in Planck units at the time 
of the bounce. From the numerical analysis performed in \cite{agullo3}, it shows that the effect of LQC on the shape of 
the primordial power spectrum of the tensor mode is significant up to $k\sim7$. This results have however been obtained 
assuming a specific duration of inflation, fitting with the WMAP 7-yrs data. This boost in the primordial power spectrum 
will inevitably translate into a boost of the CMB angular power spectrum at the largest angular scales. Two situations are 
therefore possible. First, the number of e-folds is so high that this typical scale is still super-horizontal today and distinguishing 
between LQC and standard inflation is not possible from the observation of CMB anisotropies. Second, the number of 
e-folds is small enough and one should observed a boost of CMB angular power spectra for multipoles lower than 
$\ell_\star\sim7.\times \exp(-N_\mathrm{inf}-\tilde{N})\times k^{-1}_H$.

\begin{figure}
\begin{center}
\includegraphics{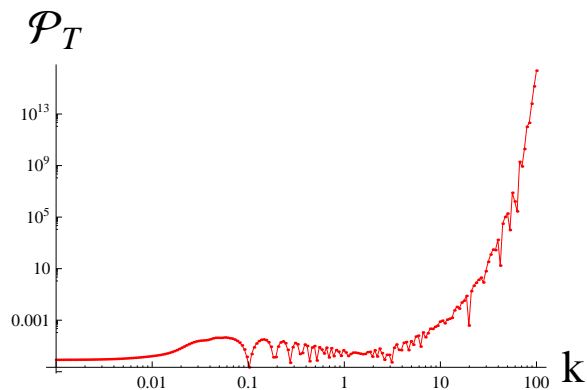}
\caption{Primordial power spectrum for tensor modes in the so-called $\Omega$-corrected approach of 
cosmological perturbations in LQC, numerically derived for $m=10^{-3}$ (from \cite{linseforspk}).}
\label{fig:pkome}
\end{center}
\end{figure}

The second approach has been presented in the previous sections. The impact of the $\Omega$-correction is 
rather drastic \cite{linseforspk}. The shape of the predicted primordial power spectrum is displayed in Fig. \ref{fig:pkome}. 
Because $-1<\Omega<1$, one can distinguish, at least, two regimes in the primordial power spectrum of tensor 
modes, the frontier between those two regimes being marked by 
$k_\star$ given by $k_\star=\mathrm{max}(z''_\mathrm{T}/z_\mathrm{T})_{t<t_i}$. For $k\ll k_\star$, the modes ``feel'' 
the curvature of space-time during the contraction and/or the bounce as well as during inflation. As a consequence, 
there is no reason for the primordial power spectrum to be identical to what is predicted in the standard inflationary 
scenario. In such a regime, the primordial power spectrum is first scale invariant for $k\to0$ and then shows some 
oscillations with a hard red envelop. For $k\gg k_\star$, the modes 'feels' the space-time curvature during inflation only. 
However, because $\Omega$ becomes negative during a short time interval around the bounce, the primordial power 
spectrum exhibits an exponential growth in this limit, $\mathcal{P}_\mathrm{T}(k\to\infty)\propto \exp(k/k_e)$ with 
$k^{-1}_e\sim\displaystyle\int^{\eta_+}_{\eta_-} \sqrt{\left|\Omega\right|}d\eta$ and $\eta_\pm$ denoting the beginning 
and the end of the time interval for which $-1<\Omega<0$. In this case, the impact of LQC on the CMB angular power 
spectrum of $B$-modes is drastic. Let us consider two extreme situations depending on the ratio value of $k_\star/k_H$. 
For $k_\star/k_H\ll1$ corresponding to an extremely long phase of inflation, the modes observable today in the CMB
all lie in the exponential regime. The angular power spectrum of $B$-modes should therefore be highly boosted and 
extremely blue as compared to the standard prediction in the inflationary paradigm. For $k_\star/k_H\gg1$, the observable 
modes now lie in the region such that $k\ll k_\star$. This situation would be rather similar to the one described in \cite{jakub} 
except that the suppression at low multipoles is quantitatively less significant.

\begin{figure}
\begin{center}
\includegraphics{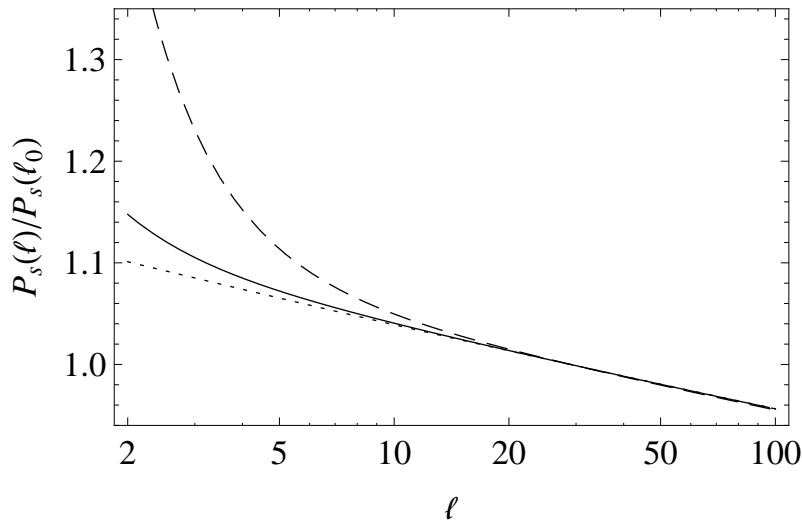}
\caption{Primordial power spectrum of tensor modes considering inverse-volume corrections. The dotted 
line corresponds to the standard power-law as predicted by inflation. The solid and dashed curves take 
into account inverse-volume corrections for $\sigma=2$ and $\delta_0=7\times10^{-5}$ and $4.8\times10^{-4}$ 
respectively. This power spectrum is depicted as a function of $\tilde{\ell}$ defined by $\tilde{\ell}=k/k_H$. 
The pivot scale is set to $k_0=0.002$ Mpc$^{-1}$ corresponding to $\ell_0=29$ (from \cite{Bojowald:2011hd}).}
\label{fig:speciv}
\end{center}
\end{figure}

Let us now consider the primordial power spectrum derived accounting for inverse-volume corrections, 
disregarding the impact of holonomies. With such a type of corrections, it is unclear how the Universe 
goes through the bounce. A consistent derivation of the impact of inverse-volume terms on the primordial power 
spectrum of the tensor mode has therefore been essentially derived assuming a purely inflationary universe. 
One could be tempted to argue that in such a setting, the LQC corrections would therefore be minor as the 
universe is assumed to be in a classical phase. Nevertheless, the inverse-volume corrections modify the 
dispersion relation for the propagation of gravity waves on such a classical background. This modification 
turns out to scale as $(a(\eta)/a_\mathrm{Pl})^{-\left|\kappa\right|}$. During inflation, the largest scales exist 
the horizon first, that is for the largest values of $(a(\eta)/a_\mathrm{Pl})^{-\left|\kappa\right|}$: one should 
therefore expects the inverse-volume correction to mainly impact  large scales. The different numerical 
and analytical calculations show that the inverse-volume corrections lead to a boost of the primordial 
power spectrum at large scales. The first attempt performed in \cite{grain_iv_ds} suggests that this boost 
is exponential, {\it i.e.} $\mathcal{P}^\mathrm{IV}_\mathrm{T}(k)\simeq \mathcal{P}^\mathrm{STD}_\mathrm{T}(k)\times(1+e^{k_\star/k})$. 
However, the refined calculations of \cite{Bojowald:2011iq,Bojowald:2010me,Bojowald:2011hd}, performing 
a Taylor expansion around the pivot scale $k_0$,  shows that such a boost is of polynomial type, {\it i.e.} 
$\mathcal{P}^\mathrm{IV}_\mathrm{T}(k)\simeq \mathcal{P}^\mathrm{STD}_\mathrm{T}(k)
\times(1+\gamma_\mathrm{T}\delta_0(k/k_0)^{-\left|\sigma\right|})$ (We notice that the same modification 
is obtained for the scalar power spectrum, replacing $\gamma_\mathrm{T}$ by $\gamma_\mathrm{S}$). 
This spectrum, normalized to unity at the pivot scale, is displayed in Fig. \ref{fig:speciv} for $\sigma=2$ and 
$\delta_0=7\times10^{-5}$ and $4.8\times10^{-4}$. This power spectrum is plotted as a function of 
$\tilde{\ell}=k/k_H$ and the pivot scale is set equal to $k_0=0.002$ Mpc$^{-1}$ leading to $\ell_0=29$. 
This shows that the CMB angular power spectra resulting from such $\mathcal{P}^\mathrm{LQC}$'s would 
exhibit a boost at the largest angular scales $\ell<5$ - $10$, depending on the value of $\delta_0$.

\subsection{From angular power spectra to primordial spectra}

Going from the observed CMB angular power spectra to constraints on the primordial power spectrum assumes 
that one is able to compute the full posterior probability of the LQC model conditionned by the knowledge of the 
observed $C_\ell$'s and their associated uncertainties. With such a strategy, it is possible either to set constraints 
on LQC using current CMB observations, or to forecast the potential ability of forthcoming CMB experiments to be 
sensible to specific distortions of the $C_\ell$'s possibility due to LQC. The later strategy has been used in Ref. 
\cite{grain_cmb} investigating what can kind of constraints could be obtained from a future precise measurement 
of the $B$-mode of the CMB anisotropies. The shape of the LQC-corrected primordial power spectrum of tensor 
modes is the one given in Eq. (\ref{eq:specpheno}). Though such an approach simplifies both the distortions of the 
primordial power spectrum and the exact shape of the posterior probabilities (assuming the likelihood to be Gaussian), 
this nevertheless highlights the global strategy one should adopt to constrain LQC from CMB observation.

The primordial component of the $B$-mode angular power spectrum is determined by the five following parameters: 
$k_\star,~R,~n_\mathrm{T},~T/S $ and the reionization optical-depth $\tau$. This set of parameters will be denoted 
$\theta_i$ hereafter. There values are not fixed as this is precisely those parameters that can be constrained by a 
potential observation of the $B$-modes. The other cosmological parameters will be fixed to the 
WMAP 7-yr  best fit, and the lensing-induced $B$-modes will be fixed to the standard prediction. The case of $\tau$ 
should be briefly discussed. Its value is already constrained by measurements of the $TT,~TE$ and $EE$ angular 
power spectra. However, this parameter is potentially degenerated with the other cosmological parameters, 
$k_\star,~R,~n_\mathrm{T},~T/S $. It is therefore worth letting this parameter free from the perspective of exploring its 
degeneracies with {\it e.g.} $k_\star$ and $R$ and evaluate how such degeneracies could affect  the estimation of 
$k_\star$ and $R$ from CMB measurements.

When compared to standard cosmology, the set of cosmological parameters is therefore enlarged by adding two 
phenomenological parameters, $k_\star$ and $R$, parametrizing the LQC-induced distortions of the primordial 
power spectrum. The parameters $k_\star,~R,~n_\mathrm{T},~T/S $ encode all the physics taking place in the 
primordial universe. They allow for a phenomenological description of the primordial power spectrum. The constraints 
that one can set on those four parameters can finally be translated into constraints on fundamental parameters of the 
model using:
\begin{eqnarray}
R &\approx& \left(\frac{m_{\text{Pl}}}{m}\right)^{0.64}, \\
k_\star&=&\frac{\frac{4\pi^{\frac{3}{2}}}{\sqrt{3}}\frac{m}{m_{\text{Pl}}}{\bar{\varphi}}_{\text{max}}}{
\exp\left(2\pi\frac{\bar{\varphi}_{\text{max}}^2}{m_{\text{Pl}}^2}\right) \frac{T_{\text{RH}}}{T_{\text{eq}}}
\left( \frac{g_{\text{RH}}}{g_{\text{eq}}} \right)^{\frac{1}{3}}(1+z_{\text{eq}})}, \label{kstar} \\
n_\mathrm{T}&=&\frac{-1}{8\pi G}\left(\frac{1}{{\bar{\varphi}}_{\text{max}}}\right)^2, \\
{T}/{S}&=&-8\times n_{\text{T}},
\end{eqnarray}
where ${\bar\varphi}_{\text{max}}$ is the maximum value of the field, $m$ is its mass, $T_{\text{RH}}$ and 
$g_{\text{RH}}$ are the reheating temperature and the corresponding number of degrees of freedom, 
respectively, and  $T_{\text{eq}}\simeq0.75$~eV, $z_{\text{eq}}\simeq3196$ and $g_{\text{eq}}\simeq3.9$ 
are the temperature, redshift and degrees of freedom at matter/radiation equality, respectively 
(see {\it e.g.},~Sec. 3.4.4 of Ref. \cite{wmap5}). In addition, numerical investigations have shown that 
${\bar\varphi}_{\text{max}}$ can be straightforwardly related with the ``initial conditions" or, more precisely, with 
the physical conditions at the bounce: 
\begin{equation}
{\bar\varphi}_{\text{max}}\approx \bar\varphi_B+m_{\text{Pl}} = \left(\frac{\sqrt{2\rho_c}}{m}\right)x_B+m_{\text{Pl}}.
\label{phimax}
\end{equation}
In this expression,  ${\bar\varphi}_B$ and $x^2_B=V({\bar\varphi}_B)/\rho_c$ correspond, respectively, to 
the value of the scalar field and the fraction of potential energy {\it at the bounce}. The number of e-folds during 
inflation is  given by $\rho_c$ and by the ratio $x/m$, through
\begin{equation}
N_{\text{inf}}\approx\frac{2\pi}{m^2_\text{Pl}}\left[\left(\frac{\sqrt{2\rho_c}}{m}\right)x_B+m_{\text{Pl}}\right]^2.
\end{equation}
The expression of $n_\mathrm{T}$ and $T/S$ are completely determined by the phase of inflation following the 
bounce. They are therefore given by the slow-roll parameters as computed in the case of a massive scalar field; 
the corrections to those parameters due to LQC being subdominant here.

In this framework, the question of a potential detection of LQC distortions of the primordial power spectrum in the 
$B$-mode anisotropies translates into the possible measurement of specific values for $R$ and $k_\star$. 
To forecast the errors on the determination of those two parameters, we used 
a Fisher analysis method, as described in Ref. \cite{fisher}. (See also Ref. \cite{stivoli} for a more refined approach.) 
The $(5\times5)$ Fisher matrix reads
\begin{equation}
F_{ij}=\displaystyle\sum_{\ell}\frac{1}{\Delta^2_\ell}\times\left.\frac{\partial 
C^B_\ell}{\partial\theta_i}\right|_{\theta_i=\bar\theta_i}\times\left.\frac{\partial 
C^B_\ell}{\partial\theta_i}\right|_{\theta_i=\bar\theta_j},
\end{equation}
where $C^B_\ell=C^{B,prim}_\ell+C^{B,lens}_\ell$ stands for the \{primordial+lensing\} $B$-mode spectrum and 
$\Delta_\ell$ is the error on the $B$-mode power spectrum recovery. (We stress that in Ref. \cite{fisher} there is an extra factor of $0.5$ in the definition of the Fisher matrix as a function of the error bars on the angular power spectrum reconstruction. This means that the computed signal-to-noise ratio are slightly pessimistic, being underestimated by a factor $\sqrt{2}$. However, this gives the right order of magnitude.) We 
consider only the sampling and noise variance, {\it i.e.} 
\begin{equation}
	\Delta^2_\ell=\frac{2}{(2\ell+1)f_{\text{sky}}}\left(C^B_\ell+\frac{N_\ell}{B^2_\ell}\right)^2,
\end{equation}
where $B^2_\ell$ and $N_\ell$ are the power spectra of the Gaussian beam and 
the instrumental noise of the experiment, respectively, and $f_{\text{sky}}$ is the fraction of the sky 
used in the analysis. For a CMBPol/B-Pol-like mission, we relied on the experimental specifications of 
{\it EPIC-2m} \cite{epic} with an 8~arcminutes beam, a noise level of 2.2~$\mu$K-arcmin, 
and a foreground separation accurate enough for a CMB power spectrum estimation 
using 70\% of the sky. 

\begin{figure}
\begin{center}
\includegraphics[scale=0.6]{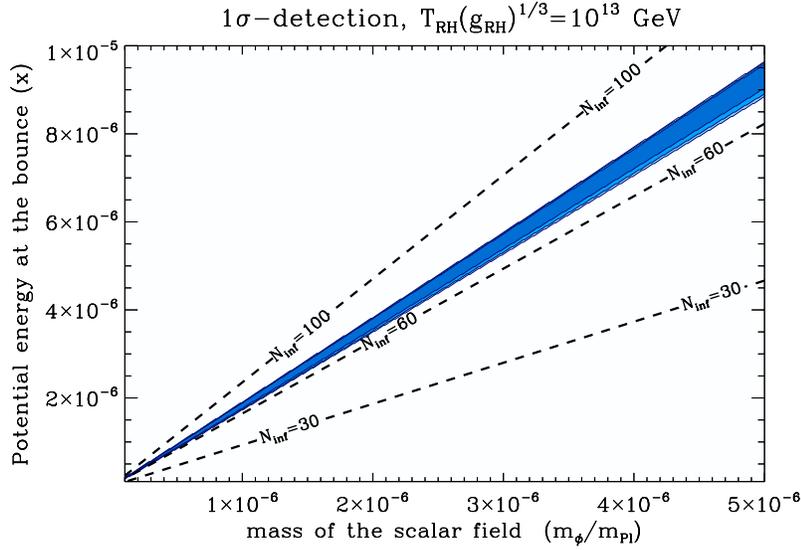}
\caption{Detectable region of the parameters $(m,x_B)$ describing a loopy universe (from \cite{grain_cmb}).}
\label{fig:detection}
\end{center}
\end{figure}

A detailed study of the impact of degeneracies between the different parameters in $\theta_i$ has been performed 
in \cite{grain_cmb}. The most conservative approach to estimate the uncertainties on {\it e.g.} the $k_\star$ measurement 
consists in marginalizing over the other parameters in order to take into account all the potential degeneracies. As a 
consequence, the uncertainty on the reconstruction of one parameter is given by $\sigma_i=\sqrt{\left[F^{-1}\right]_{ii}}$. 
(We underline that the value of $\sigma_{ii}$ explicitly depends on the specific, measured value, $\bar{\theta}_i$, of the 
parameter $\theta_i$, {\it i.e.} $\sigma_{ii}=f(\bar{\theta}_{i})$. This is inevitable in CMB physics as the signal itself, {\it i.e.} the 
$C_\ell$'s, contributes to its uncertainties via the cosmic variance.) A measurement of {\it e.g.} a specific value $\bar{k}_\star$ 
is possible if $\bar{k}_\star>\sigma_{k_\star k_\star}(\bar{k}_\star)$. The range of values of $k_\star$ which could be 
measured by a future satellite mission dedicated to the CMB  obviously depends on the values of the other 
parameters, because of the degeneracies. However, the detailed study of Ref. \cite{grain_cmb} shows that such a 
detectable range roughly goes from $10^{-4}$ Mpc$^{-1}$ to $10^{-1}$ Mpc$^{-1}$. This range can be translated into 
a range of detectable values of the couple $(m,x_B)$. The detectable values at 1-$\sigma$ of $(m,x_B)$ are depicted 
by the blue band in the right panel of Fig. \ref{fig:detection}, roughly corresponding to a detectable range of 
$k_\star\in\left[10^{-4}\right.$~Mpc$^{-1}$,$10^{-1}$~Mpc$\left.^{-1}\right]$. The upper part is not detectable as it 
corresponds to $k_\star\ll k_\mathrm{Hubble}$ making the $B$-mode power spectrum {\it undistorted} as compared to 
the standard general relativistic prediction. The lower part is not detectable as it corresponds to $k_\star\gg10^{-1}$~Mpc$^{-1}$ 
making the primordial $B$-mode systematically smaller than the lensing-induced part. Though measurements of the LQC 
parameters is not possible in this second case, a discrimination with pure general relativity is still possible as the suppression 
induced by the bounce is ``seen" via the masking of the primordial $B$-modes, that is via its {\it non-detection}. This ``suppression" could also be 
interpreted as a very low value of the tensor-to-scalar ratio $T/S$. However in the precise case of LQC, the three parameters 
$T/S$, $k_\star$ and $R$ are fully determined by {\it two} parameters, $m$ and $x$, thus breaking the degeneracy between 
$k_\star$ and $T/S$ (see \cite{grain_cmb} for details).

\begin{figure}
\begin{center}
\includegraphics[width=7cm]{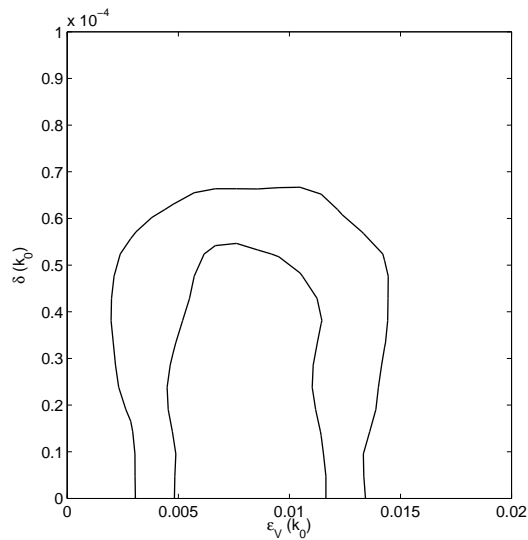}
\caption{The 2-dimensional likelihood contours in the plane $(\delta_0,\epsilon(k_0))$. 
The external and internal lines corresponds to the 68\% and 95\% confidence level, respectively (from \cite{Bojowald:2011hd}).}
\label{fig:bojolikeli}
\end{center}
\end{figure}

The case of inverse-volume corrections has also been considered in Refs. \cite{Bojowald:2011iq,Bojowald:2011hd} 
using the parametrization of the primordial power spectrum as derived in \cite{Bojowald:2010me}. In this study, the 
primordial power spectrum for scalar perturbations is also derived, allowing the authors to {\it set} constraints (and 
not only forecast constraints) using the measurements of $C^{TT}_\ell,~C^{TE}_\ell$ and $C^{EE}_\ell$ from WMAP 
7yrs, combined with large-scale structures, the Hubble constant measurement from the Hubble Space Telescope, 
supernovae type Ia, and big bang nucleosynthesis \cite{datas}. The full likelihood is finally sampled using $\delta_0$, 
$\varepsilon(k_0)$ and $\mathcal{P}^\mathrm{STD}_\mathrm{S}(k_0)$ as cosmological parameters describing the primordial 
power spectrum (in addition to those cosmological parameters describing the late-time evolution of the Universe 
such as the baryon density). The parameters $\mathcal{P}^\mathrm{STD}_\mathrm{S}(k_0),~\epsilon(k_0)$, {\it i.e.} the 
amplitude of the power spectrum and the value of the first order slow-roll parameters at the pivot scale $k_0$, 
described the standard part of the primordial power spectrum while $\delta_0$ encodes the inverse-volume distortion. The figure \ref{fig:bojolikeli} shows the 2-dimensional marginalized likelihood in the plane $(\delta_0,\varepsilon(k_0))$ 
for $\sigma=2$ and $k_0=0.002$ Mpc$^{-1}$. The boost at large scales exhibited in Fig. \ref{fig:speciv} translates into 
a boost at large angular scales in the CMB angular power spectra as compared to the standard prediction of inflation. 
This finally translates into an upper limit on $\delta_0$ as the higher $\delta_0$ the higher the boost at large scales. 
For $\sigma=2$, the parameter $\delta_0$ is bounded from above $\delta_0<6.7\times10^{-5}$ at 95\% confidence level. 
This upper limit is more stringent for higher values of $\sigma$ and $k_0$. This is intuitively understood as i) a higher 
value for $\sigma$ enhances the boost at large scales, and ii) a higher $k_0$ means that the boost covers a wider 
region in $\ell$. This inevitably enhances the inverse-volume distortions.

Sampling the full likelihood using both tensor and scalar perturbations has not been done yet for the case of holonomy 
corrections, irrespectively of the fact that the $\Omega$-correction is incorporated or not. However, the above-described 
results suggest that with the current CMB observations, it will be possible to explore the compatibility of the models 
described in Secs. \ref{sec:holo}, \ref{sec:iv}, \ref{sec:all} and \ref{sec:aan} with the datas. The latest results of the {\sc planck} 
satellite on the temperature anisotropies (see Refs. \cite{planck}) suggest that stringent constraints could be set on the 
amplitude of the distortion of the primordial power spectrum due to LQC. Basically, the prediction of inflation, {\it i.e.} 
a red-tilted power law, fits nearly perfectly to the observed $C^{TT}_\ell$ over more than two orders of magnitude in 
multipoles $\ell$. In other word, the room for modifications of a power-law is very tiny and essentially restricted to the 
largest scales, $\ell<50$. This will finally translates into a region of the parameter space of LQC models that could be 
excluded, possibly allowing discriminating between different models of the primordial universe in the LQC framework.

\section{Indirect probes: probabilities}

A clear and direct prediction from LQC which does not appear in other approaches and could be measured would be 
a smoking gun for the validity of the theory. Although, as pointed out in the previous sections, this situation might be 
reached in the next decades, it is fair to say that this is not yet the case. In the meanwhile, it is meaningful to wonder 
how ``probable'' or ``natural'' our universe is within the LQC framework. This is an ``indirect'' probe of the theory. This 
kind of questions should be handled with care: as new and more precise data are taken into account, the probability 
for a theory to predict the Universe as observed decreases, even if the model is correct. For example, the probability 
for us to write this article now is vanishing even in a correct theory, just because some aspects of the universe are 
simply contingent and not encoded in the fundamental equations. In addition, some important and well known measure 
problems arise in cosmology. This is strongly related with the choice of the surface where to impose initial conditions. 
However, in some circumstances, it is very informative to investigate the probability distribution for an a priori meaningful 
variable within a given theory. An unavoidable question to address in this framework is, as stated in the third section, the 
one of the probability for inflation to occur and last long enough. 

Obviously this question is hard to ask the standard Big Bang framework. Some arguments seem to favor a highly 
probably inflation \cite{klm}, whereas others favor an exponentially suppressed probability for inflation \cite{gt}. 
Steinhardt has summarized the prow and cons of inflation \cite{stein} in the standard paradigm and concluded that the 
initial conditions necessary for inflation to occur are everything but natural. Although the situation in LQC is not fully 
clear, as different options can be taken, the good news if that, whatever the choice made, a long enough inflation 
seems to be a clear prediction of the theory. Both approaches described below use a quadratic potential (massive 
scalar field) but most result are quite generic and would not be strongly affected by another choice. A specific type 
of field filling the universe has to be assumed as the super-inflation phase that happens  unavoidably in LQC, 
whatever the content, is not enough to account for the data (super-inflation (that is $\dot{H}$) must happen as $H=0$ 
at the bounce).

The fact that loop quantum gravity techniques "predict" inflation with a conceptual material that could very well have
been developed a few decades ago, before realizing that inflation is most probably an important part of the cosmological 
scenario, is a strong success of the model.

\subsection{Initial conditions at the bounce}

A first approach, developed in \cite{abhay}, focuses on the bounce time to set initial conditions.\\

Thanks to the presence of this canonical bounce time, one can hope to naturally resolve the ambiguity that might 
appear in the construction of a measure. Were one to try to mimic the construction used in LQC in general relativity, 
he would be led to work at the singularity in place of the bounce, where the calculation would be plugged by inconstancies. 

As usual, it is assumed that in LQC spatial geometry is encoded in the volume of a fixed, fiducial cell, $v = ({\rm
const})\times a^3$. We recall that the conjugate momentum is called $b$. The LQC-modified
 Einstein dynamics is given by
\be \label{H} H \,=\, \f{1}{2\gamma\lambda}\, \sin 2\lambda b
\,\,\approx\,\, \f{0.93}{\lp}\, \sin 2\lambda b, \ee
where $\lambda^2 \approx 5.2 \lp^2$ is the `area-gap', the smallest
non-zero eigenvalue of the area operator, whereas  $b$ ranges over
$(0, \pi/\lambda)$ and general relativity is recovered in the limit
$\lambda \rightarrow 0$.

The space of solutions $\S$ is naturally
isomorphic to a gauge fixed surface, i.e., a 2-dimensional surface
$\hat\Gamma$ of $\bar\Gamma$ which is intersected by each dynamical
trajectory only once. Since $b$ is monotonic in each
solution, the proposed strategy was to choose for $\hat\Gamma$ a
2-dimensional surface $b= b_o$ within
$\bar\Gamma$. Symplectic geometry considerations unambiguously equip
$\hat\Gamma$ with an induced
Liouville measure $\dd\hat\mu_{\L}$. As the dynamical flow
preserves the Liouville measure, $\dd\hat\mu_{\L}$ on $\S$ is
independent of the choice of $b_o$. A good choice
in LQC is to set $b_o = \pi/2\lambda$ so that $\bar\Gamma$ is just
the `bounce surface'. 

Then $\hat\Gamma$ is naturally coordinatized by
$({\bar\varphi}_{B},v_{B})$, the scalar field and the volume at the bounce.
As $b= \pi/2\lambda$, $\bar{p}_\varphi$ in unambiguously determined. The induced
measure on $\S$ can be written explicitly as
$ \dd\hat\mu_{\L} =   \f{\sqrt{3\pi}}{\lambda}\,\, \big[1-
x_{B}^2\big]^{\f{1}{2}}\,\, {\dd} {\bar\varphi}_{\B}\, {\dd} {v}_{\B}$, where $x_B^2$ is the value of $x^2$ 
at the bounce (with $x^2=m^2 {\bar\varphi}^2/(2 \rho_c))$, that is the fraction of  total energy
density in form of potential energy at the bounce. The total
Liouville volume of $\bar\Gamma \equiv \S$ is infinite because,
although ${\bar\varphi}_{B}$ is bounded for suitable potentials such as
$m^2{\bar\varphi}^2$, $v_B$ is not. However, this non-compact direction
represents gauge on the space of solutions $\S$. So, factoring out the gauge orbits is a natural prescription to
calculate fractional volumes of physically relevant sub-regions of
$\hat\Gamma$.\\

The main results of the study performed in \cite{abhay} are the following, depending on 3 different possible regimes:
\begin{itemize}
\item $x_B^2 < 10^{-4}$, that is strong kinetic energy domination
at the bounce. The number of
e-folds during this slow roll is given approximately by
\be N \approx 2\pi\, \big(1-\f{{\bar\varphi}_o^2}{{\bar\varphi}_{\rm
max}^2}\big)\, {\bar\varphi}_o^2\, \ln {\bar\varphi}_o, \ee
where ${\bar\varphi}_o$ is the value of the scalar field at the onset
of inflation and ${\bar\varphi}_{\rm max} = 1.5 \times 10^{6}$. Now, ${\bar\varphi}_o$
increases monotonically with ${\bar\varphi}_{B}$ (and is always larger
than ${\bar\varphi}_{B}$, the value to the field at the bounce). For ${\bar\varphi}_{B} = 0.99$, one has ${\bar\varphi}_o = 3.24$
and $N = 68$. Thus, there is a slow roll inflation with over $68$ e-foldings for all
${\bar\varphi}_{B}>1$, i.e., if $x_B^2 > 4.4\times 10^{-13}$.

\item $10^{-4} < x_B^2 < 0.5$, that is kinetic energy domination
at the bounce. The LQC departures from general relativity are
now  significant. The Hubble parameter is  given
to an excellent approximation by
\be \label{Ho} H_o \approx \big[\, \f{8\pi}{3}\, \rcr\, x_B^2
(1-x_B^2)\,\big]^{1/2}\, \approx \,  1.9 \, \big[\,x_B^2
(1-x_B^2)\,\big]^{1/2}\, \ee
and decreases  slowly with $\dot{H}/H^2 < 3.5 \times10^{-10}$. Thus, the Hubble parameter is 
essentially frozen at a very high value. The Hubble freezing is an LQC phenomenon: It relies on 
the fact that $H$ acquires its largest value $H_{\rm max}= 0.93\, {\rm s}_{\rm Pl}^{-1}$ at the end 
of super-inflation. Throughout this range of $x_B^2$ there are more than 68 e-foldings.

\item $0.5 <x_B^2 < 1$, that is potential energy domination at
the bounce. Now  LQC effects dominate. Again, because $\dot{\bar\varphi}
>0$, the inflaton climbs up the potential but  the turn around
($\dot{\bar\varphi} =0$) 
occurs during super-inflation.  The Hubble parameter
again freezes at the onset of inflation and the slow roll conditions are easily met as $\dot{H}/H^2$
is less than $1 \times 10^{-11}$ when $\ddot{\bar\varphi} =0$. There are many more
than 68 e-foldings already in the super-inflation phase. The
inflation exits the super-inflation phase with $H$ at its maximum
value, and little kinetic energy. The
friction term is large and the inflation enters a long (more than 68 e-folds) slow roll
inflationary phase. 

\end{itemize}

 in presence
of suitable 
potentials all LQC dynamical trajectories are funneled to conditions
which virtually guarantee slow roll inflation with more than 68
e-foldings, without any input from the pre-big bang regime. \\

This work was developed further, using analytical and numerical
methods, to calculate the a priori probability of
realizing a slow roll compatible with the CMB data (at that time, the WMAP 7-years release). It was
found that the probability is greater than 0.999997 in LQC.

\subsection{Initial conditions in the remote past}

In \cite{abhay},  the probability distribution was assumed to be flat and defined at the bounce 
(the first attempts in this direction were performed in \cite{first}). It is however possible to make 
a very different assumption: the phase of the field oscillating in the remote past can also be considered 
as a very natural random variable. As shown in \cite{Corichi:2010zp} the choice of what is a natural 
measure, and therefore the outcome of these kinds of calculations, can depend heavily on when 
one decides to define the initial conditions. It is meaningful to take seriously the meaning of an 
``initial'' condition in a Universe that extends in the past beyond the bounce, to avoid using any 
heavy machinery and rely only on very minimalistic hypotheses. Here, different conditions at the 
bounce are not ``assumed'', as in \cite{abhay}, but instead derived  explicitly as predictions of the model.

\begin{figure}
\begin{center}
\includegraphics[width=71mm]{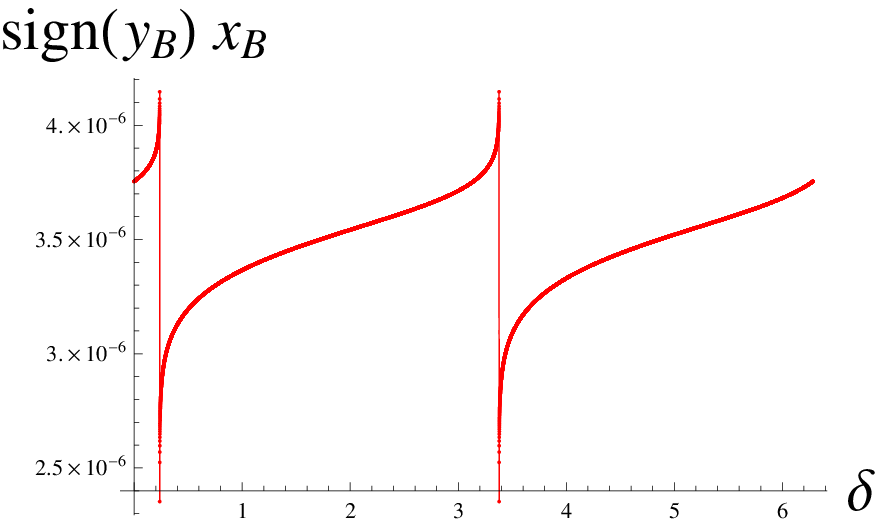}
\includegraphics[width=71mm]{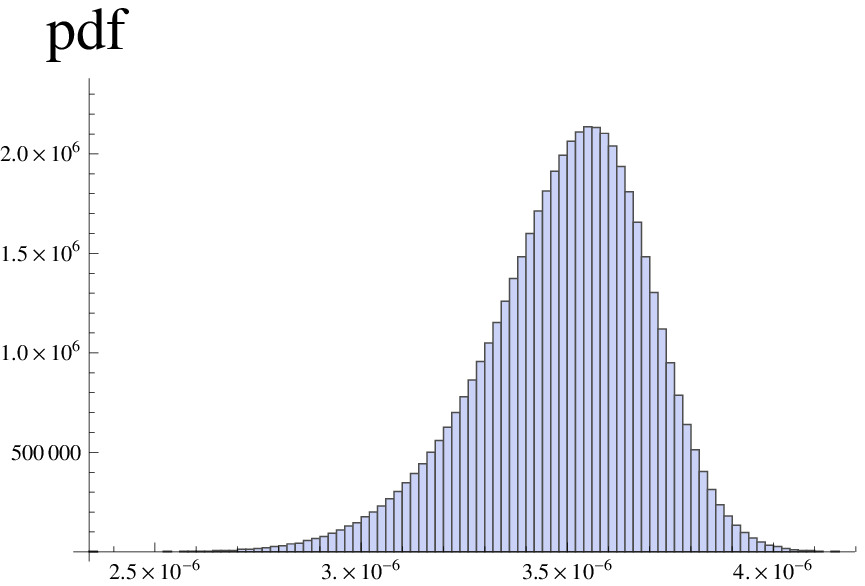}
\caption{$\sign(y_B)x_B$ as a function of $\delta$ (left plot) and its probability distribution (right plot).}
\label{xb}
\end{center}
\end{figure}

The idea is to calculate the probability density function for $x_B$, the square root of the fraction 
of potential energy at the bounce, and $N$, the number of e-folds of slow-roll inflation. This is 
done by first finding the most natural initial probability distribution, and then evolve it numerically. 
Along this line of thought, the most natural and consistent model is to set the initial probability 
distribution in the pre-bounce oscillation phase. The evolution of the Universe in this phase is described by:
\begin{equation}
\rho=\rho_0\left(1-\frac{1}{2}\sqrt{3\kappa\rho_0}\left( t+\frac{1}{2m}\sin(2mt+2\delta)\right)\right)^{-2},\label{prerho}
\end{equation}
\begin{equation}
x=\sqrt{\frac{\rho}{\rho_c}}\sin(mt+\delta)~,~ y=\sqrt{\frac{\rho}{\rho_c}}\cos(mt+\delta),  \label{prex}
\end{equation}
with parameters $\rho_0$ and $\delta$. However, the transformation
\begin{equation}
\label{transform1}
\rho_0\rightarrow \rho_1 \\
\end{equation}
corresponds to
\begin{equation}
\label{transform2}
\begin{array}{l}
\delta\rightarrow \delta-\frac{2m}{\sqrt{3\kappa\rho_1}}\left(1-\sqrt{\frac{\rho_1}{\rho_0}}\right),\\
t\rightarrow t+\frac{2}{\sqrt{3\kappa\rho_1}}\left(1-\sqrt{\frac{\rho_1}{\rho_0}}\right),
\end{array}
\end{equation}
and does not therefore generate new solutions. This allows one to take $\delta$ as the only parameter. 

In addition of being the obviously expected distribution for any oscillatory process of this kind, a flat 
probability for $\delta$ will be preserved over time within the pre-bounce oscillation phase, making it 
a very natural choice for initial conditions. This is not a trivial point as any other probability distribution 
would be distorted over time, meaning that the final result in the full numerical analysis would depend 
on the choice of $\rho_0$. 

Starting with a flat probability distribution for $\delta$, and choosing $\rho_0$ so that the solution is 
initially well approximated by Eqs. (\ref{prerho})-(\ref{prex}), the probability for different values of $x_B$ 
can be calculated numerically. At the bounce the solutions can be parametrized by $x_B$ and $\sign(y)$; 
however only the relative sign is physical. The result is therefore projected down to the physically relevant 
parameters by considering $\sign(y_B)x_B$. The value of $\sign(y_B)x_B$ as a function of $\delta$ and 
the resulting probability distribution are shown in Fig. \ref{xb}.

In \cite{abhay}, $\sign(y_B)x_B$ was taken as unknown. However, is this second approach, it can be 
shown that it is sharply peaked around $3.55\times 10^{-6}$ (this values scales with $m$ as $m\log\left(\frac{1}{m}\right)$, 
where we assumed that $m\ll1$ in Plank units). The most likely solutions are exactly those that have no slow-roll deflation. 

This result also shows that the bounce is strongly kinetic energy dominated, leading to back-reaction 
effects that can be safely neglected \cite{bojo_prl}.\\

\begin{figure}
\begin{center}
\includegraphics[width=100mm]{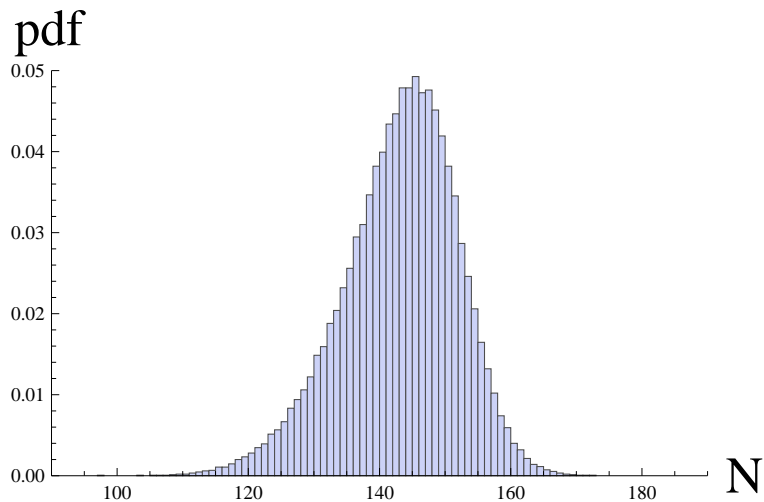}
\caption{Probability density of the number of e-folds of slow-roll inflation.}
\label{N}
\end{center}
\end{figure}

Slow-roll inflation starts when $|x|=\xmax$ where $\xmax\doteq\max_{t>t_B}(|x|)$, which is related to the length of 
slow-roll inflation by
$
N=\frac{\kappa\rho_c}{2}\left(\frac{\xmax}{m}\right)^2\simeq 5.1\left(\frac{\xmax}{m}\right)^2,
$
where $N$ is the number of e-folds during slow-roll inflation.  The probability density for $N$ is given in 
Fig. \ref{N}, showing that the model leads to a slow-roll inflation of about 145 e-folds. This becomes a 
prediction of effective LQC: inflation and its duration are not arbitrary in the approach.

In the mini-superspace homogeneous, isotropic and flat approximation used in this framework, the relevant 
problem might not be fundamentally related with the existence of infinitely many degrees of freedom or with 
divergent integrals but with the way to choose the significant measure with respect to which the probability 
distribution is flat. In this approach, our ignorance starts when the mater content begins to be well approximated 
by the (effective) scalar field. As we know neither the details of this process --which might very well be purely 
random-- nor the density at which this occurs, we translate this ignorance as a flat probability distribution for the 
most natural parameter of this phase. In addition, even if we somehow gain knowledge of the physics governing 
the ``inverse reheating", and even if this this theory predicts a non-flat probability distribution for $\delta$, unless 
this probability distribution is extremely peaked around the specific value that gives significant slow-roll deflation, 
this result will hold. \\

So far, the standard value of $\rho_c$, with a Barbero-Immirzi parameter $\gamma$ assumed to be known from 
black hole entropy (see, {\it e.g.}, \cite{bh}) was used. By instead taking $\rho_c$ as a free parameter, one can 
constrain $\rho_c$ and $\gamma$. An upper limit on $\gamma$ can be obtained by requiring a large enough 
probability for a long enough slow-roll inflation. The main results of this analysis are that $\gamma<10.1$ at 95\% 
confidence level and $\gamma<11.9$ at 99\% confidence level. This is much more stringent than previous 
cosmological constraints \cite{Mielczarek:2010ga}: $\gamma < 1100$.  \\

Finally, this study has also proved that the situation where only inflation occurs without any phase of deflation 
before the bounce is by far the most probable. Of course, deflation is possible. But its occurrence requires a 
very high degree of fine-tuning of initial conditions whereas inflation without deflation is obtained for most conditions. \\

Whatever the way initial conditions are set, it should be pointed out that the high probability for inflation 
to occur is not a truly specific feature of LQC but is fundamentally related with its attractor behavior as 
soon as initial conditions are set {\ it before} inflation. The situation might be summarized as follows. If the matter content of the model is a massive scalar field, if the surface of measure is chosen before inflation, and if one does not impose extreme fine tuning, then the probability for slow roll inflation will be extremely close to one. With this same matter content, if one defines the measure on a surface after inflation, and impose no extreme fine tuning,  the probability for slow roll inflation is going to be extremely close to zero. In LQC, there are at least two different ways to naturally define a measure based on conditions before inflation: the bounce and the remote past. In GR, there is no distinguished surface that could be the natural basis for a measure. In addition, it should be noticed that those conclusions are dependent on what type of matter is assumed. Some barotropic matter fluids will not give inflation. In itself, LQC offers no answer to the mystery of what this inflaton field is.

\subsection{Further developments}

A weakness of the previous results is related with the fact that the Universe is assumed to be isotropic. 
In bouncing cosmologies, the issue of anisotropies is crucial for a simple reason: the shear term basically 
scales as $1/a^6$ where $a$ is the scale factor of the Universe. Therefore, when the Universe is in its 
contraction phase, it is expected that the shear term eventually dominates and drives the dynamics. When 
spatial homogeneity is assumed, anisotropic hypersurfaces admit transitive groups of motion that must be 
three- or four-parameters isometry groups. The four-parameters groups admitting no simply transitive 
subgroups will not be considered here. There are nine algebraically inequivalent three-parameters simply 
transitive Lie groups, denoted Bianchi I through IX, with well known structure constants. The flat, closed 
and open generalizations of the FLRW model are respectively Bianchi I, Bianchi IX and Bianchi V. As the 
Universe is nearly flat today and as the relative weight of the curvature term in the Friedmann equation is 
decreasing with decreasing values of the scale factor, it is reasonable to focus on the Bianchi I model to 
study the dynamics around the bounce. The metric for a Bianchi I spacetime reads as:
\be
ds := -N^2 d\tau^2 + a_1^2dx^2 + a_2^2dy^2 + a_3^2dz^2,
\ee
where $a_i$ denote the directional scale factors. A dot means derivation with  respect to the cosmic time $t$, 
with $dt=Nd\tau$.\\

Many studies have already been devoted to Bianchi-I LQC \cite{bianchiI_lqc1,bianchiI_lqc2,bianchiI_lqc3,bianchiI_lqc4,bianchiI_lqc5,bianchiI_lqc6,bianchiI_lqc7,
bianchiI_lqc8,bianchiI_lqc9,bianchiI_lqc10,bianchiI_lqc11,bianchiI_lqc12,bianchiI_lqc13,bianchiI_lqc14,
bianchiI_lqc15,bianchiI_lqc16,bianchiI_lqc17,bianchiI_lqc18,bianchiI_lqc19,bianchiI_lqc20,bianchiI_lqc21}. 
In particular, it was shown that the bounce prediction is robust. As the main features of isotropic LQC are well c
aptured by semi-classical effective equations, and it is a good guess that this remains true in the extended 
Bianchi I case. The solutions of effective equations were studied into the details in \cite{Gupt:2013swa}. 
However, it is not yet clear whether the prediction of inflation still holds or not.\\

Recently \cite{linda_anisotropy}, the LQC-modified (with effective holonomy corrections) generalized Friedmann 
equation was found to be:
\be
H^2=\sigma_Q+\frac{\kappa}{3}\rho-\lambda^2\gamma^2\left(\frac{3}{2}\sigma_Q+\frac{\kappa}{3}\rho\right)^2,
\label{fried}
\ee
with  the "quantum shear"
\be
\sigma_Q:=\frac{1}{3\lambda^2\gamma^2}\left(1-\frac{1}{3}\Big[\cos(\mub_1c_1-\mub_2c_2)
+\cos(\mub_2c_2-\mub_3c_3)+\cos(\mub_3c_3-\mub_1c_1) \Big]\right),
\label{sQ}
\ee
and
\be
\mub_1 = \lambda\sqrt{\frac{p_1}{p_2p_3}}~+~\text{cyclic~terms}.
\ee

\begin{figure}
\begin{center}
\includegraphics[width=90mm]{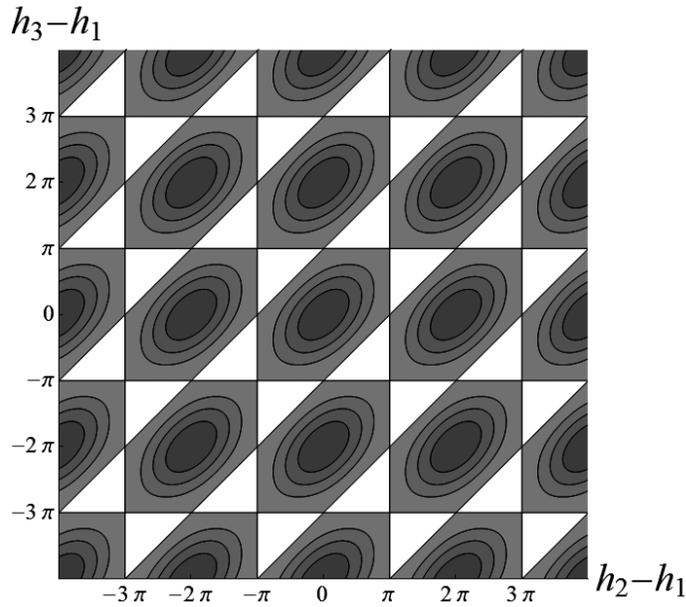}
\caption{$\sQ$ as a function of differences of the directional Hubble parameters. 
The white areas correspond to forbidden regions. The black lines are 
$\sQ=\frac{1}{4}{\sQ}_c,\frac{1}{2}{\sQ}_c,\frac{3}{4}{\sQ}_c,{\sQ}_c$.}
\label{fig:AllowedRegions}
\end{center}
\end{figure}

One can show that there exist a maximum allowed quantum shear ${\sQ}_c$. Fig. \ref{fig:AllowedRegions} 
displays the possible differences of directional Hubble parameters (there are naturally 3 Hubble parameters 
in this framework). 

The key point is that all regions but the central one have no classical limit. This raises an important question 
for LQC. If the initial conditions are to be put at the bounce, as advocated {\it e.g.} in \cite{abhay}, we face a 
delicate problem: there are infinitely many more cases leading to universes that do not resemble ours than 
cases leading to a classically expanding universe. On the other hand, if we set the initial conditions in the 
classically contracting phase, as advocated in \cite{linse}, we escape this problem. But we face another one: 
what is the ``natural'' initial shear? Or, according to which measure --and at which time-- should we assume a 
flat probability distribution function for variables quantifying the shear? In any case, this requires to investigate 
further the initial conditions problem.

\section{First attempts in LQC from the full theory}

As we have already stated, loop quantum cosmology is a theory leading to 
falsifiable predictions which can be, at least in principle, tested. This statement 
cannot be extrapolated to loop quantum gravity yet. Mainly because all the 
phenomenological considerations considered so far were obtained with  
LQC models, whose relationships with LQG are still very obscure. This situation 
would change if a procedure for recovering the cosmological sector of LQG was found. 
In this section we will review some recent attempts made in this direction. 
In particular we will discuss spinfoam cosmology and reduced loop quantum 
gravity. The issue of deformed algebra of constraints from the perspective of the 
full theory will be discussed as well. 

\subsection{Spinfoam cosmology}

LQG is a canonical approach to quantize gravity. There is however also a covariant 
formulation of LQG, based on the path integral approach, the so-called spinfoam gravity \cite{Rovelli:2011eq}. 
Spinfoam cosmology \cite{Rovelli:2008aa,Bianchi:2010zs} is an attempt to construct a truncation of 
spinfoam gravity such that the dynamics of the global degrees of freedom, e.g. the volume 
of the universe, is well captured. The truncation is performed at the level of the quantum theory by 
imposing certain restrictions on the spin network states. This is in contracts with 
loop quantum cosmology, which is obtained by symmetry reduction performed at 
the classical level and by imposing ``loop-like" quantization to the simplified classical 
system.   

In the simplest model of spinfoam cosmology, the so-called dipole model \cite{Rovelli:2008aa}, 
the spin network encoding spatial geometry is described by the graph composed 
of two nodes connected by four links. The corresponding kinematical gauge invariant 
Hilbert space is therefore given by $H_{\text{dipole}}=  L^2 \left(SU(2)^4/SU(2)^2\right)$. 
The dipole graph is dual to the triangulation of a 3-sphere obtained by gluing together faces of two 
tetrahedras. By adding more nodes to the graph, new degrees of freedom describing 
configurations at shorter scales are introduced. Roughly, this corresponds to multipole 
expansion of a 3-sphere. 

As long as we are only interested in the global dynamics of the universe, the dipole 
model is sufficient. Even more than that. Namely, it was shown that the dipole 
model is enough to describe Bianchi IX cosmology with six additional inhomogeneous 
degrees of freedom \cite{Battisti:2009kp}. However, if one is interested in recovering e.g. the structure 
of cosmological perturbations up to the very short  scales, much more complicated 
graphs must be considered. This however introduces significant technical difficulties. 

\begin{figure}
\begin{center}
\includegraphics[width=20mm]{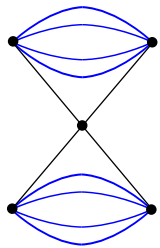}
\caption{One vertex amplitude for spin foam cosmology with dipole-type boundary states. }
\label{DipoleCosmo}
\end{center}
\end{figure}

The advantage of dipole cosmology is that its dynamics, with some additional assumption,  
can be studied analytically.  Practically, one usually restricts to non-graph-changing 
Hamiltonian constraints.  Therefore, both initial and final states are represented by 
the dipole model. Furthermore, in the lowest order of calculations, the intermediate 
state is described by a single vertex, as presented  in Fig \ref{DipoleCosmo}. 

With those assumptions one can compute transition amplitudes between two FLRW 
universes parametrized by complex phase space variable  $z= \alpha c + i \beta p$.
Here $c$ and $p$ are the standard  variables introduced earlier and $\alpha$ 
and $\beta$ are two parameters. The obtained transition amplitude between $z$
and $z'$ was found to be 
\begin{equation}
W(z,z') = N^2 z z' e^{-\frac{1}{2t\hslash}(z^2+z^{,2})}.
\end{equation}
It was shown that this amplitude is, in the large volume limit, annihilated by the 
classical FLRW Hamiltonian
\begin{equation}
\hat{H}^{\text{FLRW}}_zW(z,z') = 0.
\end{equation}
This result proves that the FLRW dynamics can be recovered from the spin foam formalism.
A similar analysis was performed in the case of a positive cosmological constant \cite{Bianchi:2011ym}. 
However, other forms of matter were not incorporated to the spinfoam cosmology picture yet. 

The achieved results are still preliminary but very promising, giving a chance to recover 
more subtle quantum effects.  In particular, the analysis of quantum corrections resulting from 
the spinfoam cosmology could fix the ambiguities present at the level of effective considerations. 
This  is also true for inhomogeneous models. However, so far,  due to computational difficulties, 
the region of small volumes, where these effects should play significant roles, is out of reach. 
These obstacles can be however overcome by using numerical techniques. \\

It should also be pointed out that recent results from the full spinfoam theory give strong indications 
in favor of singularity resolutions \cite{vidotto2013}. In addition, first attempts to derive Friedmann 
equations, at first order, from group field theory, are also very promising \cite{gielen}.

\subsection{Reduced LQG}

Another possible approach to recover LQC from LQG, so-called reduced LQG 
\cite{Alesci:2012md, Alesci:2013xd}, is also based on introducing certain restrictions to the 
kinematical Hilbert space of the full theory. In this approach, as in  spinform 
cosmology, quantization is performed prior to symmetry reduction. In this framework, it is 
in principle possible to recover the dynamics of homogeneous and as well as inhomogeneous 
LQC. Of course only if LQC truly corresponds to cosmological sector of LQG. 

Recently,  first results from the application of reduced LQG to recover the dynamics of the inhomogeneous 
vacuum Bianchi I model were announced \cite{ACTux}.  The inhomogeneities
allowed by the construction are those corresponding to space-dependent 
$\tilde{p}_i$ and $\tilde{c}_i$ variables for  Bianchi I cosmology,
\begin{equation}
E^a_i = \tilde{p}_i(t, {\bf x}) \delta^a_i, \ \  \ A^i_a = \tilde{c}_i(t, {\bf x}) \delta^i_a.  
\end{equation}
 
\begin{figure}
\begin{center}
\includegraphics[width=40mm]{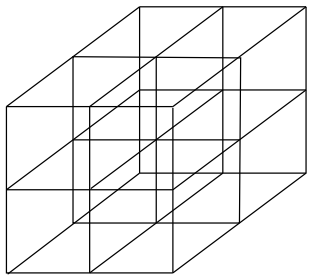}
\caption{Part of a cubic lattice state of  reduced Loop Quantum Gravity. 
The links are labelled by half integers $j=1/2,1,3/2, \dots$, encoding the anisotropy 
of space. The reduced $SU(2)$ intertwiners, defined on the nodes, are encoding
inhomogeneities.}
\label{RedNet}
\end{center}
\end{figure}

To recover the dynamics of this model, the spin network describing the spatial geometry 
was restricted to be a  cubic lattice (See Fig. \ref{RedNet}). 

Due to the imposed symmetry, the $SU(2)$ Gauss constraint effectively decomposes 
in three  $U(1)$ Gauss constraints, each for one spatial direction.  The corresponding 
quantum numbers, labeling the links, are encoding the anisotropy of space. Furthermore,
inhomogeneities are encoded  in the space dependence of $SU(2)$ intertwiners defined 
on the nodes of the spin network. 

An important advantage of the introduced symmetry reduction is that the action of the 
Hamiltonian constraint on the considered spin network states can be explicitly found. 
At least for the Eucildean part of the Hamiltonian, which was considered 
so far. It is worth stressing that the computation of the matrix elements of the Hamiltonian 
acting on general spin network states was not achieved so far. This difficulty comes 
mainly from the problem of determining the action of the volume operator entering 
the Hamiltonian. In the reduced framework, the volume operator is diagonal 
in the considered basis, making the computations possible.  As a promising result 
of this approach, it was shown that the classical Bianchi I Hamiltonian can be recovered 
in the semiclassical limit. As in the case of spinfoam cosmology, the next steps 
of these investigations should shed light on the issue of quantum corrections, 
at least in  first order in $\hbar$.  This could even be used to examine some
ambiguities present in loop quantum cosmology of  Bianchi I model.

\subsection{Deformed algebra of constraints}

The analysis of perturbations in LQC, both with holonomy and inverse-volume corrections, 
revealed a non-trivial structure of space-time encoded in the algebra of constraints (Dirac 
algebra). Namely, the hypersurface deformation algebra is predicted to be deformed with 
respect to the classical one:
\begin{eqnarray}
\left\{D[M^a],D [N^a]\right\} &=& D[M^b\partial_b N^a-N^b\partial_b M^a], \nonumber \\
\left\{D[M^a],S^Q[N]\right\} &=& S^Q[M^a\partial_a N-N\partial_a M^a],  \nonumber  \\
\left\{S^Q[M],S^Q[N]\right\} &=&
D\left[\Omega q^{ab}(M\partial_bN-N\partial_bM)\right], \nonumber
\end{eqnarray} 
where $\Omega$ is a deformation function equal to one for the classical spacetime
with Lorentzian signature.  The functional form of $\Omega$ depends on the
particular type of  quantum corrections under considerations.  

An important question to as ask is whether such a deformation of the algebra of 
constraint is present also at the level of full theory.  The derivation of the deformation 
from LQG would give a firm support  to the results of the effective analysis in LQC.
There is however a serious problem in building such a bridge. Namely, 
in LQG, the operator of spatial diffeomorphisms does not exist. Instead it is incorporated, 
by applying the procedure of group averaging, in the construction of diffeomorphism 
invariant spin network states (s-knots) forming $\mathcal{H}_{\text{kin}}^{D}$. 
The action of the Hamiltonian constraint is defined on these states in the anomaly-free manner. 
The anomaly freedom is  weak only (on-shell):  
\begin{equation}
( \phi |[\hat{S}[M],\hat{S}[N]]| \psi \rangle = 0, 
\label{onshell}
\end{equation}
where $( \phi | \in \mathcal{H}_{\text{kin}}^{D}$ and $| \psi \rangle  \in \mathcal{H}_{\text{kin}}^{G}$ 
are gauge invariant states. 

However, recents studies (See Ref. \cite{Henderson:2012ie}) suggest a new formulation 
of LQG where the off-shell algebra of quantum constraints can be studied. The analysis 
was performed in the simplified setup of 2+1 dimensional $U(1)^3$ gravity 
\cite{Henderson:2012ie, Henderson:2012cu} and four dimensional weak coupling limit 
$G_{N}\rightarrow 0$, where the gauge group $SU(2)$ is reduced to $U(1)^3$ \cite{Tomlin:2012qz}. 
In both cases, it was possible to regain an anomaly-free off-shell algebra of constraint. These new 
achievements open the analysis of the possibility of deformation of the quantum algebra 
of constraints in the general situation. It is worth stressing that this new formulation may 
allow fixing ambiguities of regularizations (Thiemann's Hamiltonian constraint). 
Namely, while the on-shell closure (\ref{onshell}) does not lead to any constraints on the 
form of the Hamiltonian constraint, the off-shell closure can be very restrictive. 
We saw this while studying anomaly-freedom for perturbations in LQC, 
where this allowed us to fix the form of the counterterms and the other free factors 
entering the effective  Hamiltonian. 

\section{Conclusion}

Loop quantum cosmology, as the symmetry reduced version of loop quantum gravity, is now a mature field. 
It is well defined and seems to be consistent. It has provided meaningful predictions for both the background 
evolution and the perturbations. It is at a very exciting stage of development: although the theory is still compatible 
with existing data -- that is predicts no strong deviation with respect to the standard model which account very 
well for observations -- it might soon predict new features that could either be a smoking gun in favor of the theory 
or could disprove it. Up to know, the resolution of singularities, the ``nearly inevitable" correct inflationary background 
behavior and the ability to calculate the propagation of perturbations give good hints in favor of the theory.\\

This is however obviously not the end of game. Several approaches still within LQC are considered and lead to different predictions. 
Each of them is internally consistent but relies on some assumptions with respect to the ``mother theory''. At this stage 
no unquestionable string prediction can be made. This is of course what should be improved in the forthcoming years.\\

\ack
A.B., J.G. and T.C. would like to thank the Perimeter Institute for Theoretical Physics for hospitality. Research at Perimeter Institute is supported by the Government of Canada through Industry Canada and by the Province of Ontario through the Ministry of Economic Development \& Innovation.


\section*{References}
\begin{thebibliography}{07}

\bibitem{carlo1} C. Rovelli and E. Bianchi, arXiv:1002.3966v3.
\bibitem{land} M. R. Douglas, JHEP  {\bf 0305} (2003) 046.
\bibitem{aurel} A. Barrau and J.-L. Nancy, {\it Dans quels mondes vivons nous ?}, Paris, Galil\'ee (2011).
\bibitem{deleuze} G. Deleuze and F. Guattari, {\it A thousand Plateaus}, (1980) trans. B. Massumi, London and New York, Continuum (2004).
\bibitem{ver} E.P. Verlinde, JHEP {\bf 04} (2011) 029.
\bibitem{jac} T. Jacobson, Phys. Rev. Lett. {\bf 75} (1995) 1260.
\bibitem{oriti} D. Oriti (ed.), {\it Quantum Gravity}, Cambridge, Cambridge University Press (2009).
\bibitem{birell} N. D. Birrell and P. C. P. W.  Davies, {Quantum fields in curved space}, Cambridge, 
Cambridge University Press (1982).
\bibitem{singul} S. Hawking and  G. F. R. Ellis, {\it The Large Scale Structure of Space-Time}, Cambridge, 
Cambridge University Press (1973).
\bibitem{kiefer} C. Kiefer, {\it Quantum Gravity (3rd edition)}, Oxford, Oxford University Publishing (2012).
\bibitem{rovelli} C. Rovelli, {\it Quantum Gravity}, Cambridge, Cambridge University Press (2007).
\bibitem{lqg_review} 
A.~Ashtekar and J.~Lewandowski, Class.\ Quant.\ Grav.\  {\bf 21} (2004) R53;
P.~Dona and S.~Speziale, arXiv:1007.0402v1;\\
A.~Perez, arXiv:gr-qc/0409061v3;\\
R. Gambini and J. Pullin, {\it A First Course in Loop Quantum Gravity}, Oxford, Oxford University Press (2011);\\
C.~Rovelli, arXiv:1102.3660v5 [gr-qc];\\
C.~Rovelli, Living Rev. Relativity {\bf 1} (1998) 1;\\
L.~Smolin, arXiv:hep-th/0408048v3;\\
T.~Thiemann, Lect. Notes Phys. {\bf 631} (2003) 41; T.~Thiemann, {\it Modern Canonical Quantum
General Relativity}, Cambridge, Cambridge University Press (2007). 
\bibitem{intro} R. Gambini and J. Pullin, {\it A First Course in Loop Quantum Gravity}, Oxford, Oxford University Press (2011).
\bibitem{lqc_review} 
A.~Ashtekar, M.~Bojowald and J.~Lewandowski, Adv.\ Theor.\ Math.\ Phys.\  {\bf 7} (2003) 233;\\
A.~Ashtekar, Gen. Rel. Grav. {\bf 41} (2009) 707;\\
A.~Ashtekar and P. Singh, Class. Quant. Grav. {\bf 28} (2011) 213001;\\
M.~Bojowald, Living Rev. Rel. {\bf 11} (2008) 4;\\
M.~Bojowald, arXiv:1209.3403 [gr-qc];\\
K. Banerjee, G. Calcagni, and M. Martin-Benito, SIGMA {\bf 8} (2012) 016.
\bibitem{Calcagni:2012vw} G.~Calcagni, Ann.\  Phys.\ {\bf 525} (2013) 323.
\bibitem{bounces}  Y.~-S.~Piao, B.~Feng and X.~-m.~Zhang,  Phys.\ Rev.\ D {\bf 69}, 103520 (2004);\\
 Y.~-S.~Piao,  Phys.\ Rev.\ D {\bf 71}, 087301 (2005);\\
 Y.~-S.~Piao,  Phys.\ Rev.\ D {\bf 70}, 101302 (2004);\\
 Z.~-G.~Liu, Z.~-K.~Guo and Y.~-S.~Piao,  Phys.\ Rev.\ D {\bf 88}, 063539 (2013)
\bibitem{agullo} I. Agullo and A. Corichi, arXiv:1302.3833 [gr-qc].
\bibitem{campi} A. Ashtekar and M. Campiglia, Class. Quantum Grav. {\bf 29} (2012) 242001.
\bibitem{abl} A.~Ashtekar, M.~Bojowald and J.~Lewandowski, Adv. Theo. Math. Phys. {\bf 7} (2003) 233.
\bibitem{acs} A.~Ashtekar, A.~Corichi and P.~Singh, Phys. Rev. D {\bf 77} (2008) 024046.
\bibitem{vt} V.~Taveras, Phys. Rev. D {\bf 78} (2008) 064072.
\bibitem{aps2} A.~Ashtekar, T.~Pawlowski and P.~Singh, Phys. Rev. D {\bf 73} (2006) 124038.
\bibitem{apsv} A.~Ashtekar, T.~Pawlowski, P.~Singh and  K.~Vandersloot, Phys. Rev. D {\bf 75} (2006) 0240035.
\bibitem{jakub} J. Mielczarek, T. Cailleteau, J. Grain and A. Barrau, Phys. Rev. D {\bf 81} (2010) 104049.
\bibitem{abhay} 
A.~Ashtekar and D. Sloan, Phys. Lett. B {\bf 694} (2012) 108;\\
A.~Ashtekar and D. Sloan, Gen. Rel. Grav. {\bf 43} (2011) 3519.
\bibitem{linse} L. Linsefors and A. Barrau, Phys. Rev. D {\bf 87} (2013) 123509.
\bibitem{edcarlo} C. Rovelli, E. Wilson-Ewing, 	arXiv:1310.8654 [gr-qc].
\bibitem{Nelson:2007um} W.~Nelson and M.~Sakellariadou, Phys.\ Rev.\ D {\bf 76} (2007) 104003.
\bibitem{bojo1} M. Bojowald and G. M. Hossain, Phys. Rev. D {\bf 77} (2007) 023508. 
\bibitem{Cailleteau:2011kr} T.~Cailleteau, J.~Mielczarek, A.~Barrau and J.~Grain, Class.\ Quant.\ Grav.\  {\bf 29} (2012) 095010.
\bibitem{Cailleteau:2012fy} T.~Cailleteau, A.~Barrau, J.~Grain and F.~Vidotto, Phys.\ Rev.\ D {\bf 86} (2012) 087301.
\bibitem{Bojowald:2011aa} M.~Bojowald and G.~M.~Paily, Phys.\ Rev.\ D {\bf 86} (2012) 104018.
\bibitem{Mielczarek:2012pf} J.~Mielczarek, arXiv:1207.4657 [gr-qc].
\bibitem{Bojowald:2007hv} M.~Bojowald and G.~M.~Hossain, Class.\ Quant.\ Grav.\  {\bf 24} (2007) 4801.
\bibitem{Mielczarek:2011ph} J.~Mielczarek, T.~Cailleteau, A.~Barrau and J.~Grain, Class.\ Quant.\ Grav.\  {\bf 29} (2012) 085009.
\bibitem{WilsonEwing:2011es} E.~Wilson-Ewing, Class.\ Quant.\ Grav.\  {\bf 29} (2012) 085005.
\bibitem{Hojman:1976vp} S.~A.~Hojman, K.~Kuchar and C.~Teitelboim, Annals Phys.\  {\bf 96} (1976) 88.
\bibitem{Mielczarek:2012tn} J.~Mielczarek, AIP Conf.\ Proc.\  {\bf 1514} (2012) 81.
\bibitem{Mielczarek2013} J.~Mielczarek, arXiv:1311.1344 [gr-qc].
\bibitem{ewe} E. Wilson-Ewing, JCAP {\bf 1308} (2013) 015; \\
E. Wilson-Ewing, JCAP {\bf 1303} (2013) 026 
\bibitem{Thiemann:1996aw} T.~Thiemann, Class.\ Quant.\ Grav.\  {\bf 15} (1998) 839.
\bibitem{Bojowald:2001vw} M.~Bojowald, Phys.\ Rev.\ D {\bf 64} (2001) 084018.
\bibitem{Bojowald:2006zi} M.~Bojowald, H.~H.~Hernandez, M.~Kagan and A.~Skirzewski, Phys.\ Rev.\ D {\bf 75} (2007) 064022.
\bibitem{Bojowald:2002ny} M.~Bojowald, Class.\ Quant.\ Grav.\  {\bf 19} (2002) 5113.
\bibitem{Bojowald:2004ax} M.~Bojowald, Pramana {\bf 63} (2004) 765.
\bibitem{Calcagni:2008ig} G.~Calcagni and G.~M.~Hossain, Adv.\ Sci.\ Lett.\  {\bf 2} (2009) 184.
\bibitem{Ashtekar:2006wn} A.~Ashtekar, T.~Pawlowski and P.~Singh, Phys.\ Rev.\ D {\bf 74} (2006) 084003.
\bibitem{Bojowald:2010me} M.~Bojowald and G.~Calcagni, JCAP {\bf 1103} (2011) 032. 
\bibitem{Bojowald:2007ra} M.~Bojowald, D.~Cartin and G.~Khanna, Phys.\ Rev.\ D {\bf 76} (2007) 064018. 
\bibitem{Bojowald:2008ik} M.~Bojowald, Class.\ Quant.\ Grav.\  {\bf 26} (2009) 075020.
\bibitem{Bojowald:2006qu} M.~Bojowald, Gen.\ Rel.\ Grav.\  {\bf 38} (2006) 1771.
\bibitem{Bojowald:2011iq} M.~Bojowald, G.~Calcagni and S.~Tsujikawa, JCAP {\bf 1111} (2011) 046.
\bibitem{Bojowald:2002nz} M.~Bojowald, Phys.\ Rev.\ Lett.\  {\bf 89} (2002) 261301. 
\bibitem{Tsujikawa:2003vr} S.~Tsujikawa, P.~Singh and R.~Maartens, Class.\ Quant.\ Grav.\  {\bf 21} (2004) 5767.
\bibitem{Bojowald:2004xq} M.~Bojowald, J.~E.~Lidsey, D.~J.~Mulryne, P.~Singh and R.~Tavakol, Phys.\ Rev.\ D {\bf 70} (2004) 043530.
\bibitem{Hossain:2004wm} G.~M.~Hossain, Class.\ Quant.\ Grav.\  {\bf 22} (2005) 2511.
\bibitem{Calcagni:2006pr} G.~Calcagni and M.~Cortes, Class.\ Quant.\ Grav.\  {\bf 24} (2007) 829. 
\bibitem{cope} E.~J.~Copeland, D.~J.~Mulryne, N.~J.~Nunes and M.~Shaeri, Phys.\ Rev.\ D {\bf 77} (2008) 023510.
\bibitem{Shimano:2009tn} M.~Shimano and T.~Harada, Phys.\ Rev.\ D {\bf 80} (2009) 063538. 
\bibitem{Bojowald:2008gz} M.~Bojowald, G.~M.~Hossain, M.~Kagan and S.~Shankaranarayanan, Phys.\ Rev.\ D {\bf 78} (2008) 063547. 
\bibitem{Bojowald:2008jv} M.~Bojowald, G.~M.~Hossain, M.~Kagan and S.~Shankaranarayanan, Phys.\ Rev.\ D {\bf 79} (2009) 043505
  [Erratum-ibid.\ D {\bf 82} (2010) 109903].
\bibitem{cope2} E.~J.~Copeland, D.~J.~Mulryne, N.~J.~Nunes and M.~Shaeri, Phys.\ Rev.\ D {\bf 79} (2009) 023508.
\bibitem{grain_iv_ds} J.~Grain, A.~Barrau and A.~Gorecki, Phys.\ Rev.\ D {\bf 79} (2009) 084015.
\bibitem{Bojowald:2011hd} M.~Bojowald, G.~Calcagni and S.~Tsujikawa, Phys.\ Rev.\ Lett.\  {\bf 107} (2011) 211302.
\bibitem{Calcagni:2011xj} G.~Calcagni, J.\ Phys.\ Conf.\ Ser.\  {\bf 360} (2012) 012027. 
\bibitem{Cailleteautoappear} T.~Cailleteau, L.~Linsefors and A.~Barrau, arXiv:1307.5238 [gr-qc]. 
\bibitem{Langlois:1994ec} D.~Langlois, Class.\ Quant.\ Grav.\  {\bf 11} (1994) 389. 
\bibitem{Cailleteau:2011mi} T.~Cailleteau and A.~Barrau, Phys.\ Rev.\ D {\bf 85} (2012) 123534.
\bibitem{grain_iv_holo_ds} J.~Grain, T.~Cailleteau, A.~Barrau and A.~Gorecki, Phys.\ Rev.\ D {\bf 81} (2012) 024040.
\bibitem{agullo1} I. Agullo, A. Ashtekar and W. Nelson, Phys. Rev. Lett. {\bf 109} (2012) 251301.
\bibitem{agullo2} I. Agullo, A. Ashtekar and W. Nelson, Phys. Rev. D {\bf 87} (2013) 043507.
\bibitem{agullo3} I. Agullo, A. Ashtekar and W. Nelson, Class. Quantum Grav. {\bf 30} (2013) 085014.
\bibitem{ashkalew} A. Ashtekar, W. Kaminski and J. Lewandowski, Phys. Rev. D  {\bf 79} (2009) 064030.
\bibitem{zalda_harari} M. Zaldarriaga and D. D. Harari, Phys. Rev. D {\bf 52} (1995) 3276.
\bibitem{zalda_seljak} U. Seljak and M. Zaldarriaga, Ap. J. {\bf 469} (1996) 437.
\bibitem{grain_cmb} J. Grain, A. Barrau, T. Cailleteau and J. Mielczarek, Phys. Rev. D {\bf 82} (2010) 123520.
\bibitem{camb} A. Lewis, A. Challinor and A. Lasenby, Ap. J. {\bf 538} (2000) 473.
\bibitem{class} D. Blas, J. Lesgourgues and T. Tram, JCAP {\bf 07} (2011) 034.
\bibitem{ma_zhao} Y.-Z. Ma, W. Zhao and M. L. Brown, JCAP {\bf 10} (2010) 007.
\bibitem{komatsu} E. Komatsu {\it et al.}, Ap. J. Supp. {\bf 192} (2011) 18.
\bibitem{bock} J. Bock {\it et al.}, arXiv:0805.4207v1 [astro-ph].
\bibitem{linseforspk} L. Linsefors, T. Cailleteau, A. Barrau and J. Grain, Phys. Rev. D {\bf 87} (2013) 107503.
\bibitem{wmap5} E. Komatsu {\it et al.}, ApJS {\bf 180}  (2009) 330.
\bibitem{fisher}
	L.~Verde, H.~Peiris and R.~Jimenez, JCAP {\bf 0601} (2006) 019; \\
	D.~Baumann {\it et al.}, AIP Conf. Proc. {\bf 1141} (2009) 10.
\bibitem{stivoli} F. Stivoli, J. Grain, S. Leach, M. Tristram, C. Baccigalupi and R. Stompor, MNRAS {\bf 408} (2010) 2319.
\bibitem{epic} J.~Bock {\it et al.}, arXiv:0805.4207v1 [astro-ph].
\bibitem{datas}
	E.~Komatsu {\it et al.}, Astrophys.\ J.\ Suppl.\ {\bf 192}  (2011) 18; \\
	B.~A.~Reid {\it et al.}, Mon.\ Not.\ Roy.\ Astron.\ Soc.\ {\bf 404} (2010) 60; \\
	A.G.~Riess {\it et al.}, Astrophys.\ J.\ {\bf 699} (2009) 539; \\
	M.~Kowalski {\it et al.}, Astrophys.\ J.\ {\bf 686} (2008) 749; \\
	S.~Burles and D.~Tytler, Astrophys.\ J.\ {\bf 499} (1998) 699.
\bibitem{planck}
	Planck Collaboration: P. A. R. Ade, N. Aghanim {\it et al.}, arXiv:1303.5075; \\
	Planck Collaboration: P. A. R. Ade, N. Aghanim {\it et al.}, arXiv:1303.5076; \\
	Planck Collaboration: P. A. R. Ade, N. Aghanim {\it et al.}, arXiv:1303.5082.
\bibitem{klm} L.A. Kofman, A. Linde and V.F. Mukhanov,  High Energy Phys. {\bf 10} (2002) 057.
\bibitem{gt} G.W. Gibbons and N. Turok, Phys. Rev. D  {\bf 77} (2008) 063516.
\bibitem{stein} P. Steinhart, Scientific American, p. 37, April 2011 Issue.
\bibitem{first} C. Germani, W. Nelson and M. Sakellariadou, Phys. Rev. D {\bf 76} (2007) 043529.
\bibitem{Corichi:2010zp}  A.~Corichi and A.~Karami,  Phys.\ Rev.\ D {\bf 83} (2011) 104006.
\bibitem{bojo_prl} M. Bojowald, Phys. Rev. Lett. {\bf 100} (2008) 221301.
\bibitem{bh} A. Ashtekar, J. Baez, A. Corichi and K. Krasnov, Phys. Rev. Lett. {\bf 80} (1998) 904.
\bibitem{Mielczarek:2010ga} J.~Mielczarek, M.~Kamionka, A.~Kurek and M.~Szydlowski, JCAP  {\bf 1007} (2010) 004.
\bibitem{bianchiI_lqc1} D. Cartin and G. Khanna, Phys. Rev. Lett. {\bf 94} (2005) 111302.
\bibitem{bianchiI_lqc2} G. Date, Phys. Rev. D {\bf 72} (2005) 067301.
\bibitem{bianchiI_lqc3} D. Cartin and G. Khanna, Phys. Rev. D {\bf 72} (2005) 084008.
\bibitem{bianchiI_lqc4} D.-W. Chiou, Phys. Rev. D {\bf 75} (2007) 024029.
\bibitem{bianchiI_lqc5} D.-W. Chiou, arXiv:gr-qc/0703010.
\bibitem{bianchiI_lqc6} D.-W. Chiou and K. Vandersloot, Phys. Rev. D  {\bf 76} (2007) 084015.
\bibitem{bianchiI_lqc7} D.-W. Chiou, Phys. Rev. D {\bf 76} (2007) 124037.
\bibitem{bianchiI_lqc8} L. Szulc, Phys. Rev. D {\bf 78} (2008) 064035.
\bibitem{bianchiI_lqc9} M. Martin-Benito, G.A. Mena Marugan and T. Pawlowski, Phys. Rev. D {\bf 78} (2008) 064008.
\bibitem{bianchiI_lqc10} D.-W. Chiou, arXiv:0812.0921 [gr-qc].
\bibitem{bianchiI_lqc11} P. Dzierzak and W. Piechocki, Phys. Rev. D {\bf 80} (2009) 124033.
\bibitem{bianchiI_lqc12} M. Martin-Benito, G. A.Mena Marugan and T. Pawlowski, Phys. Rev. D {\bf 80} (2009) 084038.
\bibitem{bianchiI_lqc13} A. Ashtekar and E. Wilson-Ewing, Phys. Rev. D {\bf 79} (2009) 083535.
\bibitem{bianchiI_lqc14} P. Dzierzak, W. Piechocki, Annalen Phys. {\bf 19} (2012) 290.
\bibitem{bianchiI_lqc15} P. Malkiewicz, W. Piechocki and P. Dzierzak, Class. Quant. Grav. {\bf 28} (2011) 085020.
\bibitem{bianchiI_lqc16} M. Martin-Benito, L.J. Garay, G.A. Mena Marugan and E. Wilson-Ewing, J. Phys. Conf. Ser. {\bf 360} (2012) 012031.
\bibitem{bianchiI_lqc17} V. Rikhvitsky, B. Saha and M. Visinescu, Astrophys. Space Sci. {\bf 339} (2012) 371.
\bibitem{bianchiI_lqc18} P. Singh, Phys. Rev. D {\bf 85} (2012) 104011.
\bibitem{bianchiI_lqc19} F. Cianfrani, A. Marchini and G. Montani, Europhys. Lett.{\bf 99} (2012) 10003.
\bibitem{bianchiI_lqc20} K.Fujio and T. Futamase, Phys. Rev. D {\bf 85} (2012) 124002.
\bibitem{bianchiI_lqc21} P. Singh, J. Phys. Conf. Ser. {\bf 360} (2012) 012008.
\bibitem{Gupt:2013swa} B.~Gupt and P.~Singh, arXiv:1304.7686 [gr-qc].
\bibitem{linda_anisotropy} L.~Linsefors and A.~Barrau, arXiv:1305.4516 [gr-qc]. 
\bibitem{Rovelli:2011eq} C.~Rovelli, PoS QGQGS {\bf 2011} (2011) 003.
\bibitem{Rovelli:2008aa} C.~Rovelli and F.~Vidotto, Class.\ Quant.\ Grav.\  {\bf 25} (2008) 225024.
\bibitem{Bianchi:2010zs} E.~Bianchi, C.~Rovelli and F.~Vidotto, Phys.\ Rev.\ D {\bf 82} (2010) 084035.
\bibitem{Battisti:2009kp} M.~V.~Battisti, A.~Marciano and C.~Rovelli, Phys.\ Rev.\ D {\bf 81} (2010) 064019.
\bibitem{Bianchi:2011ym} E.~Bianchi, T.~Krajewski, C.~Rovelli and F.~Vidotto, Phys.\ Rev.\ D {\bf 83} (2011) 104015.
\bibitem{vidotto2013} C.~Rovelli and F.~Vidotto, arXiv:1307.3228 [gr-qc].
\bibitem{gielen} S. Gielen, D.~Oriti, and L.~Sindoni, Phys.\ Rev.\ Lett. {\bf 11} (2013) 031301.
\bibitem{Alesci:2012md} E.~Alesci and F.~Cianfrani, arXiv:1210.4504 [gr-qc].
\bibitem{Alesci:2013xd} E.~Alesci and F.~Cianfrani, arXiv:1301.2245 [gr-qc].
\bibitem{ACTux} E.~Alesci and F.~Cianfrani, ``Quantum-Reduced Loop Gravity," EFI  winter conference on canonical 
and covariant LQG, 25.02-01.03.2013,  Tux, Austria.
\bibitem{Henderson:2012ie} A.~Henderson, A.~Laddha and C.~Tomlin, arXiv:1204.0211 [gr-qc].
\bibitem{Henderson:2012cu} A.~Henderson, A.~Laddha and C.~Tomlin, arXiv:1210.3960 [gr-qc].
\bibitem{Tomlin:2012qz} C.~Tomlin and M.~Varadarajan, Phys.\ Rev.\ D {\bf 87} (2013) 044039.



\end{thebibliography}
\end{document}